\documentclass [12pt]{article}
\usepackage{amsmath,amssymb,cite}
\usepackage{appendix}
\usepackage{feynmp}
\usepackage{float}
\usepackage{subfigure}
\usepackage[vcentermath,enableskew]{youngtab}
\usepackage{tabularx}
\usepackage[english]{babel}
\usepackage{graphicx}
\usepackage{indentfirst}
\usepackage{epsfig}
\usepackage{color}
\usepackage{epstopdf}
\usepackage[]{hyperref}
\usepackage[section]{placeins}
\usepackage[stable]{footmisc}
\usepackage{appendix}
\usepackage{tabularx}
\usepackage[english]{babel}
\usepackage{graphicx}
\usepackage{indentfirst}
\usepackage{epsfig}
\usepackage{slashed}
\usepackage{fancyhdr}
\numberwithin{equation}{section}

\fancypagestyle{plain}{\fancyhead{}
}

\begin{document}

\title{
Novel Phenomena in Noncommutative Field Theory: Emergent Geometry  \\
}

\author{
Badis Ydri\\
\textit{Department of Physics, Faculty of Sciences,}\\
\textit{Badji Mokhtar–Annaba University, Annaba, Algeria}
}
\date{October 28, 2024}

\maketitle
\begin{abstract}
  Noncommutative ﬁeld theory (NCFT) \cite{Ydri2017,Ydri2018} is an extension of quantum ﬁeld
theory (QFT) that redeﬁnes spacetime, replacing commuting coordinates with a
noncommutative structure. This shift fundamentally alters the way ﬁelds, interactions, and symmetries are understood. NCFT uniquely integrates with supersymmetry, making it a natural framework for unifying quantum mechanics and
gravity. It also provides a consistent mechanism for spontaneous supersymmetry
breaking.
Unlike conventional QFT, which quantizes ﬁelds on a ﬁxed spacetime, NCFT
begins by quantizing spacetime itself. This perspective reveals novel phenomena,
such as ultraviolet-infrared mixing and a natural transition from discrete to
continuous geometries. It offers insights into quantum gravity at the Planck scale.
Mathematically, NCFT bridges quantum mechanics and geometry through
operator algebras, enabling the exploration of new theories unattainable in traditional frameworks. This dual role cements NCFT as a cornerstone of modern
theoretical physics.

\noindent\textbf{Note.}

This work is published as \textbf{Chapter 18}, \emph{``Novel phenomena in noncommutative field theory: emergent geometry''}, in
\emph{A Modern Course in Quantum Field Theory, Volume 2 (Second Edition)},
IOP Publishing (2025).
ISBN~978-0-7503-5834-7.
DOI~10.1088/978-0-7503-5834-7.

\medskip
\noindent
The presentation is comprehensive and pedagogical in nature, and may contain textual overlap with earlier preprints by the author; its aim is to provide a unified and hopefully more fundamental treatment connecting results and viewpoints not previously treated together.

\medskip
\noindent
The published chapter is available at the publisher’s website:
\href{https://iopscience.iop.org/book/mono/978-0-7503-5834-7/chapter/bk978-0-7503-5834-7ch18}{IOP Science (Chapter 18)}.

  \end{abstract}

\newpage

\tableofcontents

\section{Introduction}
Noncommutative field theory is quantum field theory defined on a noncommutative spacetime which is primarily  
characterized by two fundamental features:
\begin{enumerate}
\item A Planck-like length scale (noncommutativity parameter). 
\item A discrete geometry which becomes continuous at large distances (commutative limit). 
\end{enumerate}
Noncommutative field theory and the underlying noncommutative spacetime require for their description the language of noncommutative geometry \cite{connes,Connes:1996gi}.  In this context the geometry is in a precise sense emergent (here from operator algebras) given in terms of a spectral triple $({\cal A}, \Delta, {\cal H})$ rather than in terms of a set of points. The algebra ${\cal A}$ is typically an operator operator represented on the Hilbert space ${\cal H}$  whereas $\Delta$ is the Laplace operator which defines the metric aspects of the geometry. We really should think of the algebra ${\cal A}$ as defining the topology whereas the Laplacian $\Delta$ defines the metric. And in the non-perturbative formulation of the spectral triple  $({\cal A}, \Delta, {\cal H})$ as a path or functional integral we require the language of matrix models.



A noncommutative scalar field theory is given by an action functional which can be split in the usual fashion into a kinetic part and a potential term. This action functional is defined non-perturbatively by means of a matrix/operator model (after proper regularization and Euclidean rotation). The potential term represents the algebraic structure, i.e. the algebra, the inner product, the Hilbert space, etc and as such it defines in some sense the topological aspects of the noncommutative space.  The kinetic term on the other hand, as it involves a Laplacian operator $\Delta$, plays a pivotal role in defining the metric aspects of the geometry following Fr\"{o}hlich and Gaw\c{e}dzki \cite{FroehlichGawedzki} (or Connes \cite{connes} for spin geometry).

Thus, a noncommutative scalar field theory on some underlying noncommutative spacetime (assumed to be a Groenewold-Moyal-Weyl spacetime \cite{weyl,Moyal:1949skv2,Groenewold:1946kpv2} for concreteness) is given generically by a matrix model of the form (with a single Hermitian scalar field  $\Phi=\Phi^{\dagger}\in {\cal A}$)
\begin{eqnarray}
S=a {\rm Tr}\Phi\Delta\Phi+{\rm Tr}V(\Phi)~,~\Delta(..)= \alpha [X_a,[X_a,..]].\label{matrix}
\end{eqnarray}
The trace ${\rm Tr}$ is over an infinite dimensional Hilbert space ${\cal H}$. We can generally regularize the Moyal-Weyl spacetime using noncommutative tori  \cite{rieffel,Ambjorn:2000cs} or fuzzy spheres \cite{Hoppe,Madore:1991bw}.

The operators $X_a$ entering the definition of the Laplacian operator $\Delta$ are the coordinate operators which define operationally the noncommutative spacetime. For example, the Moyal-Weyl spacetime is precisely defined in terms of the operators $X_a$ by the Heisenberg relations 
\begin{eqnarray}
[X_a,X_b]=i\Theta_{ab}~,~\Theta_{ab}={\rm constant}.\label{MY}
\end{eqnarray}
The Groenewold-Moyal-Weyl spacetime is the most important noncommutative spacetime as it plays for Poisson manifolds (via Darboux theorem) the same role played by flat Minkowski spacetime for curved manifolds, i.e. Darboux theorem is effectively the equivalence principle in this case as noted for example in \cite{Lee:2010zf1,Blaschke:2010ye}.

The most important interaction term is given by a quartic phi-four potential, viz
\begin{eqnarray}
V(\phi)=b \Phi^2+c \Phi^4.\label{phi-four}
\end{eqnarray}
The phase structure of the matrix model (\ref{matrix}), with the potential (\ref{phi-four}), was calculated non-perturbatively mostly in two dimensions on the fuzzy sphere (but also on the noncommutative torus) using the Monte Carlo method \cite{GarciaFlores:2009hf,Martin:2004un,Panero:2006bx,Das:2007gm}.   It is believed that the phase diagram of noncommutative phi-four theory in any dimension and on any noncommutative background is identical in features to the phase diagram of the potential (\ref{phi-four}) on the fuzzy sphere.    

The first result here is the emergence of a novel phase, termed the stripe or non-uniform-ordered phase, in addition to the usual commutative phases (disordered and uniform-ordered phases), and hence the emergence of a triple point in the phase diagram where the three vacuum solutions coexist. The non-uniform-ordered phase might also be called the matrix phase as it is generated in the real quartic matrix model  (\ref{phi-four}) alone.

The rich phase structure characterizing a noncommutative scalar field theory is of great implications for both particle physics, e.g. spontaneous symmetry breaking and condensed matter systems, e.g. exotic phases of matter and quantum phase transitions.

A far more important result is the emergence of geometry itself in the multitrace matrix models which were originally proposed as approximations of noncommutative scalar field theory on the fuzzy sphere  \cite{O'Connor:2007ea,Saemann:2010bw}. These models fall in the "universality class" of the real quartic matrix model  (\ref{phi-four}) and as such their phase structure is only characterized by disordered and non-uniform-ordered phases.

A priori, a generic multitrace matrix model does not require for its definition a spectral triple and thus it does not entail any geometric content. However, by tuning the parameters of the model appropriately, a uniform-ordered phase might emerge signaling the emergence of an underlying geometry. The associated critical exponents can then be used to determine its dimension while the behavior of the Wigner's semi-circle law near the origin can be used to determine its metric. Furthermore, by expanding around the uniform-ordered phase, we get a noncommutative gauge theory on the emergent space.

Thus, in the context of noncommutative scalar field theories, the emergence of the non-uniform-ordered stripe phase is important from the perspective of quantum field theory but the emergence of the uniform-ordered phase, in their multitrace matrix models, is important from the perspective of quantum gravity.

The phenomena of emergent geometry  does also occur in the Yang-Mills matrix models associated with noncommutative gauge theories.  The noncommutative Moyal-Weyl space (\ref{MY}) should be though of as a background solution of some Yang-Mills matrix model. First, we start from the operator model given by 
\begin{eqnarray}
S=\frac{\sqrt{{\Theta}^D{\rm det}(\pi B)}}{2g^2}{\rm Tr}_{\cal H}\sum_{a,b=1}^D\bigg(i[{D}_a,{D}_b]-\frac{1}{{\Theta}}B^{-1}_{ab}\bigg)^2.\label{actionBin}
\end{eqnarray}
Here, the degrees of freedom are given by the connections or covariant derivatives  $D_a$  which are Hermitian operators. The noncommutativity parameter ${\Theta}$ is of dimension length-squared so that the connection operators ${D}_a$ are of dimension $({\rm length})^{-1}$. The gauge coupling constant $g$ is of dimension $(\rm mass)^{2-\frac{D}{2}}$ while $B^{-1}$ is  a dimensionless invertible tensor which in $2$ dimensions is given by $B^{-1}_{ij}={\epsilon}^{-1}_{ij}=-{\epsilon}_{ij}$ while in higher even dimensions is given by (using Darboux theorem)
\begin{eqnarray}
B^{-1}_{ij}=\left(\begin{array}{ccccccc}
-{\epsilon}_{ij}&&&&&&\\
&.&&&&&\\
&&&&.&&\\
&&&&&&-{\epsilon}_{ij}
\end{array}\right).\label{symp}
\end{eqnarray}
The  Moyal-Weyl space (\ref{MY}) correspond to the classical configurations ${D}_a=- B^{-1}_{ab} {X}_b$ with $\Theta_{ab}=\Theta B_{ab}$. Then, by computing the effective potential in the configuration ${D}_a=-\varphi B^{-1}_{ab} {X}_b$ we verify that the Moyal-Weyl space itself, i.e. the Heisenberg algebra  (\ref{MY}) ceases to exist above a certain value $g_*$  of the gauge coupling constant.  Conversely, by moving in the phase diagram from strong coupling to weak coupling the geometry of the Moyal-Weyl spaces (star product and the Weyl map)  emerges at the critical point $g_*$.

This discussion of emergent geometry can be made more precise by considering a natural regularization of the Moyal-Weyl space which involves the matrix algebra ${\rm Mat}_N$ of $N\times N$ Hermitian matrices. In other words, we must regularize the action (\ref{actionBin}) by $i)$ replacing the operators ${D}_a$ by $N\times N$ Hermitian matrices and $ii)$ replacing the trace ${\rm Tr}_{\cal H}$ by the ordinary trace ${\rm Tr}$ on the finite-dimensional Hilbert space associated with the matrix algebra ${\rm Mat}_N$.

The regularization of choice in this work is given by fuzzy projective spaces \cite{Balachandran:2001dd}. In particular, the fuzzy sphere will regularize effectively all noncommutative Moyal-Weyl spaces as it is evident from the symplectic structure (\ref{symp}), i.e. each two-dimensional block/plane is regularized by a fuzzy sphere and we get therefore a Cartesian products of fuzzy spheres approximating even-dimensional Moyal-Weyl spaces.

In this regularization, the Heisenberg algebra  (\ref{MY}) in two dimensions will be regularized by the $SU(2)$ angular momentum algebra in the spin $s\equiv (N-1)/2$ irreducible representation given by $[L_i,L_j]=i\epsilon_{ijk}L_k$, $\sum_iL_i^2=c_2$, $c_2=(N^2-1)/2$. This gives immediately the fuzzy sphere which is defined by three $N\times N$ matrices $\hat{x}_i=RL_i/\sqrt{c_2}$ playing the role of coordinates operators and satisfying commutation relations and embedding constraint given by 
\begin{eqnarray}
[\hat{x}_i,\hat{x}_j]=i\Theta{\epsilon}_{ijk}\hat{x}_k~,~\sum_i\hat{x}_i^2=R^2~,~\Theta=\frac{R}{\sqrt{c_2}}.
\end{eqnarray}
The Yang-Mills action (\ref{actionBin}) in two dimensions will then be regularized by a genuine $D=3$ Yang-Mills matrix model of the IKKT-type \cite{Ishibashi:1996xs2} given explicitly by 
\begin{eqnarray}
  S=\frac{1}{g^2N}{\rm Tr}\sum_{i,j=1}^3\big(i[D_i,D_j]+\sum_{k=1}^3\epsilon_{ijk}D_k\big)^2.\label{actionBin1}
\end{eqnarray}
This is the basic model whose degrees of freedom are $N\times N$ Hermitian matrices $D_i$ which should be interpreted  as covariant matrix coordinates. This  model can be immediately generalized by adding mass deformation terms in the matrices $D_i$. It can also be generalized to a supersymmetric $D=3$ Yang-Mills matrix model which is the fundamental $SO(3)$symmetric theory with Euclidean action given by
\begin{eqnarray}
S=\frac{1}{g^2}{\rm Tr}\bigg[-\frac{1}{4}[X_{A},X_{B}]^2-\frac{1}{2}\bar{\Psi}{\gamma}^{A}[X_{A},\Psi]\bigg].
\end{eqnarray}
This supersymmetric model can also be defined in $D=10$, $D=6$ and $D=4$. In fact, the partition functions of these models are  convergent only in these dimensions $D=4,6,10$ \cite{Krauth:1999qw,Krauth:1998yu,Austing:2001pk,Austing:2003kd,Krauth:1998xh,Austing:2001bd,Austing:2001ib}. In $D=3$ the partition function can be made finite by adding appropriate mass deformation which consists of a positive quadratic term in the matrices $X_{A}$ which damps flat directions.

However, Yang-Mills matrix models in $D=10$ and $D=6$ have their  $SO(D)$  rotational symmetry group spontaneously broken in the large $N$ limit to $SO(3)$ \cite{Nishimura:2011xy3} while the $D=4$ case is equivalent to the $D=3$ model coupled to a scalar field $\Phi$  \cite{Ydri:2014rea}. In other words, we are very interested in the $D=3$ Yang-Mills matrix model (\ref{actionBin1}), its supersymmetric extension, generic mass deformation terms and Cartesian products thereof.

The emergent geometry transition manifests itself in the $D=3$ Yang-Mills matrix model (\ref{actionBin1}) as an exotic phase transition where the geometry disappears as the temperature is increased. The fuzzy sphere phase (low temperature phase where $D_i=\varphi L_i$ and $\varphi\sim 1$) has the background
geometry of a two dimensional spherical noncommutative manifold which macroscopically becomes a standard commutative sphere for $N\longrightarrow\infty$. The fluctuations are then of a noncommutative gauge
theory which mixes with a normal scalar field on this background. In the Yang-Mills matrix phase (high temperature phase) the order parameter $\varphi$ is not well defined and the fluctuations are
around diagonal matrices so the model is a pure matrix one corresponding to a zero-dimensional
Yang-Mills theory in the large $N$ limit. The fuzzy sphere phase occurs for $\tilde{\alpha} >\tilde{\alpha}_*$ while the
matrix phase occurs for  $\tilde{\alpha} <\tilde{\alpha}_*$ where $\tilde{\alpha}^4=1/g^2$.

The programme of emergent geometry from Yang-Mills matrix models revolves therefore around the two hypotheses:
\begin{enumerate}
\item The existence of a noncommutative structure with a Planck-like length scale corresponding to a discrete geometry. In other words, noncommutative geometry is thought of as "first quantization" of the geometry.
  \item The existence of a Yang-Mills matrix action functional where the noncommutative structure appears as its classical background solution. The Feynman path integral associated with this action defines therefore "second quantization" of the geometry where the noncommutative structure becomes dynamical, i.e. this structure can emerge/evaporates from/into a pure matrix model as some parameters in the model are varied.
  \end{enumerate}
Emergent geometry in this programme is therefore defined as "quantum noncommutative geometry".

As it turns out, emergent geometry in the scalar multitrace matrix models is more fundamental than emergent geometry in the gauge Yang-Mills matrix model. This is simply because a full-blown $D=3$ Yang-Mills matrix model can emerge in the uniform-ordered phase of a cleverly-engineered multitrace matrix model.

\begin{figure}[H]
\begin{center}
\subfigure[Landau-Lifshitz scalar field theory.]
{
\includegraphics[width=8.0cm,angle=0]{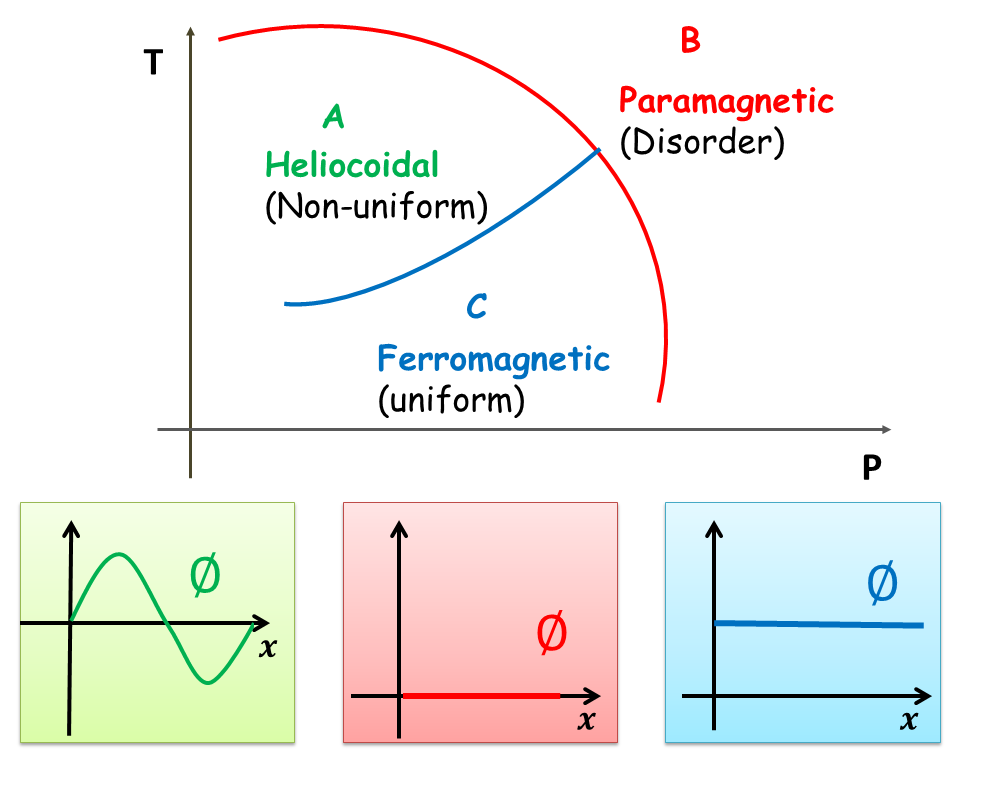}
}
\subfigure[Causal dynamical triangulation.]
{
\includegraphics[width=8.0cm,angle=0]{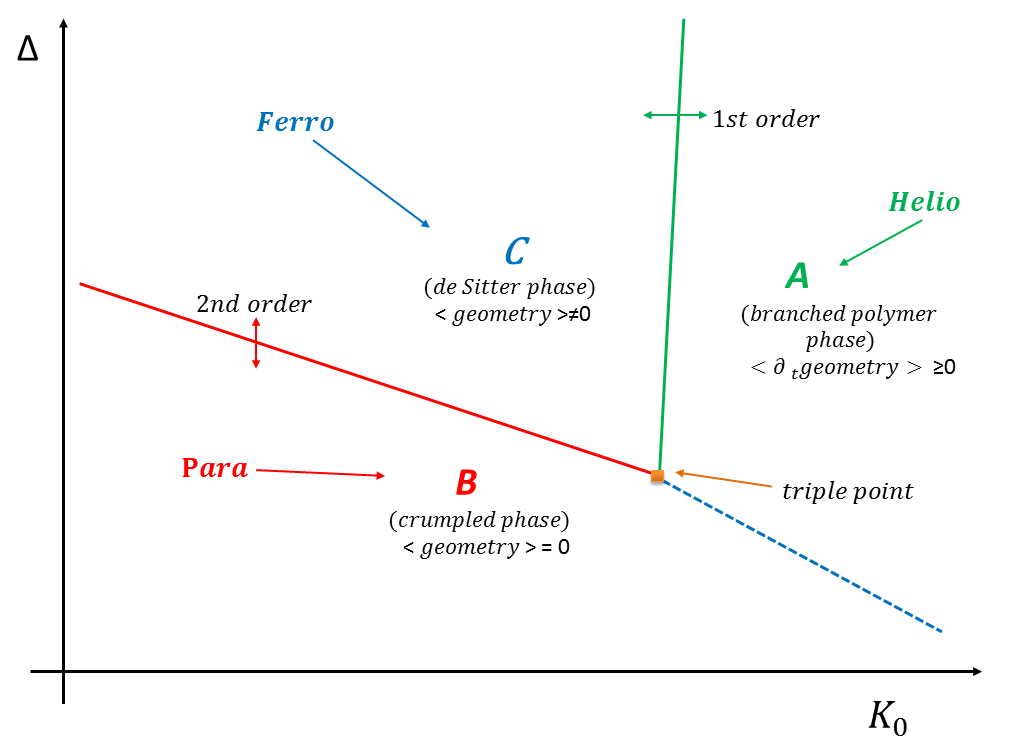}
}
\subfigure[Multitrace matrix model.]
{
\includegraphics[width=12.0cm,angle=0]{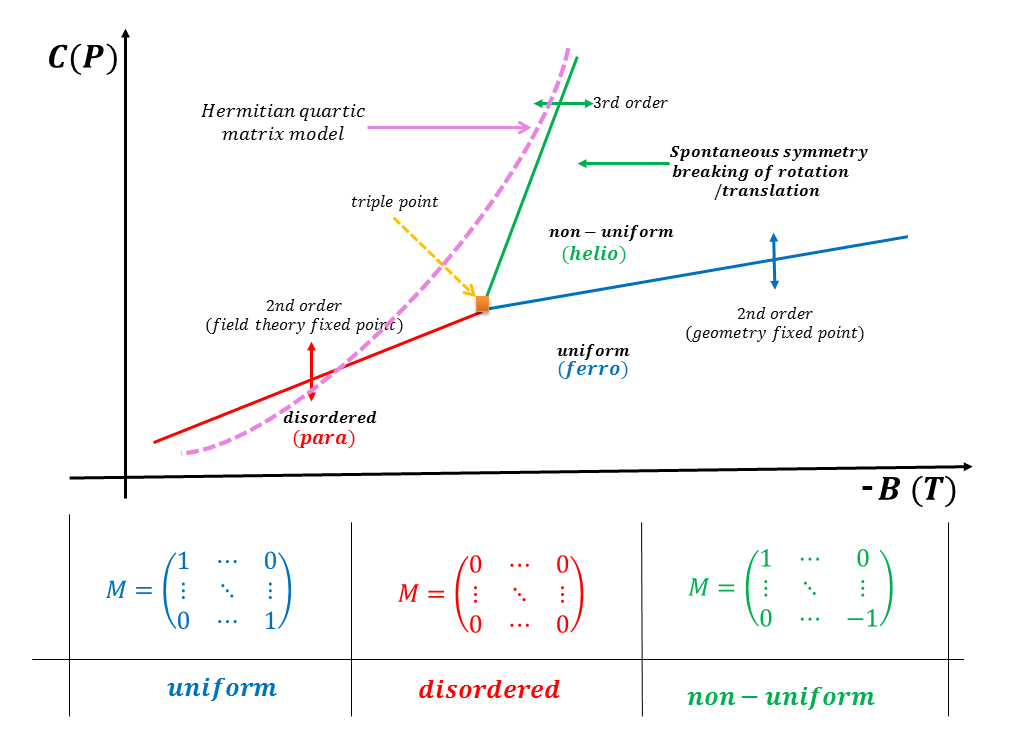}
}
\end{center}
\caption{The phase diagrams of causal dynamical triangulation, Lifshitz scalar field theory and multitrace matrix model.}
\label{CDT}
\end{figure}

This approach to emergent geometry should also be contrasted with other approaches to quantum geometry. For example, see \cite{Bombelli:1987aa,Dowker:2005tz,Thiemann:2007zz,Rovelli:2004,Maldacena:1997re,Seiberg:2006wf}. In particular, causal dynamical triangulation (CDT) \cite{Ambjorn:2005qt,Ambjorn:2006hu,Ambjorn:2007jv, Ambjorn:2010hu,Gorlich:2011ga} and Horava-Lifshitz (HL) gravity \cite{Horava:2009if,Horava:2008ih,Horava:2009uw} are very similar in structure to the multitrace matrix model approach. See figure (\ref{CDT}).

The remainder of this work is organized as follows. In section $2$, we discuss aspects of noncommutative phi-four theory on the Moyal-Weyl space. In section $3$, we discuss aspects of noncommutative phi-four theory on the fuzzy sphere. In section $4$, we discuss emergent geometry from multitrace matrix models. In section $5$, we discuss emergent geometry from Yang-Mills matrix models.

\section{Aspects of noncommutative phi-four theory on the Moyal-Weyl space}

\subsection{UV-IR mixing and stripe phase on noncommutative ${\bf R}^d_{\theta}$}

Noncommutativity is naturally described in terms of operator algebras and operators can be intuitively viewed (albeit in a rough sense) as matrix algebras. It is believed  that noncommutativity is the only non-trivial extension of supersymmetry and thus noncommutative field theory provides a natural framework for spontaneous supersymmetry breaking which is central to most particle physics beyond the standard model.

The simplest (at least formally) noncommutative spacetime is the Moyal-Weyl space ${\bf R}^d_{\theta}$ \cite{weyl,Moyal:1949skv2,Groenewold:1946kpv2}. Two nonperturbative regularizations of the Moyal-Weyl space exist. The noncommutative torus obtained via Eguchi-Kawai construction \cite{rieffel,Ambjorn:2000cs} and the fuzzy sphere obtained via co-adjoint orbit quantization \cite{Groenewold:1946kpv2,Madore:1991bw}.

The coordinates on ${\bf R}^d_{\theta}$  are operators which satisfy 
\begin{eqnarray}
[x_{\mu},x_{\nu}]=i{\theta}_{\mu\nu}.
\end{eqnarray}
The Weyl map allows us to work with a $C^*-$algebra of functions where the pointwise multiplication of function is replaced by an appropriate deformed (star) product. The Moyal-Weyl star product is given by

\begin{eqnarray}
f*g(x)=e^{\frac{i}{2}\sum_{i,j=1}^d{\theta}_{ij}\frac{\partial}{{\partial}{\xi}_i}\frac{\partial}{{\partial}{\eta}_j}}f(x+\xi)g(x+\eta)|_{\xi=\eta=0}.
\end{eqnarray}
The noncommutative phi-four theory is then given by the action
\begin{eqnarray}
S=\int d^dx\bigg[ (\partial_{\mu}\phi)^2+\mu^2\phi^2+\frac{\lambda}{4!}(\phi*\phi)^2\bigg].
\end{eqnarray}
This theory suffers from the UV-IR mixing problem which hampers perturbative renormalization of the theory \cite{Minwalla:1999px}. There seems to exist a unique renormalizable solution in which the propagator is altered by the addition of a harmonic oscillator term \cite{Grosse:2003nw}. A self-consistent Hartree-Fock analysis of the above field theory \cite{Gubser:2000cd} reveals a complicated phase structure beyond the single critical line present in the commutative theory.

\subsubsection{UV-IR mixing}
\begin{figure}[htbp]
\begin{center}
\includegraphics[width=10cm]{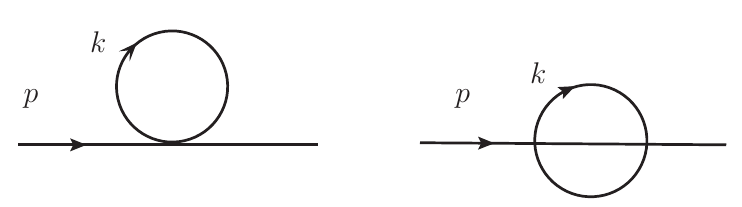}
\caption{The one-loop planar and non-planar contributions.}\label{pd0}
\end{center}
\end{figure}

The two-point function is given explicitly by
\begin{eqnarray}
{\Gamma}^{(2)}(p)=p^2+{\mu}^2+2\frac{\lambda}{4!}I_2(0,\Lambda)+\frac{\lambda}{4!}I_2(p,\Lambda).
\end{eqnarray}
The second and third terms correspond to the planar and non-planar diagrams respectively.  See figure (\ref{pd0}). These contributions are given by
\begin{eqnarray}
I_2(p,\Lambda)&=&\int \frac{d^dk}{(2\pi)^d}\frac{1}{k^2+{\mu}^2}~e^{-i{\theta}_{ij}k_ip_j}\nonumber\\
&=&\frac{1}{(4\pi)^{\frac{d}{2}}}\int_0^{\infty} \frac{d\alpha}{{\alpha}^{\frac{d}{2}}}~e^{-\frac{1}{\alpha {\Lambda}^2}}~e^{-\frac{({\theta}_{\
ij}p_j)^2}{4\alpha}-\alpha{\mu}^2}\nonumber\\
&=&(2\pi)^{-\frac{d}{2}}{\mu}^{\frac{d-2}{2}}\big(\frac{4}{{\Lambda}_{\rm eff}^2}\big)^{\frac{2-d}{4}}K_{\frac{d-2}{2}}\big(\frac{2\mu}{{\Lambda}_{\rm eff}}\big).
\end{eqnarray}
Here, $K_{\frac{d-2}{2}}$ is the modified Bessel function. The effective cutoff is defined by the equation
\begin{eqnarray}
\frac{4}{{\Lambda}_{\rm eff}^2}=\frac{4}{{\Lambda}^2}+({\theta}_{ij}p_j)^2.
\end{eqnarray}
\paragraph{$4-$Dimensions:}
\begin{eqnarray}
I_2(p,\Lambda)=\frac{\mu}{4{\pi}^2}\frac{{\Lambda}_{\rm eff}}{2}K_{1}\big(\frac{2\mu}{{\Lambda}_{\rm eff}}\big)~,~K_1(z)=\frac{1}{z}+\frac{z}{2}\ln\frac{z}{2}+...\Rightarrow I_2(p,\Lambda)=\frac{1}{16{\pi}^2}\bigg({\Lambda}_{\rm eff}^2-{\mu}^2\ln \frac{{\Lambda}_{\rm eff}^2}{{\mu}^2}+...\bigg).\nonumber\\
\end{eqnarray}

\paragraph{$2-$Dimensions:}
\begin{eqnarray}
I_2(p,\Lambda)=\frac{1}{2{\pi}}K_{0}\big(\frac{2\mu}{{\Lambda}_{\rm eff}}\big)~,~K_0(z)=-\ln\frac{z}{2}+...\Rightarrow I_2(p,\Lambda)=\frac{1}{4{\pi}}\ln \frac{{\Lambda}_{\rm eff}^2}{{\mu}^2}+...
\end{eqnarray}
In the limit $\Lambda\longrightarrow \infty$ the non-planar one-loop contribution remains finite whereas the planar one-loop contribution diverges quadratically/logarithmically as usual. However, the two-point function ${\Gamma}^{(2)}(p)$ which can be made finite in the limit $\Lambda\longrightarrow \infty$ through the introduction of the renormalized mass $m^2={\mu}^2+2\frac{\lambda}{4!}I_2(0,\Lambda)$ is singular in the limit $p\longrightarrow 0$ or $\theta\longrightarrow 0$. This is because the effective cutoff ${\Lambda}_{\rm eff}=2/|{\theta}_{ij}p_j|$ diverges as  $p\longrightarrow 0$ or $\theta\longrightarrow 0$. This is the celebrated UV-IR mixing problem discussed originally in \cite{Minwalla:1999px}. 
\subsubsection{Stripe phase}

The self-consistent $2-$point proper vertex in the disordered phase is found to be given by (with $m^2$ being the renormalized mass and $g^2=\lambda/4!$)
\begin{eqnarray}
{\Gamma}^{(2)}(p)
&=&p^2+m^2+g^2\int \frac{d^dk}{(2\pi)^d} \frac{ e^{-i{\theta}_{ij}k_ip_j}}{{\Gamma}^{(2)}(k)}.
\end{eqnarray}  
We can infer the existence of a minimum $p_c$ in  ${\Gamma}^{(2)}(p)$ from the behavior at $p\longrightarrow 0$ and at $p\longrightarrow \infty$ of ${\Gamma}^{(2)}(p)$ given by
\begin{eqnarray}
{\Gamma}^{(2)}(p)=p^2~,{\rm for}~p~{\rm large}.
\end{eqnarray}
\begin{eqnarray}
{\Gamma}^{(2)}(p)\propto \frac{g^2}{{\theta}^2p^2}~,{\rm for}~p~{\rm small}.
\end{eqnarray}
Around the minimum $p_c$ we can write  ${\Gamma}^{(2)}(p)$ as
\begin{eqnarray}
{\Gamma}^{(2)}(p)={\xi}_0^2(p^2-p_c^2)^2+r~,{\rm for}~p\simeq p_c.\label{aroundmini}
\end{eqnarray}
Explicitly, we find in the disordered phase the result (with $\alpha\propto g^{7/2}/\theta^{3/2} $, $\tau=2p_c^2+m^2$)
\begin{eqnarray}
p_c=\sqrt{\frac{g}{2\pi\theta}}~,~{\Gamma}^{(2)}(p_c)\equiv r=\tau+\frac{\alpha}{\sqrt{r}}~,~\xi_0=\frac{1}{2p_c}.
\end{eqnarray} 
Next, we consider the ordered phase. Here, we must perform perturbation around a stripe configuration $\phi_0$ which breaks translational invariance, i.e. we expand the scalar field as $\phi=\phi_0+\tilde{\phi}+X$ where $X$ is the quantum fluctuation field. The   stripe configuration $\phi_0$ is given explicitly by 
\begin{eqnarray}
{\phi}_0=A\cos  p_cx.
\end{eqnarray}
The scalar field is then expanded as $\phi=\phi_0+\varphi$.  The condition of vanishing tadpole graphs gives then either $A=0$ (disordered phase) or 
\begin{eqnarray}
\text{Tadpole}=0\Leftrightarrow r=4A^2g^2~,~{\rm ordered}~{\rm phase}.\label{16}
\end{eqnarray}
In this case the   $2-$point proper vertex is  given by 
\begin{eqnarray}
{\Gamma}^{(2)}(p_c)\equiv r=\tau+\frac{\alpha}{\sqrt{r}}+6g^2A^2.\label{17}
\end{eqnarray}
Equations (\ref{16}) and (\ref{17}) imply that there is an ordered solution iff
\begin{eqnarray}
\tau+\frac{\alpha}{\sqrt{r}}+6g^2A^2=4A^2g^2\Rightarrow 2\tau+\frac{2\alpha}{\sqrt{r_0}}+r_0=0\Rightarrow \epsilon\equiv -\frac{1}{\sqrt{3}}\frac{{\tau}^3}{{\alpha}^2}>{\epsilon}_{*1}=\sqrt{3}\frac{9}{8}.
\end{eqnarray}
This should be contrasted with the corresponding equation in the disordered phase given by
\begin{eqnarray}
\tau+\frac{\alpha}{\sqrt{r}}=r\Rightarrow \tau+\frac{\alpha}{\sqrt{r_d}}-r_d=0.
\end{eqnarray}
However, the transition between the two phases is given by the condition that the free energy difference must vanish, i.e. this is a first order transition. We have

\begin{eqnarray}
\frac{dF}{dA}=\frac{d}{d\tilde{\phi}_c}\text{Tadople}&=&2A\big(r-4g^2A^2\big)\nonumber\\
A&=&\frac{1}{2g}\sqrt{\frac{2}{3}}\sqrt{r-\tau-\frac{\alpha}{\sqrt{r}}}\nonumber\\
\Rightarrow F_d-F_o&=&\frac{1}{12g^2}(\sqrt{r_d}-\sqrt{r_o})\bigg(3\alpha+\tau(\sqrt{r_d}+\sqrt{r_o})\bigg).
\end{eqnarray}
We find then the transition point
\begin{eqnarray}
3r_o^2+6r_d^2-4\tau({r_d}-{r_o})\Rightarrow{\epsilon}_{*2} =c \sqrt{3}\frac{9}{8}.
\end{eqnarray}
The coefficient $c>1$ must be determined from numeric. 

The region ${\epsilon}_{*1}<\epsilon<{\epsilon}_{*2}$ corresponds to the disordered phase whereas $\epsilon>{\epsilon}_{*2}$ corresponds to the ordered phase.

The transition point reads in terms of $m$ and $g$ as follows
\begin{eqnarray}
{m}_{*2}^2 =-\frac{g}{\pi\theta}-c^{'}\frac{g^{\frac{7}{3}}}{\theta}.
\end{eqnarray}
The  coefficient $c^{'}$ depends on $c$ and on the proportionality factor between  $\alpha$ and $ g^{7/2}/\theta^{3/2} $.
The system avoids the  second order behaviour by precisely an amount proportional  to $g^{\frac{7}{3}}/\theta^{3/2}$. It is important to compare the second order behaviour ${m}_{*2}^2 =-{2}{g}/{\theta}$ with the critical point of real quartic matrix models. The phenomena of a phase transition of an isotropic system to a non-uniform phase was in fact realized a long time ago by Brazovkii \cite{brazovkii}.

\begin{figure}[htbp]
\begin{center}
\includegraphics[width=15cm]{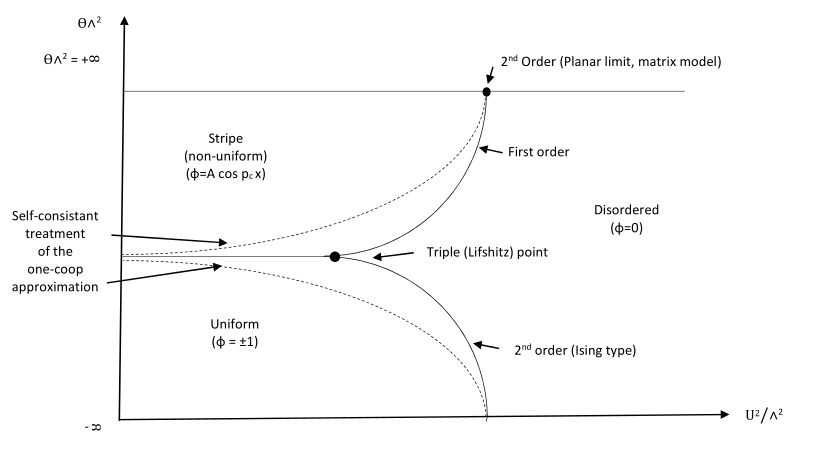}
\caption{The phase diagram of noncommutative phi-four in $d=4$ at fixed $\lambda\sim g^2$.}\label{pd1}
\end{center}
\end{figure}

For more detail see \cite{Gubser:2000cd}. See also figure (\ref{pd1}).


\subsubsection{Dispersion relations on the noncommutative torus}


The phase diagram and the dispersion relation of noncommutative phi-four theory on the noncommutative torus in  $d=3$ is also discussed using the Monte Carlo method in \cite{Bietenholz:2004xs}. Here, the noncommutative torus should be thought of as a regularization of the Moyal-Weyl space. See also  \cite{Ambjorn:2002nj,Mejia-Diaz:2014lza}. The phase diagram, with exactly the same qualitative features conjectured by Gubser and Sondhi  \cite{Gubser:2000cd}, is shown on figure $5$ of \cite{Bietenholz:2004xs}. Three phases which meet at a triple point are identified. The Ising (disordered-to-uniform) transition exists for small $\theta$ whereas transitions to the stripe phase (disordered-to-stripe and uniform-to-stripe) are favored at large $\theta$. The collapsed parameters in this case are found to be given by
\begin{eqnarray}
N^2m^2~,~N^2\lambda.
\end{eqnarray}
On the other hand, the dispersion relations are computed as usual from the exponential decay of the correlation function
\begin{eqnarray}
\frac{1}{T}\sum_t\langle \tilde{\Phi}^*(\vec{p},t)\tilde{\Phi}(\vec{p},t+\tau)\rangle.
\end{eqnarray}
This behaves as $\exp(-E(\vec{p})\tau)$ for large $\tau$ and thus we can extract the energy $E(\vec{p})$ by computing the above correlator as a function of $\tau$. In the disordered phase, i.e. for small $\lambda$ near the uniform phase, we find the usual linear behavior $E(\vec{p})=a\vec{p}^2$ and thus in this region the model looks like its commutative counterpart. As we increase $\lambda$ we observe that the rest energy $E_0\equiv E(\vec{0})$ increases, followed by a sharp dip at some small value of the momentum $\vec{p}^2$, then the energy rises  again with $\vec{p}^2$ and approaches asymptotically the linear behavior   $E(\vec{p})=a\vec{p}^2$ for $\vec{p}^2\longrightarrow \infty$. An example of a dispersion relation near the stripe phase, for $m^2=-15$ and $\lambda=50$, is shown on figure $14$ of \cite{Bietenholz:2004xs} with a fit given by
\begin{eqnarray}
E^2(\vec{p})=c_0\vec{p}^2+m^2+\frac{c_1}{\sqrt{\vec{p}^2+\bar{m}^2}}\exp(-c_2\sqrt{\vec{p}^2+\bar{m}^2}).
\end{eqnarray}
The parameters $c_i$ and $\bar{m}^2$ are given by equation $(6.2)$  of \cite{Bietenholz:2004xs}. The minimum in this case occurs around the cases $k=N|\vec{p}|/2\pi=\sqrt{2},2,\sqrt{5}$ so this corresponds actually to a multi-stripe pattern.

The above behavior of the dispersion relations stabilizes in the continuum limit defined by the double scaling limit $N\longrightarrow \infty$ (planar limit), $a\longrightarrow 0$ ($m^2\longrightarrow m_c^2=-15.01(8)$) keeping $\lambda$ and $\theta=N a^2/\pi$ fixed. In this limit the rest energy $E_0$ is found to be divergent, linearly with $\sqrt{N}\propto 1/a$, in full agreement with the UV-IR mixing.

The shifting of the energy minimum to a finite non-vanishing value of the momentum in this limit indicates the formation of a stable stripe phase in the continuum noncommutative theory. The existence of a continuum limit is also a strong indication that the theory is non-perturbatively renormalizable.

\subsubsection{The Grosse-Wulkenhaar model}


For simplicity, we consider a $2$-dimensional noncommutative space ${\bf R}^2_{\theta}$.  We introduce noncommutativity in momentum space by introducing a minimal coupling to a constant background  magnetic field $B_{ij}$. The derivation operators become
\begin{eqnarray}
&&\hat{D}_i=\hat{\partial}_i-iB_{ij}X_j~,~\hat{C}_i=\hat{\partial}_i+iB_{ij}X_j~,~X_i=\frac{\hat{x}_i+\hat{x}_i^R}{2}.
\end{eqnarray}
Instead of the conventional Laplacian ${\Delta}=(-\hat{\partial}_i^{2}+\mu^2)/2$ we will consider the generalized Laplacians
\begin{eqnarray}
{\Delta}&=&-\sigma \hat{D}_i^{2}-\tilde{\sigma}\hat{C}_i^{2}+\frac{\mu^2}{2}\nonumber\\
&=&-(\sigma+\tilde{\sigma})\hat{\partial}_i^{2}+(\sigma-\tilde{\sigma})iB_{ij}\{X_j,\hat{\partial}_i\}-(\sigma+\tilde{\sigma})(B^{2})_{ij}X_iX_j+\frac{\mu^2}{2}.\label{Delta}
\end{eqnarray}
The case $\sigma=\tilde{\sigma}$  corresponds to the Grosse-Wulkenhaar model \cite{Grosse:2003aj,Grosse:2004yu,Grosse:2003nw}, while the model $\sigma=1, \tilde{\sigma}=0$ corresponds to Langmann-Szabo-Zarembo model considered in \cite{Langmann:2002cc}.


We introduce two sets of creation and annihilation operators $(\hat{a},\hat{a}^{\dagger})$ and  $(\hat{b},\hat{b}^{\dagger})$ by the relations (with ${\theta}_0=\theta/2$)

\begin{eqnarray}
\hat{a}=\frac{1}{2}({\sqrt{{\theta}_0}}\hat{\partial}+\frac{1}{\sqrt{{\theta}_0}}{Z}^{\dagger})~,~\hat{a}^{\dagger}=\frac{1}{2}({\sqrt{{\theta}_0}}\hat{\partial}^{\dagger}+\frac{1}{\sqrt{{\theta}_0}}{Z}).
\end{eqnarray}
\begin{eqnarray}
\hat{b}=\frac{1}{2}(-{\sqrt{{\theta}_0}}\hat{\partial}^{\dagger}+\frac{1}{\sqrt{{\theta}_0}}{Z})~,~\hat{b}^{\dagger}=\frac{1}{2}(-{\sqrt{{\theta}_0}}\hat{\partial}+\frac{1}{\sqrt{{\theta}_0}}{Z}^{\dagger}).
\end{eqnarray}
In these equations we have also used the definitions
\begin{eqnarray}
Z=X_1+iX_2~,~Z^{\dagger}=X_1-iX_2~,~\hat{\partial}=\hat{\partial}_1-i\hat{\partial}_2~,~\hat{\partial}^+=-\hat{\partial}_1-i\hat{\partial}_2.
\end{eqnarray}
We expand the scalar field operator $\hat{\Phi}$ as follows
\begin{eqnarray}
\hat{\Phi}=\sum_{l,m=1}^{\infty}M_{lm}\hat{\phi}_{l,m}~,~\hat{\phi}_{l,m}=|l\rangle \langle m|.
\end{eqnarray}
The infinite dimensional matrix $M$ should be thought of  as a compact operator acting on some separable Hilbert space ${\bf H}_1$ of Schwartz sequences  with sufficiently rapid decrease \cite{Langmann:2003if}. This  in particular will guarantee the convergence of the expansion of the scalar operator $\hat{\Phi}$. 

In the operators $\hat{\phi}_{l,m}=|l\rangle \langle m|$ we can identify the kets $|l\rangle$ with the states of the harmonic oscillator operators $\hat{a}$ and $\hat{a}^{\dagger}$ whereas the bras $\langle m|$ can be identified with the states of the harmonic oscillator operators $\hat{b}$ and $\hat{b}^{\dagger}$. More precisely, the operators $\hat{\phi}_{l,m}$ are in one-to-one correspondence with the wave functions ${\phi}_{l,m}(x)=\langle x|l,m\rangle$ which are known as the Landau states \cite{Gracia-Bondia:1987ssw,Langmann:2002ai}.

The field/operator Weyl map is then given by
\begin{eqnarray}
\sqrt{2\pi{\theta}}~{\phi}_{l_1,m_1}\leftrightarrow \hat{\phi}_{l_1,m_1}.
\end{eqnarray}
\begin{eqnarray}
\int d^2x \leftrightarrow \sqrt{\det(2\pi{\theta})}{\rm Tr}_{\cal H}.
\end{eqnarray}
\begin{eqnarray}
{\Phi}=\sqrt{2\pi{\theta}}\sum_{l,m=1}^{\infty}M_{lm}{\phi}_{l,m}\leftrightarrow \hat{\Phi}.
\end{eqnarray}
\begin{eqnarray}
{\Phi}*{\Phi}^{'}=\sqrt{2\pi{\theta}}\sum_{l,m=1}^{\infty}(MM^{'})_{lm}{\phi}_{l,m}\leftrightarrow \hat{\Phi}\hat{\Phi}^{'}.
\end{eqnarray}
In other words, the star product is mapped to the operator product as it should be. Furthermore, the differential operators $\hat{D}_i^{2}$ and  $\hat{C}_i^{2}$ will be represented in the star picture by the differential operators
\begin{eqnarray}
&&D_i={\partial}_i-iB_{ij}x_j~,~{C}_i={\partial}_i+iB_{ij}x_j.
\end{eqnarray}
The Landau states are actually eigenstates of the Laplacians $D_i^{2}$, and  ${C}_i^{2}$ at the special point 
\begin{eqnarray}
B^{2}{\theta}_0^{2}\equiv \frac{B^2\theta^2}{4}=1.
\end{eqnarray}
The most general single-trace action with a phi-four interaction on a non-commutative ${\bf R}_{\theta}^d$ under the effect of a magnetic field is then given by 
\begin{eqnarray}
S&=&\int d^{2}x \bigg[{\Phi}^{\dagger}\big(-\sigma D_i^{2}-\tilde{\sigma}{C}_i^{2}+\frac{{\mu}^{2}}{2}\big){\Phi}+\frac{\lambda}{4!}{\Phi}^{\dagger}*{\Phi}*{\Phi}^{\dagger}*{\Phi}\bigg]\nonumber\\
&=&\sqrt{\det(2\pi{\theta})}{\rm Tr}_{\cal H}\bigg[\hat{\Phi}^{\dagger}\big(-\sigma \hat{D}_i^{2}-\tilde{\sigma}\hat{C}_i^{2}+\frac{{\mu}^{2}}{2}\big)\hat{\Phi}+\frac{\lambda}{4!}\hat{\Phi}^{\dagger}\hat{\Phi}~\hat{\Phi}^{\dagger}\hat{\Phi}\bigg].
\end{eqnarray}
This action enjoys a remarkable symmetry under a duality transformation which exchanges positions and momenta. Explicitly, this duality transformation under which the action retains the same form reads \cite{Langmann:2003if,Langmann:2002cc,Langmann:2003cg}

\begin{eqnarray}
x_i\leftrightarrow \tilde{k}_i=B^{-1}_{ij}k_j~,~\Phi(x)\leftrightarrow \bar{\Phi}(\tilde{k})=\sqrt{|{\rm det}\frac{B}{2\pi}|}\tilde{\Phi}(B\tilde{k}).
\end{eqnarray}
\begin{eqnarray}
\theta\leftrightarrow \bar{\theta}=-B^{-1}\theta^{-1}B^{-1}~,~
\lambda \leftrightarrow \bar{\lambda}=\frac{\lambda}{|{\rm det}B\theta|}.
\end{eqnarray}
Furthermore, we can re-express the above action in terms of the compact operators $M$ and $M^{\dagger}$ as follows (with $\alpha=1+B\theta_0$, $\beta=1-B\theta_0$)
\begin{eqnarray}
        S&=&\frac{\sqrt{\det(2\pi{\theta})}}{\theta_0}\bigg[-(\sigma+\tilde{\sigma})\alpha\beta {\rm Tr}_{{\bf H}_1}\big({\Gamma}^{+}M^{+}\Gamma M+M^{+}{\Gamma}^{+}M\Gamma \big)+(\sigma\alpha^{2}+\tilde{\sigma}\beta^{2}){\rm Tr}_{{\bf H}_1}M^{+}EM\nonumber\\
&+&(\sigma\beta^{2}+\tilde{\sigma}\alpha^{2}){\rm Tr}_{{\bf H}_1}MEM^{+}+\frac{{\mu}^{2}{\theta}_0}{2} {\rm Tr}_{{\bf H}_1}M^{+}M+\frac{\lambda {\theta}_0}{4!}Tr_{{\bf H}_1}M^{+}MM^{+}M\bigg].\label{S1}
\end{eqnarray}
The infinite dimensional matrices $\Gamma$ and $E$ are defined by
\begin{eqnarray}
(\Gamma)_{lm}=\sqrt{m-1}{\delta}_{lm-1}~,~(E)_{lm}=(l-\frac{1}{2}){\delta}_{lm}.
\end{eqnarray}
Now, we regularize the theory by taking $M$ to be an $N\times N$ matrix.   The states ${\phi}_{l,m}(x)$, with  $l,m < N$, where $N$ is some large integer, correspond to simultaneous  cut-offs in position and momentum spaces \cite{Grosse:2003nw}. The infrared cut-off is found to be proportional to $R=\sqrt{2\theta N}$ while the UV cut-off is found to be proportional to $\Lambda=\sqrt{8N/\theta}$. 

The so-called Grosse-Wulkenhaar model corresponds to the values $\sigma=\tilde{\sigma}\neq 0$ so that the mixing term in (\ref{Delta})  cancels. This contains, compared with the usual case, a harmonic oscillator term in the Laplacian which modifies and thus allows us to control the IR behavior of the theory. This model is perturbatively renormalizable which makes it the more interesting. We consider, without any loss of generality,  $\sigma=\tilde{\sigma}=1/4$.  We obtain therefore the action 
\begin{eqnarray}
S&=&\int d^{2}x \bigg[{\Phi}^{\dagger}\bigg(-\frac{1}{2}{\partial}_i^2+\frac{1}{2}{\Omega}^2\tilde{x}_i^2+\frac{{\mu}^{2}}{2}\bigg){\Phi}+\frac{\lambda}{4!}{\Phi}^{\dagger}*{\Phi}*{\Phi}^{\dagger}*{\Phi}\bigg]\nonumber\\
&=&\sqrt{\det(2\pi{\theta})}{\rm Tr}_{\cal H}\bigg[\hat{\Phi}^{\dagger}\bigg(-\frac{1}{2}\hat{\partial}_i^2+\frac{1}{2}\Omega^2\tilde{X}_i^2+\frac{{\mu}^2}{2}\bigg)\hat{\Phi}+\frac{\lambda}{4!}\hat{\Phi}^{\dagger}\hat{\Phi}~\hat{\Phi}^{\dagger}\hat{\Phi}\bigg].
\end{eqnarray}
Here,  $\tilde{x}_i=2({\theta}^{-1})_{ij}x_j$ and the parameter $\Omega$ is defined by 
\begin{eqnarray}
B{\theta}=2{\Omega}.
\end{eqnarray}
The above action is also covariant under a duality transformation which exchanges positions and momenta as $x_i\leftrightarrow \tilde{p}_i=B^{-1}_{ij}p_j$. The value ${\Omega}^2=1$ gives an action which is actually invariant under this duality transformation.

In the Landau basis, the above action reads


\begin{eqnarray}
S&=&\frac{\nu_2}{\theta}\bigg[(\Omega^2-1) Tr_{H}\big({\Gamma}^{+}M^{+}\Gamma M+M^{+}{\Gamma}^{+}M\Gamma \big)+(\Omega^2+1)Tr_{H}(M^{+}EM+MEM^{+})\nonumber\\
&+&\frac{{\mu}^{2}{\theta}}{2} Tr_{H}M^{+}M+\frac{\lambda {\theta}}{4!}Tr_{H}M^{+}MM^{+}M\bigg].\label{S2GW}
\end{eqnarray}
This is a special case of (\ref{S1}). Equivalently
\begin{eqnarray}
S&=&\nu_2~\sum_{m,n,k,l}\bigg(\frac{1}{2}(M^+)_{mn}G_{mn,kl}M_{kl}+\frac{\lambda }{4!}(M^+)_{mn}M_{nk}(M^+)_{kl}M_{lm}\bigg).\label{S2e}
\end{eqnarray}
\begin{eqnarray}
G_{mn,kl}&=&\big({\mu}^2+{\mu}_1^2(m+n-1)\big){\delta}_{n,k}{\delta}_{m,l}-{\mu}_1^2\sqrt{\omega (m-1)(n-1)}~{\delta}_{n-1,k}{\delta}_{m-1,l}\nonumber\\
&-&{\mu}_1^2\sqrt{\omega  m  n}~{\delta}_{n+1,k}{\delta}_{m+1,l}.
\end{eqnarray}
The parameters of the model are $\mu^2$, $\lambda$ and
\begin{eqnarray}
\nu_2=\sqrt{\det(2\pi{\theta})}~,~\mu_1^2=2(\Omega^2+1)/\theta~,~\sqrt{\omega}=(\Omega^2-1)/(\Omega^2+1).
\end{eqnarray}
However, there are only three independent coupling constants in this theory which we can take to be  $\mu^2$, $\lambda$, and $\Omega^2$.

\subsection{Wilson renormalization group recursion formula}

\subsubsection{The Wilson-Fisher fixed point in noncommutative phi-four}
In this section we will apply the renormalization group recursion formula of Wilson \cite{Wilson:1973jj} to noncommutative phi-four theory on the Moyal-Weyl space ${\bf R}^d_{\theta}$. This formula was applied in \cite{Ferretti:1995zn,Nishigaki:1996ts} to vector and hermitian matrix models in the large $N$ limit. Their method can be summarized as follows:
\begin{itemize}
\item[$1)$]We split the field into a background and a fluctuation and then integrate the fluctuation obtaining therefore an effective action for the background field alone. 

\item[$2)$]We keep, following Wilson, only induced corrections to the terms that are already present in the classical action. Thus, we will only need to calculate quantum corrections to the $2$- and $4$-point functions. 
\item[$3)$]We perform the so-called Wilson contraction which consists in estimating momentum loop integrals using the following three approximations or rules:
\begin{itemize}
\item{}{\bf Rule} $1$: All external momenta which are wedged with internal momenta will be set to zero.
\item{}{\bf Rule} $2$: We approximate every internal propagator $\Delta({k})$ by $\Delta({\lambda})$ where ${\lambda}$ is a typical momentum in the range $\rho\Lambda\leq\lambda\leq \Lambda$. 
\item{}{\bf Rule} $3$: We replace every internal momentum loop integral $\int_{{k}}$ by a typical volume.
 \end{itemize}
The two last approximations are equivalent to the reduction of all loop integrals to their zero-dimensional counterparts. These two approximations are quite natural in the limit $\rho\longrightarrow 1$. 

 As it turns out we do not need to use the first approximation in estimating the $2$-point function. In fact rule $1$ was proposed first in the context of a non-commutative  $\Phi^4$ theory in \cite{Chen:2001an} in order to simply the calculation of the $4$-point function. In some sense the first approximation is equivalent to taking the limit $\bar{\theta}=\theta\Lambda^2\longrightarrow 0$.

\item[$4)$]The last step in the renormalization group program of Wilson consists in rescaling the momenta so that the cutoff is restored to its original value. We can then obtain renormalization group recursion equations which relate the new values of the coupling constants to the old values.
\end{itemize}  

The action we will study is given by
\begin{eqnarray}
S[\Phi]=\int d^dx \bigg[\Phi(-{\partial}_i^2+{\mu}^2)\Phi+\frac{\lambda}{4!}\Phi_*^4\bigg]=S_0+S_4.
\end{eqnarray}
We introduce a sharp momentum cut-off $\Lambda$. We introduce the modes with low and high momenta  by (with  $0\leq b\leq1$)
\begin{eqnarray}
\phi(k)\equiv {\phi}_L(k)~,~k\leq b\Lambda.
\end{eqnarray}
\begin{eqnarray}
\phi(k)\equiv {\phi}_H(k)~,~b\Lambda\leq k\leq \Lambda.
\end{eqnarray}
By integrating over the modes ${\Phi}_H(k)$ in the path integral we obtain 

\begin{eqnarray}
Z=\int d\phi ~e^{-S_0[\phi]-S_4[\phi]}&=&\int d{\phi}_L~d{\phi}_H~e^{-S_0[{\phi}_L]-S_0[{\phi}_H]-S_4[{\phi}_L,{\phi}_H]}\nonumber\\
&=&\int d{\phi}_L~e^{-S_0[{\phi}_L]}~e^{-S_4^{'}[{\phi}_L]}\nonumber\\
e^{-S_4^{'}[{\phi}_L]}&=&\int d{\phi}_H~e^{-S_0[{\phi}_H]-S_4[{\phi}_L,{\phi}_H]}\propto  \langle e^{-S_4[{\phi}_L,{\phi}_H]}\rangle_{0H}.
\end{eqnarray}
The expectation value $\langle ...\rangle_{0H}$ is taken with respect to the probability distribution $e^{-S_0[{\phi}_H]}$. We verify the identity (cumulant expansion)
\begin{eqnarray}
\langle e^{-S_4[{\phi}_L,{\phi}_H]}\rangle_{0H}&=& \exp\bigg[-\langle S_4[{\phi}_L,{\phi}_H]\rangle_{0H}+\frac{1}{2}\bigg(\langle S_4^2[{\phi}_L,{\phi}_H]\rangle_{0H}-\langle S_4[{\phi}_L,{\phi}_H]\rangle _{0H}^2\bigg)\bigg]\nonumber\\
&=&\exp\bigg[-S_4[{\phi}_L]-\langle \delta S[{\phi}_L,{\phi}_H]\rangle_{0H}+\frac{1}{2}\bigg(\langle {\delta S}^2[{\phi}_L,{\phi}_H]\rangle _{0H}-\langle \delta S [{\phi}_L,{\phi}_H]\rangle _{0H}^2\bigg)\bigg].\nonumber\\
\end{eqnarray}
The action $\delta S$ is defined by $\delta S[{\phi}_L,{\phi}_H]={\delta}S_1+\delta S_2+\delta S_3+\delta S_4$ where $\delta S_i$ involves $i$ fields $\phi_H$ and $4-i$ fields $\phi_L$.

The correction to the $2$-point function is found to be given by (with $n=(d-2)/2$, $K_d=S_d/(2\pi)^d$,  ${\mu}^2=r{\Lambda}^2$ and $g=\lambda {\Lambda}^{d-4}$)
\begin{eqnarray}
\Delta {\Gamma}_2(p)
&=&\frac{gK_d{\Lambda}^2}{12(1+r)}(1-b) \bigg[1+2^{n-1}n!\frac{J_{n}(\theta \Lambda p)}{(\theta \Lambda p)^{n}}\bigg].\label{zz}
\end{eqnarray}
The corrected mass parameter and its renormalization group equation are then given by (with the mass parameter $r$ assumed to be near $0$ whereas $b$ is by construction near $1$)
\begin{eqnarray}
r^{'}&=&\frac{r}{b^2}-\frac{gK_d}{12}\frac{1}{1+r}\frac{\ln b}{b^2} \bigg[1+2^{n-1}n!\frac{J_{n}(b\theta \Lambda p)}{(b\theta \Lambda p)^{n}}\bigg]_{p=0}\nonumber\\
&=&\frac{r}{b^2}-\frac{gK_d}{8}(1-r)\frac{\ln b}{b^2}\nonumber\\
\Rightarrow b\frac{d r}{db}&=&(-2+\frac{g^{}K_d}{8})r^{'}-\frac{g^{}K_d}{8}
\end{eqnarray}
In this equation we have also employed the renormalization group transformations of the action of the background field $\phi_L$ which are given by 

\begin{eqnarray}
&&p\longrightarrow p^{'}=\frac{p}{b}\nonumber\\
&&{\phi}_L(p)\longrightarrow {\phi}_L^{'}(p^{'})=b^{\frac{d}{2}+1}{\phi}_L(p)\nonumber\\
&&{\mu}^{'2}=\frac{{\mu}^2}{b^2}\Leftrightarrow r^{'2}=\frac{r^2}{b^2}\nonumber\\
&&{\lambda}^{'}=b^{-4+d}\lambda\Leftrightarrow g^{'}=b^{-4+d}g\nonumber\\
&&{\theta}^{'}=b^{2}\theta.
\end{eqnarray}
The above result (\ref{zz}) gives also the wave function renormalization which is contained in the $p$-dependent part of the quadratic term. Explicitly, we have
\begin{eqnarray}
\big(1+\frac{gK_d(\theta\Lambda^2)^2}{192}\ln b\big)\int_ {p\leq b{\Lambda}}{\phi}_L(p){\phi}_L(-p) p^2=\int_ {p^{'}\leq {\Lambda}}{\phi}_L^{'}(p^{'}){\phi}_L^{'}(-p^{'}) p^{'2}~,~{\phi}_L^{'}(p^{'})=b^{\frac{d+2-\gamma}{2}}{\phi}_L^{}(p^{}).\nonumber\\
\end{eqnarray}
Thus, we have obtained a negative wave function renormalization which signals an instability in the theory. Indeed, the anomalous dimension $\gamma$ is negative given explicitly by
\begin{eqnarray}
\gamma=-\frac{gK_d(\theta\Lambda^2)^2}{192}<0.
\end{eqnarray}
This is also $\theta$-dependent. This result also implies a novel behavior for the $2$-point function which must behave as
\begin{eqnarray}
\langle \phi(x)\phi(0)\rangle \sim \frac{1}{|x|^{d-2+\gamma}}.
\end{eqnarray}
This should vanish for large distances as it should be as long as $d-2+\gamma>0$. Thus, for large values of $\theta$ we get an instability because $d-2+\gamma$ becomes negative. The critical value of $\theta$ is precisely given by
 \begin{eqnarray}
d-2+\gamma_c=0\Rightarrow (\theta_c\Lambda^2)^2=\frac{196(d-2)}{gK_d}.
\end{eqnarray}
The corrected quartic coupling and its renormalization group equation are found to be given by (by assuming the external momenta to be very small compared to the cutoff)
\begin{eqnarray}
g^{'}&=&b^{d-4}\bigg(g +g^2\frac{3K_d}{8(1+r)^2}\ln b \bigg)\nonumber\\
&=&b^{d-4}\bigg(g +g^2\frac{3K_d}{8}(1-2r)\ln b \bigg)\nonumber\\
\Rightarrow b\frac{dg^{'}}{db}&=&(d-4)g^{'}+\frac{3g^{'2}K_d}{8}.
\end{eqnarray}
The fixed points are then given by the equation
\begin{eqnarray}
0=(-2+\frac{g_*K_d}{8})r_*-\frac{g_*K_d}{8}.
\end{eqnarray}
\begin{eqnarray}
0=(d-4)g_*+\frac{3g_*^{2}K_d}{8}.
\end{eqnarray}
We get immediately the two solutions (in dimension $d<4$ with small $\epsilon=4-d$)

\begin{eqnarray}
r_*=g_*=0~,~\text{trivial (Gaussian) fixed point},
\end{eqnarray}
\begin{eqnarray}
r_*=-\frac{\epsilon}{6}~,~g_*=\frac{64\pi^2\epsilon}{3}~,~\text{Wilson-Fisher~fixed~point}.
\end{eqnarray}
The critical exponent $\nu$ is given by the usual value whereas the critical exponent $\eta$ is now $\theta$-dependent given by
\begin{eqnarray}
\eta=\gamma|_{*}=-\frac{g_*K_d(\theta\Lambda^2)^2}{192}=-\frac{(\theta\Lambda^2)^2\epsilon}{72}.
\end{eqnarray}
This is proportional to $\epsilon$ (and not $\epsilon^2$) and is negative. The behavior of the $2$-point function is now given by

\begin{eqnarray}
\langle \phi(x)\phi(0)\rangle\sim \frac{1}{|x|^{2-\epsilon(1+(\theta\Lambda^2)^2/72)}}.
\end{eqnarray}
We obtain now the critical point
\begin{eqnarray}
\theta_c\Lambda^2=\frac{12}{\sqrt{\epsilon}}.
\end{eqnarray}
The noncommutative Wilson-Fisher fixed point is only stable for $\theta<\theta_c$.

The above negative anomalous dimension, which is due to the non-locality of the theory,  leads immediately to  the existence of a first order transition to a modulated phase via the Lifshitz scenario \cite{Kleinert:2001hr}. Indeed, we can show that below the critical value $\theta_c$ the coefficient of $k^2$ is positive whereas above $\theta_c$ the coefficient of $k^2$ becomes negative and thus one requires, in order to maintain stability,  the inclusion of the term proportional to $k^4$  which turns out to have a positive coefficient as opposed to the commutative theory. We can show explicitly that at $\theta=\theta_c$ the dispersion relation changes from $k^2$ to $k^4$. Thus, the effective action is necessarily of the form (with positive $a$ and $b$)
\begin{eqnarray}
\int \frac{d^dk}{(2\pi)^d}\phi(k)\phi(-k)\big[(1-ag\theta\Lambda^2)k^2+bk^4\big]+{\rm interaction}.
\end{eqnarray}
The Lifshitz point is a tri-critical point in the phase diagram where the coefficient of $k^2$  vanishes exactly and that of $k^4$ is positive. In this case, this point is given precisely by the value $\theta=\theta_c$, and the transition is a first order transition because it is not related to a change of symmetry. In this transition the system develops a soft mode associated with the minimum of the kinetic energy and as a consequence the ordering above $\theta_c$ is given by a modulating order parameter. A more thorough discussion of this point can be found in \cite{Chen:2001an}.

\subsubsection{The noncommutative $O(N)$ Wilson-Fisher fixed point}


A non perturbative study of the fixed point in noncommutative $O(N)$ model can be carried out along the above lines \cite{Ydri:2012nw,Ydri:2015yta}. In this case the analysis is exact in $1/N$. It is found that the Wilson-Fisher fixed point makes good sense only for sufficiently small values of $ \theta$ up to a certain maximal noncommutativity. This fixed point describes the transition from the disordered phase to the uniform ordered phase, i.e. the Ising universality class.

As it turns out, there is a $\theta$-dependent fixed point, termed the noncommutative  Wilson-Fisher fixed point, which interpolates between the commutative Wilson-Fisher fixed point of the Ising universality class, which is found to lie at $t=1$ and $a_*=0$, and a novel strongly interacting fixed point which lies at $t=2$ and $a_*\longrightarrow\infty$. Here,  $t$ is the effective noncommutativity parameter and $a$ is the coupling constant of the zero-dimensional reduction of the theory. Thus, the point $t=2$ corresponds to maximal noncommutativity and the corresponding fixed point is identified with the transition between non-uniform and uniform orders.

Indeed, the noncommutative Wilson-Fisher fixed point does not scale to zero when we send the dilation parameter $\rho$ to zero, which corresponds to a single step of the renormalization group transformation, in contrast with the commutative Wilson-Fisher fixed point which still scales to zero in the limit $\rho\longrightarrow 1$. The main obstacle comes from the fact that the critical coupling constant $a_*$ which starts from $0$ at $\rho^{\epsilon}=1$, and then increases to $\infty$ at $\rho^{\epsilon}=1-t/2$, does not return to zero as we decrease $\rho^{\epsilon}$ back from $\rho^{\epsilon}=1-t/2$ to $0$. In other words, the non-perturbative sheet shrinks as we increase the non-commutativity until it disappears at $t=2$.

\subsubsection{The $\theta=\infty$ matrix model fixed point}

As discussed above, in the Wilson recursion formula we perform the usual truncation but also we perform a reduction to zero dimension which allows explicit calculation, or more precisely estimation, of Feynman diagrams. This method was also applied in \cite{Ydri:2013zya,Ydri:2015yta} to noncommutative phi-four theory, with a harmonic oscillator term at the self-dual point, on a degenerate Moyal-Weyl space with two strongly noncommuting coordinates, viz  ${\bf R}^{d}_{\theta}={\bf R}^D\times {\bf R}^2_{\theta}$. In the matrix basis this theory becomes, after appropriate non-perturbative definition, an $N\times N$ matrix model where $N$  is a regulator in the noncommutative directions which is directly connected to the noncommutativity parameter $\theta$ itself. The action is explicitly given by

\begin{eqnarray}
S[M]=\int d^Dx {\rm Tr}_N\bigg[\frac{1}{2}(\partial_{\mu}M)^2+\frac{1}{2}\mu^2M^2+r^2EM^2+\frac{u}{N}M^4\bigg].
\end{eqnarray}
The external (diagonal) matrix $E$ originates from the harmonic oscillator term and $r^2=4/\theta$.
 
Thus, in order to compute the effective action we employ, following \cite{Ferretti:1996tk,Ferretti:1995zn,Nishigaki:1996ts}, a combination of two methods given by 
\begin{itemize}
\item[$1)$] The Wilson  approximate renormalization group recursion formula.
\item[$2)$] The solution of the zero-dimensional large $N$ counting problem (given in this case by the Penner matrix model $ {\rm Tr}_N\big[\frac{1}{2}\mu^2M^2+r^2EM^2+\frac{u}{N}M^4\big]$ which can be turned into  a multitrace  matrix model for large values of $\theta$). 
\end{itemize}
As discussed neatly in  \cite{Ferretti:1996tk} the virtue and power of combining these two methods lies in the crucial fact that all leading Feynman diagrams in $1/N$ will be counted correctly in this scheme including the so-called "setting sun" diagrams. Here, we choose to take the limit $\theta\longrightarrow\infty$ first and then $N\longrightarrow\infty$ so that the $1/N$ expansion of the theory is still given by that of the $D$-dimensional hermitian matrix model, i.e. the action with $r^2=0$.

The obtained matrix model fixed point describes the transition from the one-cut (disordered) phase to the two-cut (non-uniform ordered, stripe) phase  in the same way that the noncommutative Wilson-Fisher fixed point describes  transition from the disordered phase to the uniform ordered phase.

Thus, the analysis of phi-four theory on noncommutative spaces using a combination of the  Wilson renormalization group recursion formula and the solution to the zero-dimensional vector/matrix models at large $N$ suggests the existence of three fixed points. The matrix model $\theta=\infty$ fixed point which describes the disordered-to-non-uniform-ordered transition, the Wilson-Fisher fixed point at $\theta=0$ which describes the disordered-to-uniform-ordered transition, and a noncommutative Wilson-Fisher fixed point at a maximum value of $\theta$ which is associated with the transition between non-uniform-order and uniform-order phases.

\section{Noncommutative phi-four on the fuzzy sphere ${\bf S}^2_N$}

\subsection{UV-IR mixing and renormalizability}
A real scalar field $\Phi$ on the fuzzy sphere  ${\bf S}^2_N$ is an $N\times N$ hermitian matrix where $N=L+1$. Let $L_a$ be the generators of $SU(2)$ in the spin $s=L/2$ irreducible representation. The noncommutative phi-four theory on the fuzzy sphere  ${\bf S}^2_N$ is then given by 

\begin{eqnarray}
S&=&\frac{1}{N}{\rm Tr}\bigg[\Phi[L_a,[L_a,\Phi]]+m^2 {\Phi}^2 + \lambda {\Phi}^4\bigg]\nonumber\\
&=&\frac{1}{N}{\rm Tr}\big({\Phi}{\Delta}{\Phi}+m^2{\Phi}^2+\lambda{\Phi}^4\big)~,~{\Delta}={\cal L}_a^2.\label{lkp}
\end{eqnarray}
This model has the correct commutative large $N$ limit, viz 
\begin{eqnarray}
S=\int \frac{d\Omega}{4\pi}\bigg[\Phi{\cal L}_a^2\Phi+m^2 {\Phi}^2 + \lambda {\Phi}^4\bigg].
\end{eqnarray}
Quantum field theories on the fuzzy sphere were proposed originally in \cite{Grosse:1995ar,Grosse:1995pr}. In perturbation theory of the matrix model (\ref{lkp}) only the tadpole diagram can diverge in the limit $N\longrightarrow \infty$ \cite{Vaidya:2001bt,Chu:2001xi}. See also \cite{Vaidya:2003ew,Vaidya:2002qj}. On the fuzzy sphere, and similarly to the Moyal-Weyl plane, the planar and non-planar tadpole graphs are different and their difference is finite in the limit. This is the infamous UV-IR mixing. This problem can be removed in this case by standard normal ordering of the interaction \cite{Dolan:2001gn}.

We use the background field method to quantize this model. We write $\Phi ={\Phi}_0+{\Phi}_1$ where ${\Phi}_0$ is a background field which satisfies the classical equation of motion and ${\Phi}_1$ is a fluctuation. By integrating over $\Phi_1$ in the path integral we obtain the effective action

\begin{eqnarray}
S_{{\rm eff}}[{\Phi}_0]=S[{\Phi}_0]+\frac{1}{2} {\rm TR}~{\log}~{\Omega}~,~\Omega = {\Delta}+m^2+4\lambda {\Phi}_0^2+2\lambda {\Phi}_0{\Phi}_0^R.\label{lkp1}
\end{eqnarray}
The $2$-point function is deduced from
\begin{eqnarray}
S_{{\rm eff}}^{\rm
  quad}=\frac{1}{N}{\rm Tr}{\Phi}_0\bigg({\Delta}+m^2\bigg){\Phi}_0+\lambda {\rm TR}\bigg(\frac{2}{{\Delta}+m^2}{\Phi}_0^2+\frac{1}{{\Delta}+m^2}{\Phi}_0{\Phi}_0^R\bigg).\label{deduce}
\end{eqnarray}
The free propagator is given explicitly by 

\begin{eqnarray}
\bigg(\frac{1}{{\Delta}+m^2}\bigg)^{AB,CD}=\sum_{k,k_3}\frac{1}{{\Delta}(k)+m^2}{T}_{kk_3}^{AB}({T}_{kk_3}^{+})^{DC}.
\end{eqnarray}
The planar and non-planar contributions are found explicitly to be given by 
\begin{eqnarray}
{\rm TR}
\frac{2}{{\Delta}+m^2}{\Phi}_0^2&=&2\sum_{p,p_3}~|{\phi}(pp_3)|^2{\Pi}^P~,~{\Pi}^P=\frac{1}{N}\sum_{k}\frac{2k+1}{k(k+1)+m^2}.\label{ppp}
\end{eqnarray}
\begin{eqnarray}
{\rm TR} \frac{1}{{\Delta}+m^2}{\Phi}_0{\Phi}_0^R&=&\sum_{p,p_3}~|{\phi}(pp_3)|^2{\Pi}^{N-P}(p)~,~{\Pi}^{N-P}(p)=\sum_{k}\frac{2k+1}{k(k+1)+m^2}(-1)^{p+k+2s}\left\{\begin{array}{ccc}
                   p & s & s \\
            k & s  & s
                 \end{array}\right\}.\nonumber\\
\end{eqnarray}
The UV-IR mixing is measured by the difference of the two contributions, viz
\begin{eqnarray}
{\Pi}^{N-P}-{\Pi}^P=\frac{1}{N}\sum_{k}\frac{2k+1}{k(k+1)+m}\bigg[N(-1)^{p+k+2s}\left\{\begin{array}{ccc}
                   p & s & s \\
            k & s  & s
                 \end{array}\right\}-1\bigg].
\end{eqnarray}
Remark that the non-planar contribution depends on the external momentum $p$. When $p$ is small compared to $2s=N-1$
one is justified to use the following approximation for the $6j$ symbols \cite{Varshalovich:1988ye}
\begin{eqnarray}
\left\{ \begin{array}{ccc}
            p&s  & s\\
	    k& s&s
         \end{array}\right\}
{\approx}\frac{(-1)^{p+k+2s}}{N} P_{p}(1-\frac{2k^2}{N^2}),
~s{\rightarrow}{\infty},~p\ll 2s,~0{\leq}k{\leq}2s.
\end{eqnarray}
Since $P_{p}(1)=1$ for all $p$, only $k\gg 1$ contribute in the above sum, and therefore it
can be approximated by an integral as follows
\begin{eqnarray}
{\Pi}^{N-P}-{\Pi}^P&=&\frac{1}{N}\sum_{k}\frac{2k+1}{k(k+1)+m^2}\bigg[P_{p}(1-\frac{2k^2}{N^2})-1\bigg]\nonumber\\
&=&\frac{1}{N}h_p~,~h_p=\int_{-1}^{+1}\frac{dx}{1-x+\frac{2m^2}{N^2}}\bigg[P_p(x)-1\bigg]\nonumber\\
&=&-\frac{2}{N}\sum_{n=1}^p\frac{1}{n}.
\end{eqnarray}
This is non-zero in the continuum limit. It has also the correct
planar limit on the Moyal-Weyl plane. See for example \cite{Ydri:2016dmy}.

The planar contribution ${\Pi}^P$ is given explicitly
by $\frac{1}{N}\log \frac{N^2}{m^2}$ (if
we replace the sum in (\ref{ppp}) by an integral). Thus, the total quadratic effective
action is given by 
\begin{eqnarray}
S_{{\rm eff}}^{\rm
  quad}=\frac{1}{N}Tr{\Phi}_0\bigg({\Delta}+m^2+3\lambda \log \frac{N^2}{m^2}-2
  \lambda Q\bigg){\Phi}_0.
\end{eqnarray}
The operator $Q=Q({\cal L}^2)$ is defined by its eigenvalues $ Q(p)$
given by
\begin{eqnarray}
Q\hat{Y}_{pm}=Q(p)\hat{Y}_{pm}~,~Q(p)=\sum_{n=1}^p\frac{1}{n}.
\end{eqnarray}
The quartic effective action is also deduced from (\ref{lkp1}). Formally, we obtain
\begin{eqnarray}
S_{\rm eff}^{\rm
  quart}&=&\frac{\lambda}{N}{\rm Tr}{\Phi}_0^4-{\lambda}^2{\rm TR}\bigg[2\bigg(\frac{1}{{\Delta}+m^2}{\Phi}_0^2\bigg)^2+2\bigg(\frac{1}{{\Delta}+m^2}{\Phi}_0^2\bigg)\bigg(\frac{1}{{\Delta}+m^2}({\Phi}_0^R)^2\bigg)+\bigg(\frac{1}{{\Delta}+m^2}{\Phi}_0{\Phi}_0^R\bigg)^2\nonumber\\
&+&
4\bigg(\frac{1}{{\Delta}+m^2}{\Phi}_0^2\bigg)\bigg(\frac{1}{{\Delta}+m^2}{\Phi}_0{\Phi}_0^R\bigg)\bigg].
\end{eqnarray}
Explicitly, we have
\begin{eqnarray}
S_{\rm eff}^{\rm
  quart}&=&\frac{\lambda}{N}{\rm Tr}{\Phi}_0^4-{\lambda}^2\sum_{jj_3}\sum_{ll_3}\sum_{qq_3}\sum_{tt_3}{\phi}(jj_3){\phi}(ll_3){\phi}(qq_3){\phi}(tt_3)\bigg[2V_{P,P}+2\bar{V}_{P,P}+V_{NP,NP}+4V_{P,NP}\bigg].\nonumber\\
\end{eqnarray}
We introduce the interaction vertex (with the  notation $\vec{k}=(kk_3)$)

\begin{eqnarray}
v(\vec{k},\vec{j},\vec{p},\vec{q})={\rm Tr}
T^+_{kk_3}T_{jj_3}T_{pp_3}T_{qq_3}.
\end{eqnarray}
The two planar-planar contributions are given by
\begin{eqnarray}
&&V_{P,P}(\vec{j},\vec{l},\vec{q},\vec{t})=\sum_{kk_3}\sum_{pp_3}\frac{v(\vec{k},\vec{j},\vec{l},\vec{p})}{k(k+1)+m^2}\frac{v(\vec{p},\vec{q},\vec{t},\vec{k})}{p(p+1)+m^2}\nonumber\\
&&\bar{V}_{P,P}(\vec{j},\vec{l},\vec{q},\vec{t})=\sum_{kk_3}\sum_{pp_3}\frac{v(\vec{k},\vec{j},\vec{l},\vec{p})}{k(k+1)+m^2}\frac{v(\vec{p},\vec{k},\vec{q},\vec{t})}{p(p+1)+m^2}.
\end{eqnarray}
The non-planar-non-planar and  planar-non-planar contributions are given by
\begin{eqnarray}
V_{NP,NP}(\vec{j},\vec{l},\vec{q},\vec{t})=\sum_{kk_3}\sum_{pp_3}\frac{v(\vec{k},\vec{j},\vec{p},\vec{l})}{k(k+1)+m^2}\frac{v(\vec{p},\vec{q},\vec{k},\vec{t})}{p(p+1)+m^2}.
\end{eqnarray}

\begin{eqnarray}
V_{P,NP}(\vec{j},\vec{l},\vec{q},\vec{q},\vec{t})=\sum_{kk_3}\sum_{pp_3}\frac{v(\vec{k},\vec{j},\vec{l},\vec{p})}{k(k+1)+m^2}\frac{v(\vec{p},\vec{q},\vec{k},\vec{t})}{p(p+1)+m^2}.
\end{eqnarray}
In the commutative limit the planar-planar contribution $V_{P,P}$
remains finite and tends in the commutative large $N$ limit to the result
\begin{eqnarray}
V_{P,P}(\vec{j},\vec{l},\vec{q},\vec{t})=\frac{1}{N}\sum_{kk_3}\sum_{pp_3}\frac{w(\vec{k},\vec{j},\vec{l},\vec{p})}{k(k+1)+m^2}\frac{w(\vec{p},\vec{q},\vec{t},\vec{k})}{p(p+1)+m^2}~,~w(\vec{k},\vec{j},\vec{p},\vec{q})=\int \frac{d\Omega}{4\pi}
Y^+_{kk_3}Y_{jj_3}Y_{pp_3}Y_{qq_3}.\nonumber\\
\end{eqnarray}
Furthermore, it can be shown that all other contributions become
equal in the commutative large $N$ limit to the above
result \cite{Dolan:2001gn}. In other words, there is no difference between planar and non-planar
graphs and the UV-IR mixing is absent in this case.

Hence, to remove the UV-IR mixing from this model a standard prescription of
normal ordering which amounts to the substraction of tadpoloe
contributions will be sufficient. We consider therefore the action
\begin{eqnarray}
S=\frac{1}{N}Tr \bigg[\Phi \bigg(\Delta +m^2-3N\lambda {\Pi}^P+2
  \lambda {Q}\bigg)\Phi +\lambda {\Phi}^4\bigg].
\end{eqnarray}
In above ${Q}={Q}({\cal L}_a^2)$ is given for any $N$ by the expression
\begin{eqnarray}
{Q}\hat{Y}_{pp_3}={Q}(p)\hat{Y}_{pp_3}~,~{Q}(p)&=&-\frac{1}{2}\sum_{k}\frac{2k+1}{k(k+1)+m}\bigg[N(-1)^{p+k+2s}\left\{\begin{array}{ccc}
                   p & s & s \\
            k & s  & s
                 \end{array}\right\}-1\bigg].\nonumber\\
\end{eqnarray}
The first substraction is the usual tadpole substraction which renders the
 limiting commutative theory finite. The second substraction is
 to cancel the UV-IR mixing. Although this action does not have the correct commutative limit (due
 to the non-local substraction) the corresponding quantum theory is
  standard phi-four in $2$ dimensions.
\subsection{The phase diagram}


The coordinates operators $\hat{x}_a$ on the fuzzy sphere are $N\times N$ matrices defined in terms of the angular momentum generators by

\begin{eqnarray}
\hat{x}_a=\frac{\theta}{R}L_a~,~\theta=\frac{2R^2}{\sqrt{N^2-1}}.
\end{eqnarray}
Here, $L_a$ are the angular momentum generators in the spin $s\equiv (N-1)/{2}$ irreducible representation of $SU(2)$. We have then
\begin{eqnarray}
\hat{x}_1^2+\hat{x}_2^2+\hat{x}_3^2=R^2~,~[\hat{x}_a,\hat{x}_b]=\frac{\theta}{R}i{\epsilon}_{abc}\hat{x}_c.
\end{eqnarray}

In the limit $N\longrightarrow \infty$ we recover the commuting sphere.

In the limit $N\longrightarrow \infty$ and $R\longrightarrow \infty$ keeping $\theta$ fixed we get the Moyal-Weyl plane.

We can construct a Weyl map and a star product which allows us to work with  the $C^*$-algebra of functions on the sphere  $C^{\infty}({\bf S}^2)$  and thus functions can be expanded in the basis of spherical harmonics. However, functions on the fuzzy sphere are $N\times N$ matrices which can be expanded in $SU(2)$ polarization tensors. The Weyl map of the polarization tensors are precisely the spherical harmonics.

A scalar field on the fuzzy sphere is then an $N\times N$ matrix which can be expanded in $SU(2)$ polarization tensors.  The action and partition function of a phi-four theory on the fuzzy sphere is given by

\begin{eqnarray}
S=-aTr[L_a,\Phi]^2+Tr \big(b\Phi^2+c\Phi^4\big).
\end{eqnarray}
\begin{eqnarray}
Z[J]=\int d\Phi~exp(-S(\Phi)-TrJ\Phi).
\end{eqnarray} 
We can choose

\begin{eqnarray}
a=\frac{2\pi}{N}~,~b=\frac{2\pi rR^2}{N}~,~c=\frac{\pi\lambda R^2}{6N}.
\end{eqnarray}
In the commutative limit $N\longrightarrow \infty$ we obtain the phi-four theory on  the ordinary sphere ${\bf S}^2$, viz
\begin{eqnarray}
S=\int_{S^2} d\Omega ({\cal L}_a\Phi)^2+\int_{S^2}d\Omega \big(\frac{rR^2}{2}\Phi^2+\frac{\lambda R^2}{4!}\Phi^4\big).
\end{eqnarray}
However, in the limit $N\longrightarrow \infty$ and $R\longrightarrow \infty$ keeping $\theta$ fixed we get noncommutative phi-four on the Moyal-Weyl plane.

A Monte Carlo study of the $d=2$ noncommutative phi-four theory on the fuzzy sphere established the existence of an extra phase (known variously as the non-uniform phase, the stripe phase, the matrix phase) together with the two usual phases observed in the commutative theory \cite{GarciaFlores:2009hf,Martin:2004un,Panero:2006bx}. Similarly, the existence of the stripe phase in $d=2$ noncommutative phi-four theory on the noncommutative torus is shown in \cite{Ambjorn:2002nj}. This structure is expected to hold true in all dimensions, e.g. a Monte Carlo study of the $d=3$ phi-four theory on the noncommutative torus  revealed the existence of a minimum in the dispersion relation corresponding to a stripe phase \cite{Bietenholz:2004xs}. This phase structure is also observed in condensed matter physics (Lifshitz triple points) \cite{CL}.

The collapsed (scaled) variables of the model are given by 
\begin{eqnarray}
\tilde{b}=\frac{b}{aN^{3/2}}~,~\tilde{c}=\frac{c}{a^2N^2}.
\end{eqnarray}  
The Ising (disordered-to-uniform) transition is found to lie at the critical point
\begin{eqnarray}
\tilde{c}=-0.23\tilde{b}.
\end{eqnarray}  
This is a second order phase transition which can be located at the peaks of the specific heat and susceptibility defined by
\begin{eqnarray}
C_v=\langle S^2\rangle-\langle S\rangle^2~,~\chi=\langle |{\rm Tr} {\Phi}|^2\rangle-\langle |{\rm Tr} {\Phi}|\rangle^2.
\end{eqnarray}
The non-uniform-to-uniform transition is found to lie at the critical point 
\begin{eqnarray}
\tilde{c}=-0.2\tilde{b}+0.07.
\end{eqnarray}  
This seems also to be a second order phase transition which is in fact a continuation of the Ising transition to larger negative values of the mass parameter.

The disorder-to-non-uniform transition for small $\tilde{b}$ is located at the critical point 
\begin{eqnarray}
\tilde{c}=-2.29\tilde{b}-4.74.
\end{eqnarray}  
The behavior for large $\tilde{b}$ must be of the form (prediction of the real quartic matrix model)
\begin{eqnarray}
\tilde{c}=\tilde{b}^2/4.
\end{eqnarray}
This is a third order transition which can be located at the point where the eigenvalue distribution of $\Phi$ splits into two disjoint sets. At this point the first derivative of $C_v$ has a finite discontinuity.

The magnetization is defined by 
\begin{eqnarray}
M=\langle |Tr\Phi|\rangle.
\end{eqnarray}
The magnetization seems to change smoothly across the disorder-to-uniform transition point but it jumps at the uniform-to-uniform transition point. The magnetization seems to go through a maximum across the disorder-to-non-uniform transition.  

The intersection of the above three lines yields an estimation of the tripe point of the noncommutative phi-four theory on the fuzzy sphere ${\bf S}^2_N$. We get
\begin{eqnarray}
(\tilde{b}_T,\tilde{c}_T)=(-2.3,0.52).
\end{eqnarray}
In summary, the noncommutative phi-four theory on the fuzzy sphere ${\bf S}^2_N$ exhibits the following three known phases: 
\begin{itemize}
\item The usual $2$nd order Ising phase transition between disordered $\langle \Phi\rangle =0$ and uniform-ordered $\langle \Phi\rangle \sim {\bf 1}_N$ phases. This appears for small values of $c$. This is the only transition observed in commutative phi-four, and thus, it can be accessed in a small noncommutativity  parameter expansion.

Thus, the  uniform-ordered phase $\langle \Phi\rangle\sim {\bf 1}_N$, is stable in this theory. This is in contrast with the case of the real quartic matrix model given by 
\begin{eqnarray}
V={\rm Tr} ({b}M^2+{c}M^4).
\end{eqnarray} 
Indeed, in this case the uniform-ordered phase becomes unstable for all values of the couplings. The source of this stability is obviously the addition of the kinetic term (geometry) to the action.

\item  A matrix transition between disordered $\langle \Phi\rangle =0$ and non-uniform-ordered $\langle \Phi\rangle\sim \gamma$ phases with $\gamma^2={\bf 1}_N$. This transition coincides, for very large values of $c$,  with the $3$rd order transition of the real quartic matrix model, i.e. the model with $a=0$, which occurs at 
\begin{eqnarray}
b=-2\sqrt{Nc}. 
\end{eqnarray}
This is therefore a transition from a one-cut (disc) phase to a two-cut (annulus) phase.
\item  A transition between uniform-ordered  $\langle \Phi\rangle \sim {\bf 1}_N$ and non-uniform-ordered $\langle \Phi\rangle\sim \gamma$ phases.  Some of  the properties of the non-uniform phase are: 
\begin{itemize}
\item The non-uniform phase, in which translational/rotational invariance is spontaneously broken, is absent in the commutative theory. 
\item The non-uniform phase is essentially  the stripe phase observed originally on  Moyal-Weyl spaces. 
\item The non-uniform ordered phase is a full blown nonperturbative manifestation of the perturbative  UV-IR mixing effect which is due to the underlying highly non-local matrix degrees of freedom of the noncommutative scalar field.
\end{itemize}
\end{itemize}

\begin{figure}[htbp]
\begin{center}
\includegraphics[width=16cm]{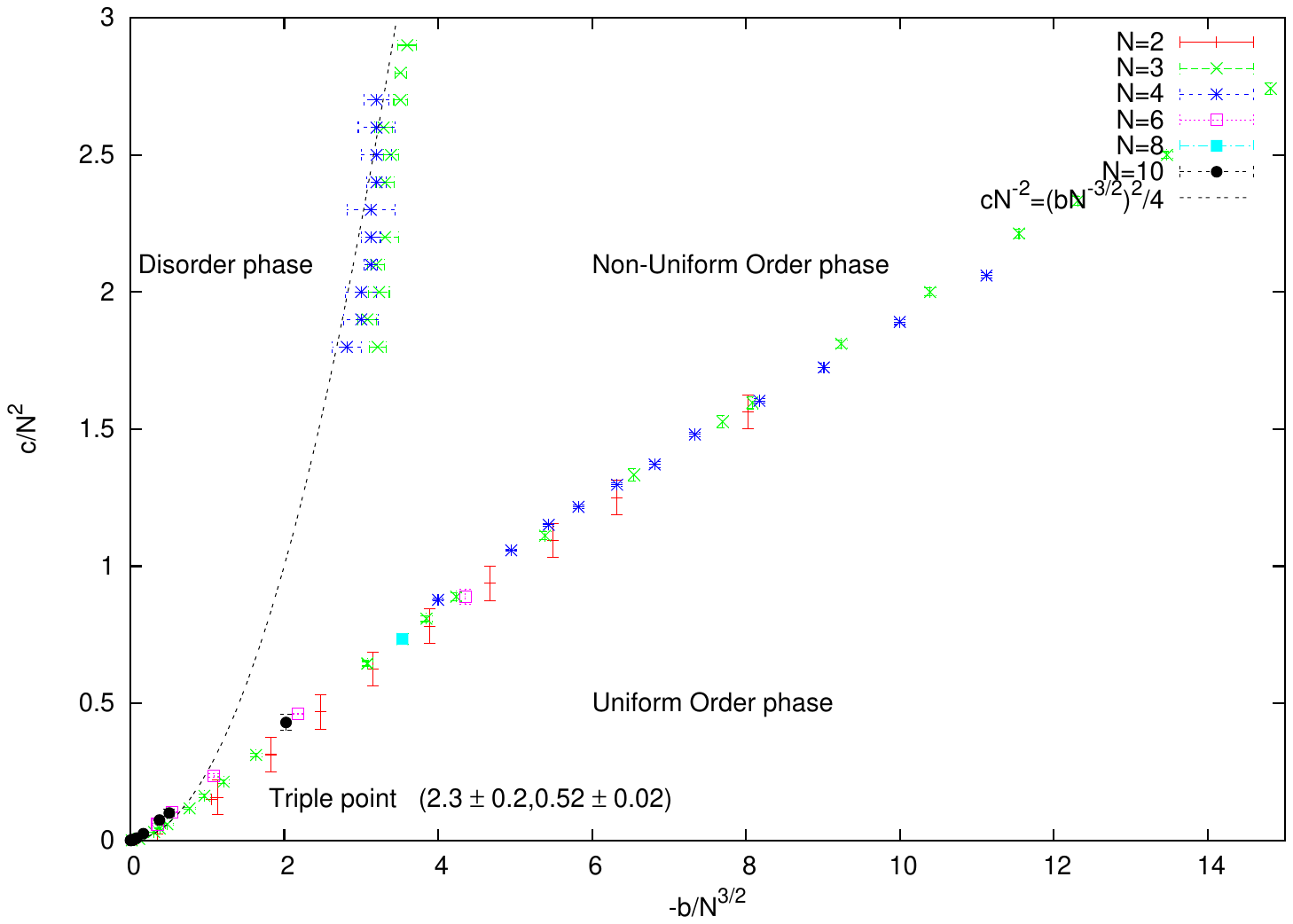}
\caption{The phase diagram of phi-four theory on the fuzzy sphere.}\label{pds2}
\end{center}
\end{figure}

The phase diagram is shown in figure (\ref{pds2}).

\subsection{On the phase diagram of the real quartic matrix model}

The basic model is given by
\begin{eqnarray}
V=B{\rm Tr} \Phi^2+C{\rm Tr} \Phi^4~,~\Phi^{\dagger}=\Phi.
\end{eqnarray}
The ground state configurations of this classical pure potential model are given by the matrices:
\begin{eqnarray}
{\Phi}_0=0.
\end{eqnarray}
\begin{eqnarray}
{\Phi}_{\gamma}=\sqrt{-\frac{B}{2C}}U\gamma U^+~,~{\gamma}^2={\bf 1}_N~,~UU^+=U^+U={\bf 1}_N.
\end{eqnarray}
The first configuration corresponds to the disordered (one-cut) phase whereas the second solution, valid only for $B\leq 0$,  corresponds to the ordered (two-cut) phase. The idempotent $\gamma$ can be always chosen such that $\gamma=\gamma_k={\rm diag}({\bf 1}_{k},-{\bf 1}_{N-k})$.

From energy consideration the configurations $\Phi_{\gamma}$ are degenerate. If we include the effect of the kinetic energy then the configurations corresponding to $k=0$ and $k=N$ become favorable. This is the Ising configuration $\Phi_{\gamma}\propto \pm {\bf 1}$.

The orbit of $\gamma_k$ is the Grassmannian manifold $U(N)/(U(k)\times U(N-k))$ whose dimension is $d_k=2kN-2k^2$. It is not difficult to show that this dimension is maximum at $k=N/2$ (assuming that $N$ is even). Thus, from entropy consideration the most important two-cut solution is the so-called stripe configuration given by $\gamma={\rm diag}({\bf 1}_{{N}/{2}},-{\bf 1}_{{N}/{2}})$.

In summary, we have therefore three possible phases:

\begin{eqnarray}
\langle {\Phi}\rangle=0~:~\text{disordered phase}.
\end{eqnarray}
\begin{eqnarray}
\langle {\Phi}\rangle =\pm\sqrt{-\frac{B}{2C}}{\bf 1}_N~:~\text{uniform phase}.
\end{eqnarray}
\begin{eqnarray} 
\langle {\Phi}\rangle =\pm\sqrt{-\frac{B}{2C}}\gamma~:~\text{non-uniform phase}.
\end{eqnarray}
However, quantum mechanically,  there are only two stable phases  in this model \cite{Brezin:1977sv,Shimamune:1981qf}

\begin{itemize}
\item{} {\bf Disordered phase (one-cut) for $B\geq B_c$:}

\begin{eqnarray}
\rho(\lambda)&=&\frac{1}{N\pi}(2C\lambda^2+B+C\delta^2)\sqrt{\delta^2-\lambda^2}~,~-\delta\leq \lambda\leq \delta.\nonumber\\
\end{eqnarray}
\begin{eqnarray}
\delta^2&=&\frac{1}{3C}(-B+\sqrt{B^2+12 NC}).
\end{eqnarray}
\item{} {\bf Non-uniform ordered phase (two-cut) for $B\le B_c$:}

\begin{eqnarray}
\rho(\lambda)&=&\frac{2C|\lambda|}{N\pi}\sqrt{(\lambda^2-r_{-}^2)(r_{+}^2-\lambda^2)}~,~r_{-}\leq |\lambda|\leq r_{+}.
\end{eqnarray}
\begin{eqnarray}
r_{\mp}^2&=&\frac{1}{2C}(-B\mp 2\sqrt{NC}).
\end{eqnarray}
\end{itemize}
Here, $\rho(\lambda)$ is the eigenvalue distribution of the matrix $\Phi$.

Remark:
\begin{itemize}
\item{}{\bf Critical point:}
A third order transition between the above two phases occurs at the critical point 
\begin{eqnarray}
B_c^2=4NC \leftrightarrow B_c=-2\sqrt{NC}.\label{cr}
\end{eqnarray}
\item{} {\bf Specific heat:} The behavior of the specific heat across the matrix transition is given by (with $ \bar{B}=B/B_c$)

\begin{eqnarray}
\frac{C_v}{N^2}=\begin{cases}
\frac{1}{4} & \bar{B}<-1 \\
\frac{1}{4}+\frac{2\bar{B}^4}{27}-\frac{\bar{B}}{27}(2\bar{B}^2-3)\sqrt{\bar{B}^2+3} & \bar{B}>-1
\end{cases}.
\end{eqnarray}
\item{}{\bf Uniform ordered phase:}
The real quartic matrix model admits also a solution with ${\rm Tr} M\neq 0$ corresponding to a possible uniform-ordered (Ising) phase. This $U(N)$-like solution can appear only for negative values of the mass parameter.

The density of eigenvalues in this case is given by
\begin{eqnarray}
\rho(z)=\frac{1}{\pi N}(2C z^2 +2\sigma C z+B +2C\sigma^2+C\tau^2)\sqrt{((\sigma+\tau)-z)(z-(\sigma-\tau))}.
\end{eqnarray}
This is a one-cut solution centered around $\tau$ in the interval $[\sigma-\tau,\sigma+\tau]$ where $\sigma$ and $\tau$ are given by
\begin{eqnarray}
\sigma^2=\frac{1}{10C}(-3B +2\sqrt{B^2-15NC})~,~\tau^2=\frac{1}{15C}(-2B -2\sqrt{B^2-15NC}).
\end{eqnarray}
This solution makes sense only for 
\begin{eqnarray}
B\leq B_c=-\sqrt{15}\sqrt{NC}.
\end{eqnarray}
\end{itemize}

\subsection{A brief outline of the multitrace approach}
We consider now phi-four theory on the fuzzy sphere ${\bf S}^2_N$ given by the action 

\begin{eqnarray}
S=-a{\rm Tr}[L_a,\Phi]^2+{\rm Tr} \big(b\Phi^2+c\Phi^4\big).
\end{eqnarray}
By diagonalizing the scalar field $\Phi$ as $\Phi=U\Lambda U^{\dagger}$we can put the  partition function in the form
 \begin{eqnarray}
Z=\int d\Lambda \Delta (\Lambda)\exp\big(-{\rm Tr}(b\Lambda^2+c\Lambda^4)\big)\int dU  \exp\big( a{\rm Tr} [U^{\dagger}L_aU,\Lambda]^2\big).
\end{eqnarray} 
Expanding in powers of $a$ and integrating over $U(N)$ leads to a multitrace matrix model of the noncommutative phi-four on the fuzzy sphere \cite{O'Connor:2007ea,Saemann:2010bw}.  For example, we find up to order $a^2$ the multitrace matrix model \cite{Ydri:2014uaa,Ydri:2015zsa}
\begin{eqnarray}
S_{\rm eff}&=&\sum_{i}(b\lambda_i^2+c\lambda_i^4)-\frac{1}{2}\sum_{i\neq j}\ln(\lambda_i-\lambda_j)^2\nonumber\\
&+&\bigg[\frac{aN}{4}v_{2,1}\sum_{i\ne j}(\lambda_i-\lambda_j)^2+\frac{a^2N^2}{12}v_{4,1}\sum_{i\ne j}(\lambda_i-\lambda_j)^4-\frac{a^2}{6}v_{2,2}\big[\sum_{\
i\ne j}(\lambda_i-\lambda_j)^2\big]^2+...\bigg].\label{ft}\nonumber\\
\end{eqnarray}
The logarithmic potential arises from the Vandermonde determinant $\Delta(\Lambda)$, i.e. from diagonalization. The coefficients $v_{2,1}$, $v_{4,1}$ and $v_{2,2}$ are given by  $v_{2,1}=+1~,~v_{4,1}=0~,~v_{2,2}={1}/{8}$. It is not difficult to convince ourselves that the above action is indeed a multitrace matrix model since it can be expressed in terms of various moments $m_n={\rm Tr} M^n$ of the matrix $M\equiv\Phi$. Indeed, the above multitrace matrix model takes the form
\begin{eqnarray}
V&=&{B}{\rm Tr} M^2+{C}{\rm Tr} M^4+D(Tr M^2)^2\nonumber\\
&+&B^{'} ({\rm Tr} M)^2+C^{'} {\rm Tr} M {\rm Tr} M^3+D^{'}({\rm Tr} M)^4+A^{'}{\rm Tr} M^2 ({\rm Tr} M)^2+....\label{fundamental0m}
\end{eqnarray}
\begin{eqnarray}
B=b+\frac{a^2N}{2}~,~C=c~,~D=-\frac{a^2N^2}{12}.\label{336}
\end{eqnarray}
\begin{eqnarray}
B^{'}=-\frac{aN}{2}~,~C^{'}=0~,~D^{'}=-\frac{a^2}{12}~,~A^{'}=\frac{a^2N}{6}.\label{337}
\end{eqnarray}
The saddle point method can be applied to the above multitrace matrix model to give us an estimation of the non-uniform-to-uniform transition line and the triple point. Furthermore, the whole phase diagram of the noncommutative phi-four on the fuzzy sphere can be constructed from Monte Carlo simulation of the multitrace matrix model \cite{Ydri:2015zsa,Ydri:2015vba}.

\begin{figure}[htbp]
\begin{center}
\includegraphics[width=15cm]{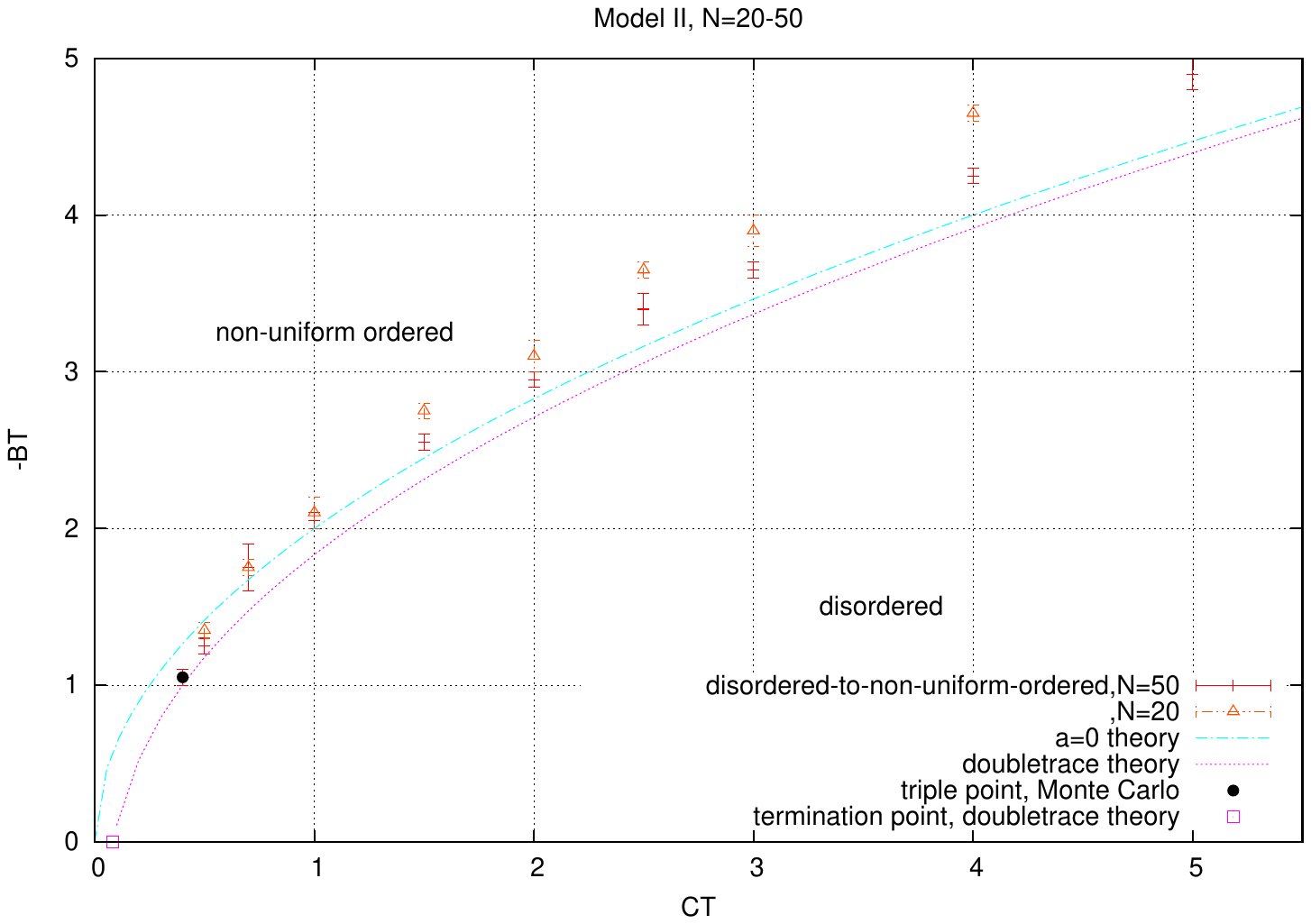}
\caption{The termination point of the doubletrace matrix model as the triple point.}\label{tp}
\end{center}
\end{figure}

Let us consider the example of the $({\rm Tr}M^2)^2$ multitrace matrix model. Remark that in the above model all multitrace terms depend on the odd moment ${\rm Tr} M$ with the exception of the doubletrace term  $({\rm Tr}M^2)^2$. The  $({\rm Tr}M^2)^2$ multitrace matrix model  is given by the model (\ref{fundamental0m}) with all odd moments set to zero, i.e. by imposing the symmetry $M\longrightarrow -M$. We obtain a doubletrace matrix model given by 
\begin{eqnarray}
V=B{\rm Tr} M^2+C{\rm Tr} M^4+D ({\rm Tr} M^2)^2.
\end{eqnarray}
The two stable phases are given by the disordered (one-cut) phase and the non-uniform-ordered (two-cut) phase separated by a deformation of the line $\tilde{B}_*=-2\sqrt{\tilde{C}}$ given by \cite{Polychronakos:2013nca,Ydri:2014uaa}
\begin{eqnarray}
\tilde{B}_*=-2\sqrt{\tilde{C}}-\frac{2\tilde{D}}{\sqrt{\tilde{C}}}.\label{pre0}
\end{eqnarray}
Here, $\tilde{D}=D/N=-2v_{2,2}\tilde{a}^2/3$. There exists a termination point in this  model since the critical line does not extend to zero  \cite{Ydri:2014uaa,Ydri:2015vba}. Indeed, in order for the critical value $\tilde{B}_*$ to be negative one must have $\tilde{C}$ in the range 
\begin{eqnarray}
\tilde{C}\geq \tilde{C}_*=\frac{\tilde{a}^2}{12}.
\end{eqnarray}
Thus, the termination point is located at (for $\tilde{a}\equiv a\sqrt{N}=1$)
\begin{eqnarray}
(\tilde{B},\tilde{C})=(0,1/12).\label{pre}
\end{eqnarray}
The triple point is identified as a termination point of the one-cut-to-two-cut transition line and is located in the Monte Carlo data of the full model, i.e. with the odd terms included, at $(-1.05,0.4)$ which compares favorably with previous Monte Carlo estimate. See figure (\ref{tp}).

Thus, the phase diagram of the multitrace matrix model (\ref{fundamental0m})  does not contain the uniform-ordered phase. In fact,  this phase diagram will not contain the uniform-ordered phase even after including the terms depending on the odd moment ${\rm Tr} M$. This means in particular that the above multitrace matrix model (\ref{fundamental0m}), with the parameters (\ref{336}) and (\ref{337}), lies in the universality class of the real quartic matrix model. It is expected that the correct phase diagram of the noncommutative phi-four on the fuzzy sphere ${\bf S}^2_N$ can be reproduced by considering higher order multitrace corrections. The inclusion of the terms depending on the odd moment  ${\rm Tr} M$ is absolutely necessary in order to correctly reproduce the uniform-ordered phase. For example, in \cite{Ydri:2015zsa,Ydri:2015vba} the multitrace matrix model (\ref{ft}) is considered with the parameters  \cite{O'Connor:2007ea}
\begin{eqnarray}
v_{2,1}=-1~,~v_{4,1}=\frac{3}{2}~,~v_{2,2}=0.
\end{eqnarray}
Explicitly, we have the multitrace action
\begin{eqnarray}
V={B}{\rm Tr} M^2+{C}{\rm Tr} M^4+D\big[ {\rm Tr} M^2\big]^2+B^{'} ({\rm Tr} M)^2+C^{'} {\rm Tr} M {\rm Tr} M^3.\label{example}
\end{eqnarray}
\begin{eqnarray}
B=b-\frac{aN^2}{2}~,~C=c+\frac{a^2N^3}{4}~,~D=\frac{3a^2N^2}{4}~,~B^{'}=\frac{aN}{2}~,~C^{'}=-a^2N^2.\label{example1}
\end{eqnarray}
The phase diagram of this model is shown to contain the three phases of the noncommutative phi-four on the fuzzy sphere ${\bf S}^2_N$. Indeed, the uniform-ordered phase exists in this model only with the terms depending on the odd moment  ${\rm Tr} M$ included. If we assume the symmetry $M\longrightarrow -M$ then the second line of (\ref{fundamental0m}) becomes identically zero and the uniform ordered phase disappears.

In \cite{Ydri:2017riq} an even simpler multitrace matrix model is shown to exhibit the correct phase structure of the noncommutative phi-four on the fuzzy sphere ${\bf S}^2_N$.  This model includes a single term depending on the odd moment  ${\rm Tr} M$ given by ${\rm Tr}M{\rm Tr}M^3$. Explicitly, the action of this  multitrace matrix model is given by  
\begin{eqnarray}
V={B}{\rm Tr} M^2+{C}{\rm Tr} M^4+C^{'} {\rm Tr} M {\rm Tr} M^3.\label{example2}
\end{eqnarray}
\begin{eqnarray}
B=b~,~C=c~,~C^{'}=-a^2N^2.\label{example2}
\end{eqnarray}


\subsection{Critical exponents}

The critical exponents  of the noncommutative phi-four theory on the fuzzy sphere ${\bf S}^2_N$ can be computed from its first multitrace approximation which reproduces the correct phase diagram, i.e. a phase diagram which includes the Ising phase and a triple point. A priori, the emergence of the Ising phase in any matrix model is rather surprising but in these multitrace matrix models it is guaranteed that the Ising phase appears at some higher order of the multitrace approximation.

 A simpler approach is to consider a simpler multitrace matrix model which involves the odd moment ${\rm Tr} M$ and hence is likely to include in its phase diagram an Ising phase and a triple point representing therefore the universality class of the  noncommutative phi-four theory on the fuzzy sphere ${\bf S}^2_N$. This is the route taken in \cite{Ydri:2015zsa,Ydri:2015vba} with the simple multitrace matrix model (\ref{example})+(\ref{example1}) and also taken in \cite{Ydri:2017riq} with the even much simpler multitrace matrix model (\ref{example2})+(\ref{example2}).

\begin{figure}[htbp]
\begin{center}
\includegraphics[width=16cm]{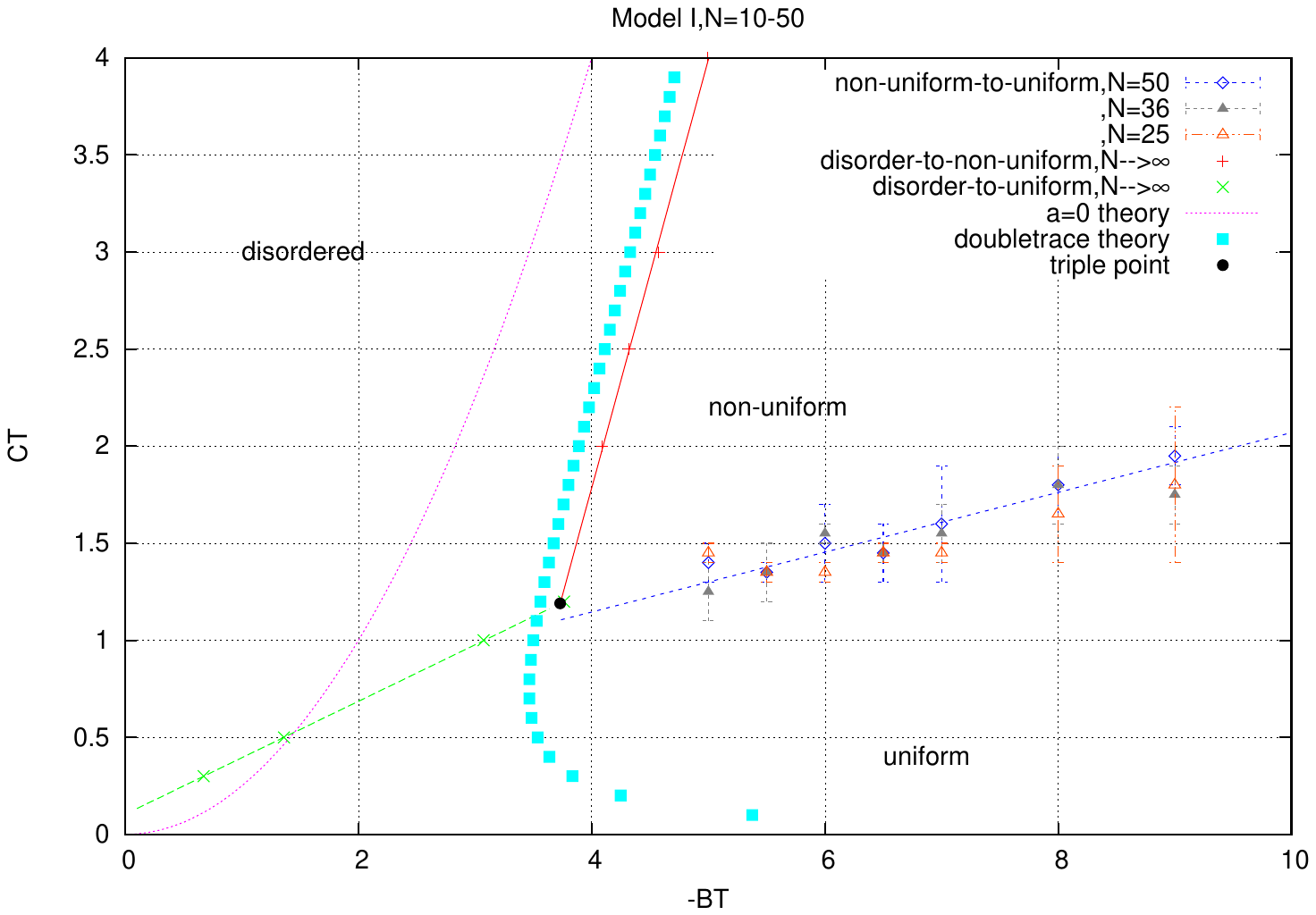}
\caption{The phase diagram of the multitrace model  (\ref{example})+(\ref{example1}).}\label{pdRT}
\end{center}
\end{figure}

In the remainder we discuss in some detail the phase structure and the critical exponents of the  multitrace matrix model (\ref{example})+(\ref{example1}) discussed in  \cite{Ydri:2015zsa,Ydri:2015vba}. We claim that this model contains three stable phases: disorder, uniform order and non-uniform order. See figure (\ref{pdRT}).
\begin{itemize}
\item{}The Ising and the matrix phase transitions:
\begin{itemize}
\item {}{\bf Ising}: The critical point is measured at the peak of the susceptibility $\chi=\langle |Tr M|^2\rangle-\langle |Tr M|\rangle^2$. The fit for the extrapolated critical value is given by
\begin{eqnarray}
\tilde{C}=0.291(0).(-\tilde{B})+0.104(1).
\end{eqnarray}

\item{}{\bf Matrix}: The critical point is determined at the point where the eigenvalue distributions go from one-cut, in the disorder phase, to  two-cut, in the non-uniform phase. The splitting of the distribution is considered to have been occurred  when the height of the distribution at $\lambda=0$ is less than some tolerance.   The fit for the extrapolated critical value is given by
\begin{eqnarray}
\tilde{C}=2.206(67).(-\tilde{B})-7.039(301).
\end{eqnarray}
The behavior of the specific heat across this transition is effectively that of the pure quartic matrix model $a=0$. However, the critical line is better approximated with the doubletrace matrix model prediction.
\end{itemize}

\item{}The stripe phase transition:

\begin{itemize}
\item{} We  approach the critical boundary by fixing $\tilde{B}$ and changing $\tilde{C}$ starting from small values, i.e. inside the uniform ordered phase, until the curves for the total and  zero powers start to diverge, marking the transition to the non-uniform ordered phase. 

\item{} Since this is a very delicate transition we do not perform any extrapolation of the critical point and the critical boundary is given by the fit of the largest value of $N$. In any case we observe no strong dependence on $N$ of the measured critical value $\tilde{C}$.  The stripe critical line is  approximated by the fit for $N=50$ given by
\begin{eqnarray}
\tilde{C}=0.154(22).(-\tilde{B})+0.530(131)~,~N=50.
\end{eqnarray}
\end{itemize}

\end{itemize}

The critical exponents in the context of the multitrace model  (\ref{example})+(\ref{example1}) are computed using the Monte Carlo method in  \cite{Ydri:2015vba}.

\begin{itemize}
\item {\bf The critical exponent $\nu$}:
  
\begin{itemize}
\item{}  The correlation length should behave as
 \begin{eqnarray}
&&\xi\sim|B-B_c|^{-\nu}\sim N~,~\nu_{\rm Onsgaer}=1.\label{cb}
\end{eqnarray} 
\item{}This is very delicate to check explicitly in Monte Carlo. Indeed, since we must necessarily deal with the critical region we must face the two famous problems: 1) finite size effects and 2) critical slowing down. The critical slowing down problem can be shown to start appearing in Monte Carlo simulations around $N>60$ so we will keep below this value and employ very large statistics of the order of $2^{20}$.  The problem of finite size effects is also very serious for the measurement of the critical exponents since the above behavior (\ref{cb}) is supposed to hold only for large $N$. This problem can be avoided by including values of $N$ larger or equal than  $20$.
\item{} We choose $\tilde{C}=1.0$ which is relatively large, but well within the Ising phase, before  the appearance of the transition between the disordered and non-uniform-ordered phases around $\tilde{C}=1.5$.

\item{}We plot the critical point $\tilde{B}_c$ versus $N$. We get the $N=\infty$ critical point $\tilde{B}_*$ and the critical exponent $\nu$:
  \begin{eqnarray}
&&\tilde{B}_c=-1.061(168).N^{-0.926(83)}-3.074(6) \Rightarrow~,~\nu=0.926(83).\label{nu}
  \end{eqnarray}
  Also, we obtain
 \begin{eqnarray}
 \tilde{B}_*=-3.074(6).\label{me1}
\end{eqnarray} 
This prediction for $\nu$ agrees reasonably well with the Onsager calculation $\nu=1$. 
\end{itemize}

\item {\bf The critical exponent $\beta$}:
\begin{itemize}
\item{}The magnetization is defined by
\begin{eqnarray}
m=\langle |{\rm Tr} M|\rangle.
\end{eqnarray}
\item{} The critical behavior is
  \begin{eqnarray}
m/N=\langle |{\rm Tr} M|\rangle/N \sim (B_c-B)^{\beta}\sim N^{-\beta/\nu}~,~\beta_{\rm Onsager}=\frac{1}{8}.
\end{eqnarray} 
\item{} Measurements of the magnetization $m/N$ were performed near the extrapolated critical point, $\tilde{B}=-3.07$, for $\tilde{C}=1.0$, but inside the uniform-ordered phase. These are then used to compute the critical exponent $\beta$ by searching for a power law behavior.

\item {} We measure $\ln(m/N)$ versus $\ln N$, for each value of $\tilde{B}$ very near and around  $\tilde{B}=-3.10$, fit to a straight line in the range $20\leq N\leq 60$, and compute the slope $\beta$, then search for the flattest line, i.e. the smallest slope $\beta$. 
\item{} Deep inside the Ising phase the slope should approach the mean field value $-1/4$ which can be shown from the scaling behavior of the dominant configuration. 
\item {} After determining the critical value we then consider the value of $\tilde{B}$ nearest to it but within the Ising phase and take the slope there to be the value of the critical exponent $\beta$. 
\item{} In our example, here, with $\tilde{C}=1.0$, the flattest line occurs at $\tilde{B}=-3.13$, with slope $-0.088(10)$, thus the measurement of the critical value from the magnetization is 
 
\begin{eqnarray}
\tilde{B}_*=-3.13.\label{me2}
\end{eqnarray}

\item{} After this value the slope becomes $-0.109(11)$ at $\tilde{B}=-3.14$. The slope goes fast to the mean field value $-0.25$ as we keep decreasing $\tilde{B}$.

\item Our measured value, for $\tilde{C}=1.0$, of the critical exponent $\beta$  is:

  \begin{eqnarray}
\ln\frac{m}{N}=-0.109(11).\ln N-1.423(43)\Rightarrow \beta=-0.109(11).\label{beta}
\end{eqnarray} 

\end{itemize}

\item {\bf The critical exponent $\gamma$}:

  \begin{itemize}
\item{}  The critical behavior of the susceptibility is given by  
\begin{eqnarray}
&&\chi=\langle |{\rm Tr} M|^2\rangle-\langle|{\rm Tr} M|\rangle^2\sim (B-B_c)^{-\gamma}\sim N^{\gamma/\nu}\sim N^{2-\eta}.
\end{eqnarray} 
\item{} If we try to fit the values of the susceptibility at its maximum, i.e. at the peak, which keeps slowly moving with $\tilde{B}$, then we will obtain a very bad underestimate of the critical exponent $\gamma$ given by
   \begin{eqnarray}
\ln \chi_{\rm max}=0.515(08).\ln N-0.652(30)\Rightarrow \gamma=0.515(08).
\end{eqnarray}  
This is due in part to the dependence of $\tilde{B}_c$ on $N$, and in another part, is an indication of the critical slowing down problem showing up in the measurement of this second moment, i.e. the size of the fluctuations is observed to grow with $N$ at the critical point but not at the correct rate indicated by the independent measurements of the zero moment and the magnetization.

\item{} The measurement of the critical exponent $\gamma$ is thus, quite delicate, and will be done indirectly as follows. We rewrite the susceptibility in terms of the zero power and magnetization as
 \begin{eqnarray}
\chi&=&\langle |{\rm Tr} M|^2\rangle-\langle |{\rm Tr}M|\rangle^2\nonumber\\
&=&N^2P_0-m^2.
\end{eqnarray} 
\begin{eqnarray}
P_0=\langle \big(\frac{1}{N}{\rm Tr} M)^2\rangle.
\end{eqnarray}
The critical exponent $\gamma$ in terms of the critical exponent $\gamma^{'}$ of $P_0$ is then given by
 \begin{eqnarray}
\gamma=2+\gamma^{'}.
\end{eqnarray}
\item{}  We can check that the second term in the  susceptibility behaves using the result (\ref{beta}) as
 \begin{eqnarray}
\ln m^2=1.782(22).\ln N-2.846(86)\Rightarrow \gamma=1.782(22).
\end{eqnarray} 
This measurements of the critical exponent $\gamma$ agree reasonably well with the Onsager values.


\item{} From the results, at $\tilde{B}=-3.14$, we obtain the exponent

\begin{eqnarray}
\ln N^2 P_0=1.648(10).\ln N-2.289(36)\Rightarrow \gamma=1.648(10).
\end{eqnarray} 
  \end{itemize}

\item {\bf The critical exponent $\alpha$}:

  \begin{itemize}
\item{} The sepcific heat is defined by
\begin{eqnarray}
C_v=\langle S^2\rangle-\langle S\rangle^2.
\end{eqnarray}
The critical point $\tilde{B}_*$ as measured from the specific heat is identified by the intersection point of the various curves with different $N$. We get
 \begin{eqnarray}
&& \tilde{B}_*=-3.08.
\end{eqnarray}  
This measurement is contrasted very favorably with the independent measurement obtained from the extrapolated value of $\tilde{B}_c$ shown in equation (\ref{me1}) but should also be contrasted with the measurement obtained from the magnetization shown in equation (\ref{me2}).

\item{} The critical behavior is
\begin{eqnarray}
&&C_v/N^2 \sim (B-B_c)^{-\alpha}\sim N^{\alpha/\nu}~,~\alpha_{\rm Onsager}=0.
\end{eqnarray} 
\item From the results, at the critical point $\tilde{B}=-3.08$, we obtain 
   \begin{eqnarray}
\ln \frac{C_v}{N^2}=0.024(9).\ln N-0.623(31)\Rightarrow \alpha=0.024(9).
\end{eqnarray} 
  \end{itemize}
\end{itemize}




  

  

\subsection{Coupling to a Yang-Mills theory}

It is very hard to observe the phase transition from the non-uniform-ordered to the uniform-ordered phases using the ordinary Metropolis algorithm due to the existence of an infinite number of non-uniform-ordered vacuum states that can be occupied by the system. In fact, there is a tendency for the system to become stuck in one of these non-uniform-ordered vacuum states, and as a consequence, tunneling to other vacuum states is very hard to observe in practice.

The correct sampling of the tunneling between vacuum states is essential for any Monte Carlo method to work properly. The only known method which was quite successful in overcoming this problem is due to Garcia Flores \cite{GarciaFlores:2009hf,GarciaFlores:2005xc} which uses together with the Metropolis algorithm elements from the annealing algorithm. His algorithm does however break detail balance.

We claim here that an exact Metropolis algorithm using gauge invariance is sufficient to probe the transition from the non-uniform-ordered to the uniform-ordered phases.    

The construction of this algorithm goes as follows \cite{Ydri:2014rea}. By coupling the scalar field to a $U(1)$ gauge field, in a particular way, we can use gauge symmetry to completely diagonalize the scalar sector and reduce it thus to an eigenvalue problem. 

In more detail, coupling the noncommutative scalar field $\Phi$ on the fuzzy sphere ${\bf S}^2_N$ to a Yang-Mills theory is done through the substitution $L_a\longrightarrow X_a/\alpha$. The parameter $\alpha$ plays the role of the gauge coupling constant. The matter action, i.e. the action of the noncommutative scalar field becomes
\begin{eqnarray}
S_m=-\frac{a}{\alpha^2}Tr[X_a,\Phi]^2+Tr \big(b\Phi^2+c\Phi^4\big).
\end{eqnarray}
The dynamics of the covariant matrix coordinates $X_a$ is given by a Yang-Mills matrix model in $D=3$ dimensions. In order for the underlying geometry to be a fuzzy sphere it is absolutely necessary to include also a Chern-Simons term. The gauge action reads then
\begin{eqnarray}
S_g=N Tr\big(-\frac{1}{4}[X_a,X_b]^2+\frac{2i\alpha}{3}\epsilon_{abc} X_aX_bX_c\big).
\end{eqnarray}
For $b=c=0$ the  relevant solutions of the classical equations of motion are $X_a=\alpha L_a$ and $\Phi=0$. In fact, for $b=c=0$ the action $S_g+S_m$ describes  a Yang-Mills matrix model in $D=4$ dimensions where the fourth covariant matrix coordinate is proportional to the scalar field, viz $\Phi=\alpha \sqrt{N/2a} X_4$  \cite{Ydri:2012bq}.

The gauge field $A_a$ is introduced as $X_a=\alpha(L_a+A_a)$. The scalar field $\Phi$ is in the adjoint representation of the $U(1)$ gauge group. 
 Thus, in the commutative limit $N\longrightarrow \infty$ the scalar and gauge fields decouple and the  $\Phi$-dynamics of the action $S_m+S_g$ reduces to the dynamics of phi-four theory on the sphere . 

It is well established that the theory with $b=c=0$ suffers from an emergent geometry transition. The fuzzy sphere is only stable  for \cite{Ydri:2012bq} 
\begin{eqnarray}
\tilde{\alpha}=\sqrt{N}\alpha\geq \tilde{\alpha}_*=2.55.
\end{eqnarray}
We must choose always $\tilde{\alpha}\gg \tilde{\alpha}_*$ to avoid this transition.

We diagonalize the scalar field as before, viz $\Phi=U\Lambda U^+$. However, we can now use the available gauge invariance to perform the integral over $U$. The resulting partition function is  
  \begin{eqnarray}
Z=\int \prod_a dX_a\int d\Lambda~\exp\bigg[N Tr\big(\frac{1}{4}[X_a,X_b]^2-\frac{2i\alpha}{3}\epsilon_{abc} X_aX_bX_c\big)+S[\Lambda]\bigg].
\end{eqnarray}
The scalar action is given by
 \begin{eqnarray}
S[\Lambda]
=N\sum_i(X_a^2)_{ii}\lambda_i^2-N\sum_{ij}(X_a)_{ij}(X_a)_{ji}\lambda_i\lambda_j+\sum_i(r\lambda_i^2+u\lambda_i^4)-\sum_{i\neq j}\ln|\lambda_i-\lambda_j|.
\end{eqnarray}
We use an exact Metropolis algorithm to generate statistically independent matrices $X_a$ and  eigenvalues $\lambda_i$. This is done for a large value of $\tilde{\alpha}=\sqrt{N}\alpha$ where we know that the scalar field propagates on a fuzzy sphere. 


In summary, instead of sampling the unitary matrix $U$  in the noncommutative phi-four theory on the fuzzy sphere (which is very inefficient) we  have now to sample the covariant matrix coordinates $X_a$ (which is known to be very efficient). 


\section{Emergent geometry from multitrace matrix models}
\subsection{The ${\rm Tr}M{\rm Tr}M^3$ multitrace matrix model}

The multitrace approach to noncommutative scalar field theories was initiated in \cite{O'Connor:2007ea,Saemann:2010bw} on the fuzzy sphere.  For an earlier approach see \cite{Steinacker:2005wj} and for a similar non-perturbative approach see \cite{Polychronakos:2013nca,Tekel:2014bta,Nair:2011ux,Tekel:2013vz} and \cite{Tekel:2015uza,Tekel:2015zga,Subjakova:2020haa}.

In this approach, as we have seen, we diagonalize the Hermitian scalar matrix $\Phi$ as $\Phi=U M U^{\dagger}$ and then integrate over the unitary matrix $U$ using the group theoretic structure and properties of $SU(2)$ and $SU(N)$ extensively. In effect,  in this approach the kinetic term is expanded while the potential term is treated exactly. This is in fact a hopping-parameter-like expansion. The end result is to convert the kinetic term into a multitrace matrix model, which to the lowest non-trivial order, is of the form
\begin{eqnarray}
&&\int dU \exp\bigg(a{\rm Tr}[L_a,UMU^{\dagger}]^2-{\rm Tr}V(M)\bigg)\nonumber\\
&=&\exp\bigg(B{\rm Tr}M^2+C{\rm Tr}M^4+D({\rm Tr}M^2)^2+B^{\prime}({\rm Tr}M)^2\nonumber\\
&+&C^{\prime}{\rm Tr}M{\rm Tr}M^3+D^{\prime}({\rm Tr}M)^4+A^{\prime}{\rm Tr}M^2({\rm Tr}M))^2+...\bigg).
\end{eqnarray}
The parameters $B$ and $C$ have shifted values with respect to the original parameters of the potential $b$ and $c$ respectively while the primed parameters and $D$ have purely quantum values coming from the hopping-parameter-like expansion of the kinetic term.

The basic statement is that the Laplacian operator on the fuzzy sphere (which is the example considered here) is exactly equivalent to this multitrace matrix model with these particular value of the coefficients. And each Laplacian operator, i.e. any other noncommutative space comes with its own set of multitrace coefficients. 

It has been shown in   \cite{Ydri:2015zsa,Ydri:2017riq} that the essential features of the phase diagram of noncommutative phi-four theory in two dimensions can be captured by a truncated multitrace matrix model depending on the cubic moment ${\rm Tr}M^3$, i.e. a multitrace matrix model given simply by the potential

\begin{eqnarray}
V_{\rm trunc}=
B{\rm Tr}M^2+C{\rm Tr}M^4+C^{\prime}{\rm Tr}M{\rm Tr}M^3.\label{multitrace}
\end{eqnarray}
Naturally, stability requirement of this multitrace matrix model constrains the range of the quartic coupling $C$ in a particular way. The statistical physics of this multitrace matrix model interpolates between the statistical physics of the noncommutative field theory (\ref{matrix}) and (\ref{phi-four}) and the statistical physics of the real quartic matrix model given by 
\begin{eqnarray}
V_{\rm pure}=
B{\rm Tr}M^2+C{\rm Tr}M^4.\label{purematrix}
\end{eqnarray}
Indeed, the random multitrace matrix model (\ref{multitrace}) lies in the universality class of noncommutative phi-four theory for $C^{\prime}$ negative whereas it lies in the universality class of the real quartic matrix model for $C^{\prime}$ positive. 


The above random multitrace matrix model  should be thought of as a generalization of the discretization of random Riemannian surfaces with regular polygons such as dynamical triangulation \cite{DiFrancesco:1993cyw,Zarembo:1998uk}. Indeed, and as we will discuss shortly, random multitrace matrix models can sustain emergent geometry as well as growing dimensions and topology change. For example, the multitrace matrix model (\ref{multitrace}) works in two dimensions but also it works away from two dimensions where it can generate a large class of spaces, such as fuzzy projective space $\mathbb{C}{\mathbb P}_N^n$ \cite{Balachandran:2001dd}, which admit finite spectral triples that can be captured by the multitrace term ${\rm Tr}M {\rm Tr}M^3$.


In terms of the collapsed or scaled parameters  $\tilde{B}=B/N^{3/2}$, $\tilde{C}=C/N^2$ and $\tilde{C}^{\prime}=C^{\prime}/N$ the multitrace matrix model (\ref{multitrace}) is rewritten (with the scaled field $\tilde{M}=N^{1/4}M$) as
\begin{eqnarray}
V_{\rm trunc}=
N\tilde{B}{\rm Tr}{M}^2+N\tilde{C}{\rm Tr}{M}^4+\tilde{C}^{\prime}{\rm Tr}{M}{\rm Tr}{M}^3.\label{multitrace1}
\end{eqnarray}
In the large $N$ limit the saddle point equation controls the statistical quantum physics of the theory. The saddle point equation and the corresponding  free energy read explicitly
\begin{eqnarray}
&&2\tilde{B}\lambda+4\tilde{C}\lambda^3+\tilde{C}^{\prime}{m}_3+3\tilde{C}^{\prime}{m}_1\lambda^2-2\int d\lambda^{\prime}\rho(\lambda^{'})\frac{1}{\lambda-\lambda^{\prime}}=0.\label{spe}
\end{eqnarray}
\begin{eqnarray}
\frac{E}{N^2}&=&\int d\lambda \rho(\lambda)(\tilde{B}\lambda^2+\tilde{C}\lambda^4)+\tilde{C}^{\prime}\int d\lambda\rho(\lambda)\int d\lambda^{\prime}\rho(\lambda^{\prime})\lambda^{\prime 3}\nonumber\\
&-&\frac{1}{2}\int d\lambda\rho(\lambda)\int d\lambda^{\prime}\rho(\lambda^{\prime})\log (\lambda-\lambda^{\prime})^2.
\end{eqnarray}
The last term in both equations comes from the Vandermonde determinant which causes the eigenvalues to repel each other and become spread evenly around zero. In the saddle point equation (\ref{spe}) there appears also the moments $m_q=\int d\lambda\rho(\lambda)\lambda^q$ which modify the behavior of the real quartic matrix model through the multitrace coupling $C^{\prime}$.


The matrix model (\ref{matrix})+(\ref{phi-four}) without kinetic term ($a=0$) can be solved exactly to obtain a phase structure consisting of
\begin{itemize}
\item A) a disordered (symmmetric, one-cut, disk) phase, and
\item B) a non-uniform ordered (stripe, two-cut, anuulus) phase,
\end{itemize}
separated by a third order transition line \cite{Brezin:1977sv,Shimamune:1981qf}.

This is the so-called matrix line and it corresponds to the matrix model fixed point of noncommutative field theory at infinite noncommutativity $\theta=\infty$ \cite{Bietenholz:2004as,Becchi:2003dg,Grosse:2003nw,Ydri:2013zya}.

The matrix model (\ref{matrix})+(\ref{phi-four}) with a kinetic term ($a\neq 0$) is much harder to solve but much more interesting. The phase diagram, in addition to the above two phases A) and B), involves
\begin{itemize}
\item a C) uniform (asymmetric, one-cut, Ising) ordered phase.
\end{itemize}
The three phases meet at a triple point where the three co-existence curves intersect. This phase diagram is also observed in the truncated multitrace matrix model (\ref{multitrace}).

The main difference between the real quartic matrix model (\ref{purematrix}) from the one hand, and the noncommutative phi-four theory (\ref{matrix})+(\ref{phi-four}) on the other hand, lies in the fact that the uniform-ordered phase is stable in the latter.

\begin{figure}[htbp]
\begin{center}
\includegraphics[width=9.0cm,angle=-0]{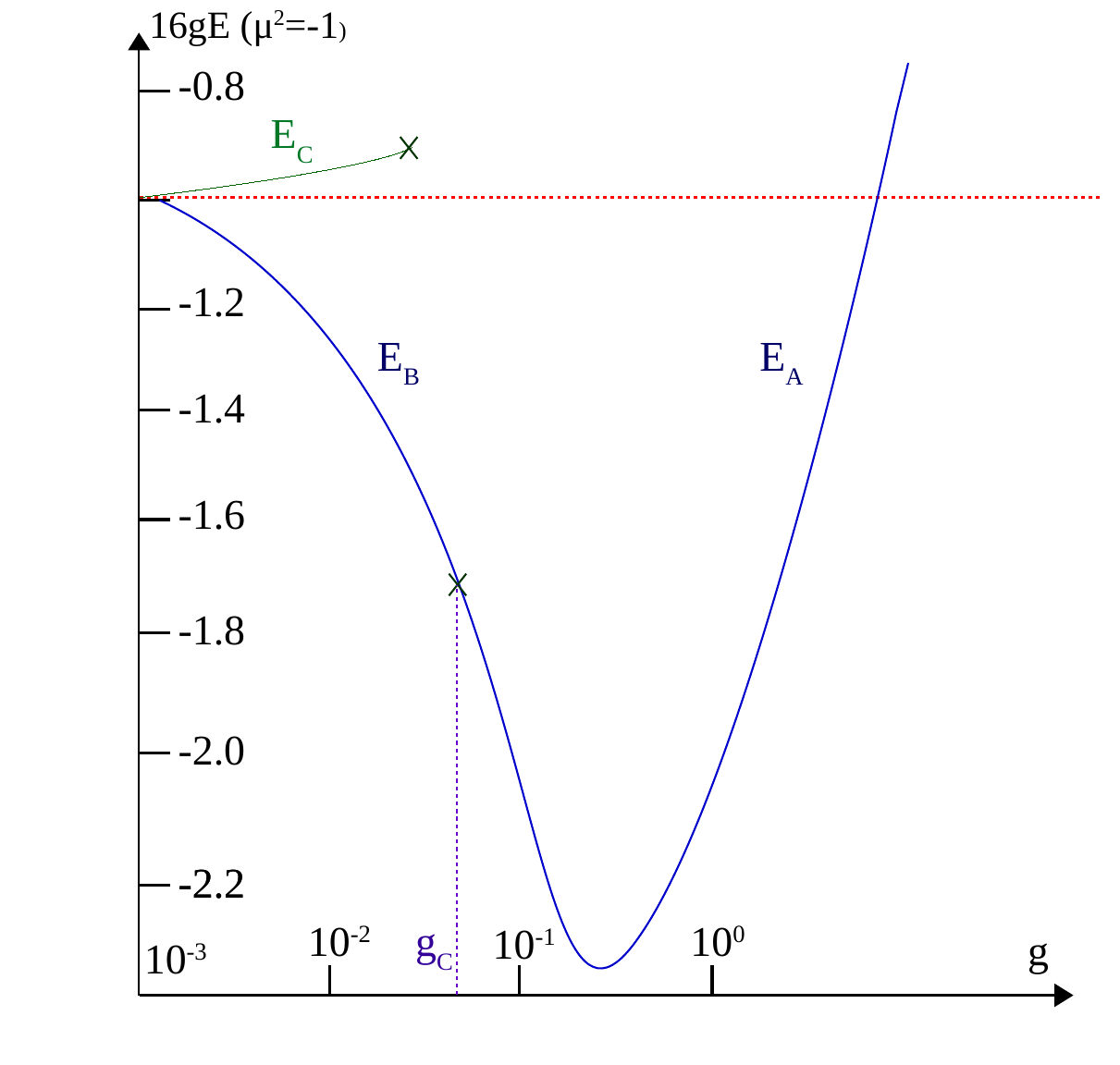}
\caption{The free energy of the real quartic matrix model (\ref{purematrix}) in the disordered, uniform-ordered non-uniform-ordered phases for $\mu^2=-1$. The coupling constants are identified as $g\equiv NC$ and $\mu^2\equiv 2B$. The uniform-ordered phase is metastable in this model \cite{Shimamune:1981qf}. }\label{free_energy}
\end{center}
\end{figure}

The three phases A), B) and C) in the real quartic matrix model are given explicitly by the following density eigenvalues  \cite{Shimamune:1981qf}
\begin{eqnarray}
A)~:~\rho(z)&=&\frac{1}{N\pi}(2Cz^2+B+C\delta^2)\sqrt{\delta^2-z^2},\nonumber\\
&&-B\leq -B_c=2\sqrt{NC}.
\end{eqnarray}
\begin{eqnarray}
B)~:~\rho(z)&=&\frac{2C|z|}{N\pi}\sqrt{(z^2-\delta_1^2)(\delta_2^2-z^2)},\nonumber\\
&&-B\geq -B_c=2\sqrt{NC}.
\end{eqnarray}
\begin{eqnarray}
C)~:~\rho(z)&=&\frac{1}{\pi N}(2C z^2 +2\sigma C z +B +2C\sigma^2+C\tau^2)\nonumber\\
&\times &\sqrt{((\sigma+\tau)-z)(z-(\sigma-\tau))},\nonumber\\
&&-B\geq -B_c=\sqrt{15}\sqrt{NC}.
\end{eqnarray}
The expressions of the various cuts $\delta$, $\delta_1$, $\delta_2$, $\sigma$ and $\tau$ in terms of the parameters of the model can be found in \cite{Shimamune:1981qf}. Following Shimamune it is then not very difficult to show that the free energy $E_C$ in the uniform-ordered phase is always higher than the free energies $E_A$ and $E_B$ in the disordered and non-uniform-ordered phases and hence the uniform-ordered phase is metastable in this model. In particular, the energies $E_B$ and $E_C$ are given explicitly by \cite{Shimamune:1981qf}
\begin{eqnarray}
E_B&=&-\frac{B^2}{4NC}+\frac{1}{4}\ln \frac{NC}{4B^2}-\frac{3}{8}\nonumber\\
E_C&=&-\frac{B^2}{4NC}+O(C^0)~,~C\longrightarrow 0.
\end{eqnarray}
In the limit $C\longrightarrow 0$ we have $E_B<E_C$ because of the logarithmic contribution. See figure (\ref{free_energy}).

Similarly, the main difference between the real quartic matrix model (\ref{purematrix}) from the one hand, and the multitrace matrix model (\ref{multitrace}) on the other hand, lies in the fact that the uniform-ordered phase is stable in the latter. This important fact can be seen in both the saddle point equation and the Monte Carlo algorithm.  For example, the saddle point equation (\ref{spe}) of the multitrace matrix model (\ref{multitrace}) can be solved exactly to determine the boundaries between the three phases A), B) and C) and the location of the triple point by following the remarkable work of \cite{Saemann:2010bw} who solved a much larger class of multitrace matrix models.

The important point to stress here is the fact that the uniform-ordered phase becomes stable within the multitrace matrix model (\ref{multitrace}), in contrast to the case of the pure matrix model (\ref{purematrix}), which can be shown using a mean-field-like analysis as follows.

First, by using the invariance of the partition function of the multitrace matrix model (\ref{multitrace}) under $M\longrightarrow (1+\epsilon)M$ we can derive the Schwinger-Dyson identity
\begin{eqnarray}
\frac{\langle V\rangle}{N^2}=\frac{1}{4}+\frac{\tilde{B}}{2}\langle \frac{1}{N}{\rm Tr}M^2\rangle.
\end{eqnarray}
Next we can use the classical configurations $M=0$,  $M=m_B\gamma$ and $M=m_C{\bf 1}$ deep inside the disordered, non-uniform-ordered and uniform-ordered phases to estimate the energies $E_A$, $E_B$ and $E_C$ respectively.

Thus, we calculate $m_C^2=-\tilde{B}/2(\tilde{C}+\tilde{C}^{\prime})$ which indicates that the model for $\tilde{B}<0$ is only stable if $\tilde{C}>-\tilde{C}^{\prime}$. The energy $E_C$ in the uniform configuration  is then estimated to be given by $E_C=(\tilde{C}+\tilde{C}^{\prime}-\tilde{B}^2)/4(\tilde{C}+\tilde{C}^{\prime})$. Similarly, we calculate $m_B^2=-\tilde{B}/2\tilde{C}$ using the fact that $\gamma^2={\bf 1}$.  The energy $E_B$ in the non-uniform configuration  is then estimated to be given by $E_B=(\tilde{C}-\tilde{B}^2)/4\tilde{C}$. It is not difficult to check that $E_B>E_C$ (and hence the uniform-ordered phase is much more stable than the non-uniform-ordered phase) if and only if $\tilde{C}^{\prime}<0$ and vice versa.

Remark that the effect of the multitrace term in the Ising phase is then only to shift the quartic coupling as $\tilde{C}\longrightarrow \tilde{C}+\tilde{C}^{\prime}$ whereas there is no effect in the stripe phase  from the multitrace term. Hence, the energies $E_B$ and $E_C$ in this mean-field-like approximation are estimated as follows 
\begin{eqnarray}
E_B&=&-\frac{B^2}{4NC}+\frac{1}{4}\ln \frac{N C}{4B^2}-\frac{3}{8}\nonumber\\
E_C&=&-\frac{B^2}{4N(C+C^{'})}+O((C+C^{'})^0)~,~C+C^{'}\longrightarrow 0.\nonumber\\
\end{eqnarray}
Since we can not reach the point $C=0$ (the model becomes unstable there) we can only take the limit  $C\longrightarrow -C^{'}$. In this case we have instead $E_C<E_B$.

\subsection{The phase diagram}
As we have said, the noncommutative phi-four theory (\ref{matrix})+(\ref{phi-four}) and the random multitrace matrix model (\ref{multitrace}) share similar phase diagrams up to and including the triple point. Indeed, the multitrace matrix model (\ref{multitrace}) captures, in fact surprisingly very well, the  phase structure of noncommutative phi-four theory. This phase structure can be probed non-perturbatively using the Monte Carlo algorithm. 

A direct simulation of the noncommutative phi-four theory (\ref{matrix})+(\ref{phi-four}) is very involved for various physical and technical reasons  \cite{GarciaFlores:2009hf,Martin:2004un,Panero:2006bx,Das:2007gm,Ydri:2014rea}. Thus, the numerical results reported here are simply obtained by applying the Monte Carlo method or more precisely  the Metropolis algorithm to the multitrace matrix model (\ref{multitrace}) with the value $C^{'}=-N$ which is the value obtained for the multitrace interaction ${\rm Tr}M{\rm Tr}M^3$ in the multitrace expansion of the kinetic term in the noncommutative phi-four theory (\ref{matrix})+(\ref{phi-four}) on the fuzzy sphere \cite{O'Connor:2007ea,Saemann:2010bw}.

\begin{figure}[htbp]
\begin{center}
\includegraphics[width=15.0cm,angle=-0]{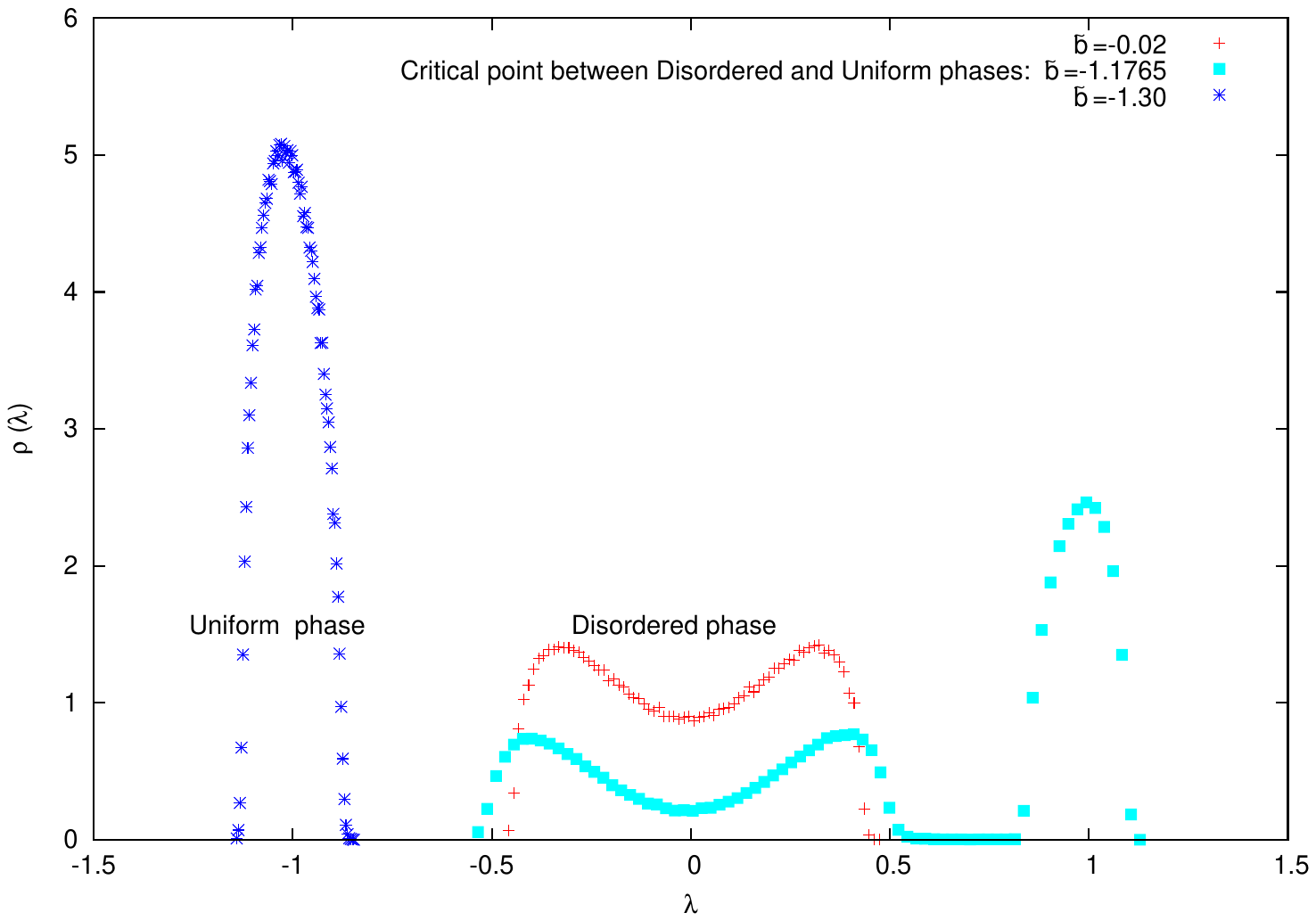}
\caption{The eigenvalue distribution as it transits between the disordered and uniform-ordered phases in the multitrace matrix model (\ref{multitrace}). }\label{commutative}
\end{center}
\end{figure}

\begin{figure}[htbp]
\begin{center}
\includegraphics[width=13.0cm,angle=-0]{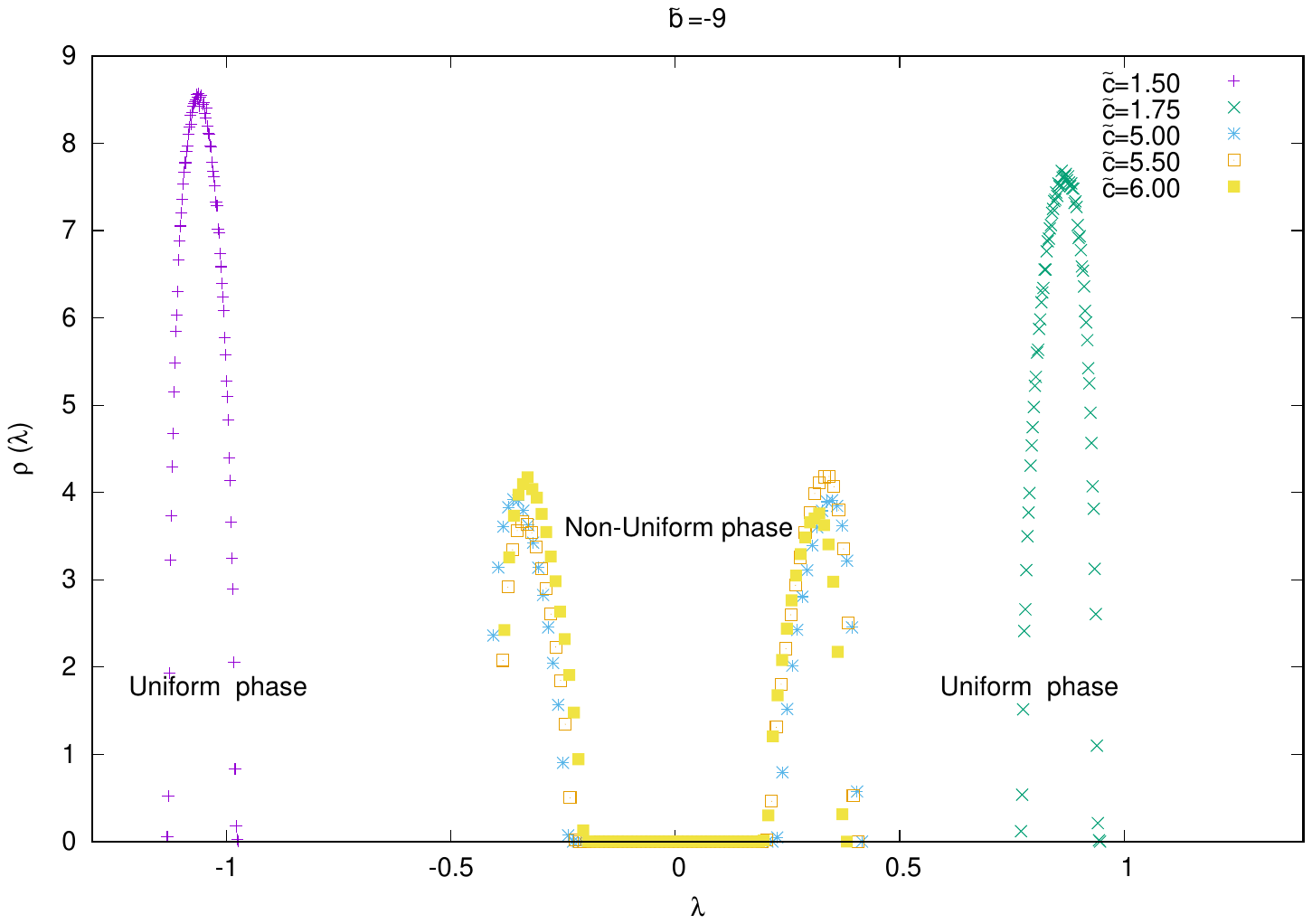}
\caption{The eigenvalue distribution as it transits between the uniform-ordered and non-uniform-ordered phases  in the multitrace matrix model (\ref{multitrace}). }\label{noncommutative}
\end{center}
\end{figure}

\begin{figure}[htbp]
\begin{center}
\includegraphics[width=13.0cm,angle=-0]{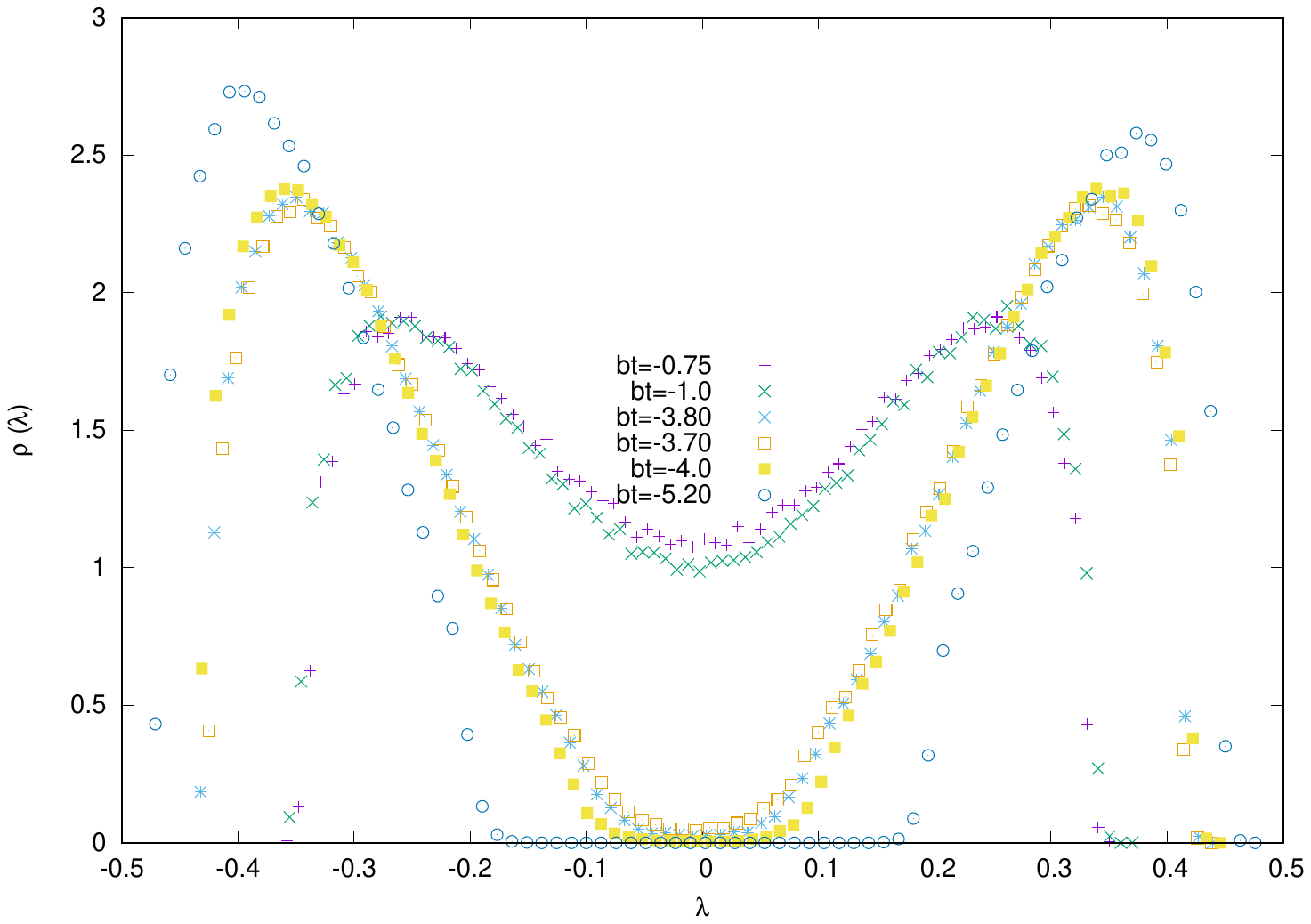}
\caption{The eigenvalue distribution as it transits between the disordered and non-uniform-ordered phases  in the multitrace matrix model (\ref{multitrace}). }\label{matrix_model}
\end{center}
\end{figure}

\begin{figure}[htbp]
\begin{flushleft}
\includegraphics[width=20.0cm,angle=-0]{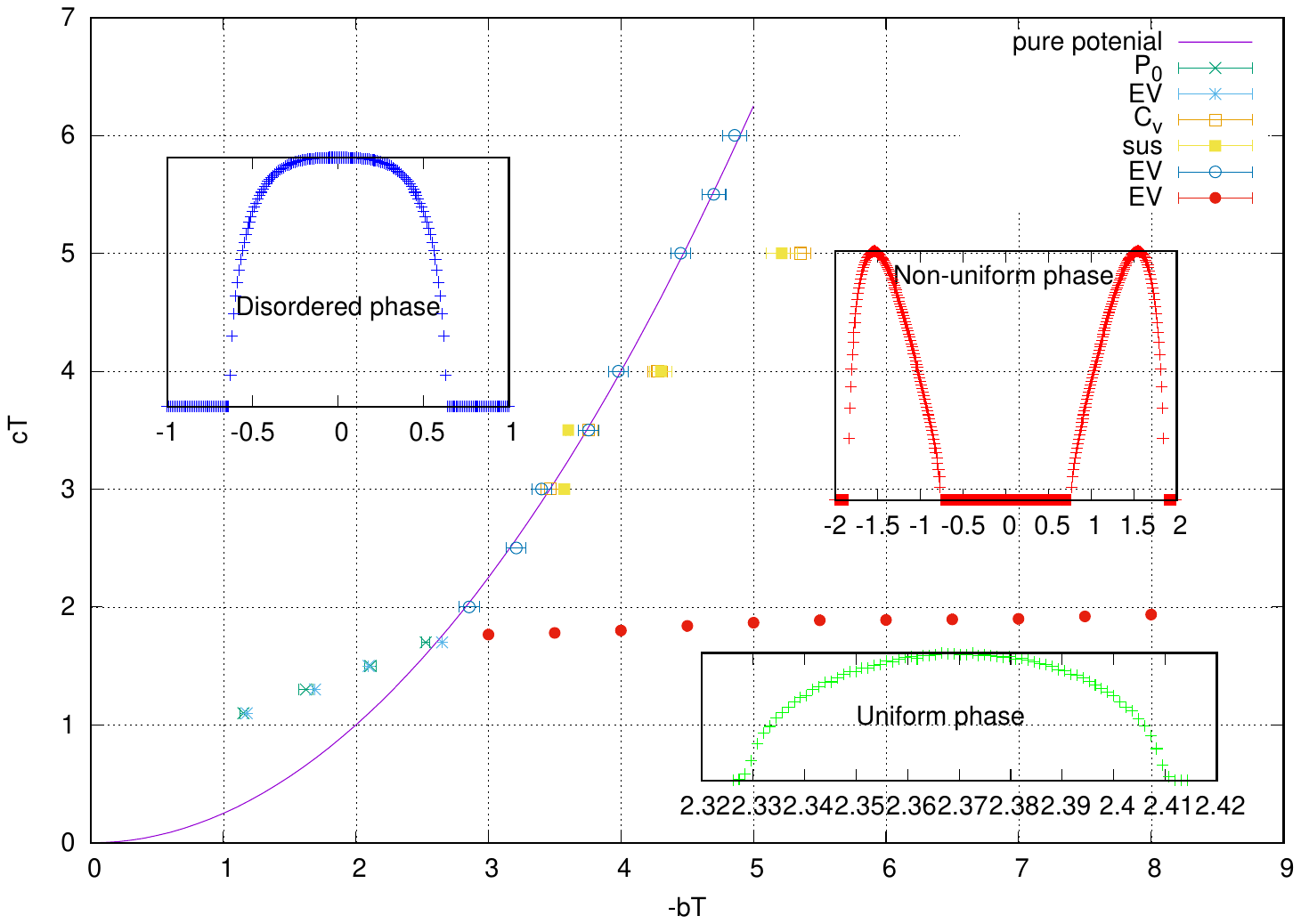}
\caption{The phase diagram of the noncommutative-phi four theory (\ref{matrix})+(\ref{phi-four}) on the fuzzy sphere as approximated by the multitrace matrix model (\ref{multitrace}). }\label{phase_diagram}
\end{flushleft}
\end{figure}

We have for both cases, i.e. for the  noncommutative phi-four theory (\ref{matrix})+(\ref{phi-four}) and for the random multitrace matrix model (\ref{multitrace}), the following phase structure:
\begin{enumerate}
\item A $2$nd order phase transition between disordered ($\Phi\sim 0$) and uniform-ordered ($\Phi\sim 1$) phases at small values of the quartic coupling. This is the usual Ising phase transition \cite{Onsager:1943jn}. The eigenvalue distribution $\rho(\lambda)$ as it transits between the  disordered and uniform-ordered phases in the multitrace matrix model (\ref{multitrace}) is shown in figure (\ref{commutative}).
\item A $2$nd order phase transition between non-uniform-ordered ($\Phi\sim \gamma$ with $\gamma^2=1$) and uniform-ordered ($\Phi\sim 1$) phases which is a continuation  of the Ising transition  to large values of the quartic coupling \cite{brazovkii}. The eigenvalue distribution $\rho(\lambda)$ as it transits between the  uniform-ordered and non-uniform-ordered phases in the multitrace matrix model (\ref{multitrace}) is shown in figure (\ref{noncommutative}).
\item A $3$rd order phase transition between disordered and non-uniform ordered phases. This is the matrix phase transition observed previously in the real quartic matrix model (\ref{purematrix}) (model without kinetic term). The eigenvalue distribution $\rho(\lambda)$ as it transits between the  disordered and non-uniform-ordered phases in the multitrace matrix model (\ref{multitrace}) is shown in figure (\ref{matrix_model}).
\item The three co-existence curves intersect at a triple point. The full phase diagram of the multitrace matrix model (\ref{multitrace}) in the plane $(-\tilde{B},\tilde{C})$, as measured by the Monte Carlo method, is shown in figure (\ref{phase_diagram}) where the coexistence line $\tilde{C}=\tilde{B}_c^2/4$ of the real quartic matrix model (\ref{purematrix}) is also included for comparison. The agreement with the phase diagram of noncommutative phi-four (\ref{matrix})+(\ref{phi-four}) on the fuzzy sphere is extremely favorable.

The observables used in the Monte Carlo measurement of the phase diagram include, in addition to the eigenvalue (EV) distribution  $\rho(\lambda)$, the specific heat $C_v=\langle S^2\rangle-\langle S\rangle^2$ where $S\equiv V_{\rm trunc}$, the susceptibility (${\rm sus}$) defined by $\chi=\langle |{\rm Tr}M|^2\rangle-\langle |{\rm Tr}M|\rangle^2$  and the power in the zero mode $P_0=\langle({\rm Tr}M)^2/N^2\rangle$. We also measure the magnetization $m=\langle |{\rm Tr} M|\rangle$, the total power $P_T=\langle {\rm Tr}M^2\rangle/N$ and the quartic coupling $\langle {\rm Tr}M^4\rangle$. 
\end{enumerate}

\subsection{The uniform-ordered phase $\Rightarrow$ Geometry}
Now, the existence of  the uniform-ordered phase and the Ising phase transition between this uniform-ordered phase and disordered phase signal as usual the spontaneous symmetry breaking of the discrete symmetry $\Phi\longrightarrow -\Phi$. The fundamental underlying hypothesis in this chapter is the statement that the existence of an Ising uniform-ordered phase in a "pure matrix model" (such as the multitrace matrix model (\ref {multitrace}) which does not come with a pre-defined Laplacian operator) is an unambiguous signal for a lurking underlying geometry. In other words, an emergent geometry transition in this scenario (obtained by allowing the value of the coefficient $C^{\prime}$ of the multitrace term to change) is identified with the existence of a stable uniform-ordered phase and a corresponding Ising transition for some values of  the coefficient $C^{\prime}$.

On the other hand, the existence of the non-uniform-ordered phase signals the spontaneous symmetry breaking of translation symmetry (see for example \cite{Mejia-Diaz:2014lza}) which is quite remarkable for two reasons. First,  this breaking is possible even in two dimensions in contrast to the Coleman-Mermin-Wagner theorem \cite{Mermin:1966fe,Coleman:1973ci} and it is due to the fact that noncommutative field theories are non-local by construction. Second, this breaking can be extended to supersymmetric models allowing us to obtain spontaneous supersymmetry breaking which is usually quite hard to obtain otherwise (see \cite{Volkholz:2007kva} for a courageous attempt). The non-uniform-ordered phase is precisely the so-called stripe phase and it is the non-perturbative manifestation of the celebrated phenomena of the UV-IR mixing in noncommutative field theories \cite{Minwalla:1999px}.

 \subsection{Critical exponents $\Rightarrow $ Dimension}
 The uniform-ordered phase is called the Ising phase precisely because we believe that the corresponding transition to the disordered phase is characterized by the universal critical exponents of the Ising model in two dimensions given by the Onsager solution. These critical exponents are defined as usual by the  behavior near the critical point, viz
   \begin{eqnarray}
&&m/N=<|Tr M|>/N \sim (B_c-B)^{\beta}\sim N^{-\beta/\nu}\nonumber\\
&&C_v/N^2 \sim (B-B_c)^{-\alpha}\sim N^{\alpha/\nu}\nonumber\\
&&\chi=<|Tr M|^2>-<|Tr M|>^2\sim (B-B_c)^{-\gamma}\sim N^{\gamma/\nu}\sim N^{2-\eta}\nonumber\\
&&\xi\sim|B-B_c|^{-\nu}\sim N.\label{cb}
\end{eqnarray}
There are in total six critical exponents, the above five plus the critical exponent $\delta$ which controls the equation of state, but only two are truly in
dependent because of the so-called scaling laws. The Onsager solution of the Ising model in two dimensions gives the following celebrated values \cite{Onsager:1943jn}

  \begin{eqnarray}
\nu=1~,~\beta=1/8~,~\gamma=7/4~,~\alpha=0~,~\eta=1/4~,~\delta=15.
\end{eqnarray}
This fundamental result can be established for the multitrace matrix model (\ref{multitrace}) following the same steps used in deriving the critical exponents of the multitrace model  (\ref{example})+(\ref{example1}). This is a very delicate exercise which is essential in order to confirm the existence of the uniform-ordered phase beyond any doubt and thus confirm the fact that we are really dealing with a two-dimensional space geometry.

\subsection{Wigner's semi-circle law $\Rightarrow $ Free propagator (Metric)}

As it turns out, a free noncommutative scalar field theory with mass parameter $m^2$ (the theory given by (\ref{matrix})+(\ref{phi-four}) with zero interaction) is characterized by a Wigner's semi-circle law  in stark contrast to free commutative scalar field theory. This stems from the fact that planar diagrams dominate over the non-planar ones in the limit of infinite cutoff \cite{Steinacker:2005wj,Nair:2011ux}.

Explicitly, a noncommutative phi-four on a $d$-dimensional noncommutative spacetime ${\bf R}^d_{\theta}$ reads in position representation
\begin{eqnarray}
S&=&\int d^dx \big(\frac{1}{2}\partial_i \Phi\partial_i {\Phi}+\frac{1}{2}m^2{\Phi}^2+\frac{g}{4}{\Phi}_*^4\big).
\end{eqnarray} 
The first step is to regularize this theory in terms of  a finite $N_0$-dimensional matrix $\Phi$ and rewrite the theory in matrix representation. Then, we diagonalize the matrix $\Phi$, i.e. we write $\Phi=U^{\dagger}\Lambda U$ where $U\in U(N_0)$ and $\Lambda$ is the matrix of eigenvalues $\lambda_i$. The measure becomes $\int \prod_i d\lambda_i\Delta^2(\Lambda)\int dU$ where  $dU$ is the Haar measure and $\Delta^2(\Lambda)=\prod_{i<j}(\lambda_i-\lambda_j)^2$ is the Vandermonde determinant. 

The effective probability distribution of the eigenvalues $\lambda_i$ can be determined uniquely from the behavior of the expectation values $\langle \int  d^dx \Phi_*^{2n}(x)\rangle$. These objects clearly depend only on the eigenvalues $\lambda_i$ and are computed using a sharp UV cutoff $\Lambda$. 

If we are only interested in the eigenvalues of the scalar matrix $\Phi$ then the free theory $g=0$ can be replaced by the effective matrix model \cite{Steinacker:2005wj} 
\begin{eqnarray}
S=\frac{2{\cal N}}{\delta^2}Tr\Phi^2.
\end{eqnarray} 
This result can be traced to the fact that planar diagrams dominates over the non-planar ones in the limit $\Lambda\longrightarrow\infty$. This means in particular that the eigenvalues $\lambda_i$ are distributed according to the famous Wigner semi-circle law with $\delta$ being the largest eigenvalue, viz
\begin{eqnarray}
\rho(x)=\frac{2}{\pi\delta^2}\sqrt{\delta^2-x^2}~,~-\delta\leq x\leq +\delta.\label{pred1}
\end{eqnarray} 
 The most important case is $d=2$ since the case $d=4$ is eliminated by the results of the critical exponents. In this case we find the radius \cite{Steinacker:2005wj} 
\begin{eqnarray}
\delta(m,\Lambda)=\frac{1}{\pi}\ln(1+\frac{\Lambda^2}{m^2})~,~d=2.\label{pred2}
\end{eqnarray}
In two dimensions the regulator $\Lambda$ originates either from the noncommutative torus or from the fuzzy sphere.

The behavior of the squared-radius $\delta^2$ on the noncommutative torus ${\bf T}^2_{\theta}$ (with cutoff $\Lambda=\sqrt{{N\pi}/{\theta}}$) is found to be different from the above sharp UV cutoff result (\ref{pred2}) due to the different behavior of the propagator for large momenta. This behavior can also be excluded in our Monte Carlo data and by hindsight we know that this should be indeed the case since the original multitrace approximation was obtained from noncommutative phi-four theory on the fuzzy sphere.


We are left therefore with the case of the fuzzy sphere ${\bf S}^2_N$ where we have  $N_0\equiv N$ since the scalar field $\Phi$ is an $N\times N$ matrix $\phi$ given by $\phi=\sqrt{2\pi/Na}\Phi$. In this case the cutoff $\Lambda$ is given in terms of the matrix size $N$ and the radius $R$ of the sphere by the relation $\Lambda={N}/{R}$ and the mass parameters $B$ and $m^2$ are related by $m^2={B}/{aR^2}$. By using $\tilde{B}=B/N^{3/2}$, and choosing $a=2\pi/N$ so that $\Phi=\phi$, we obtain
 \begin{eqnarray}
\frac{\Lambda^2}{m^2}=\frac{2\pi N}{B}=\frac{2\pi}{\sqrt{N}\tilde{B}}.
\end{eqnarray}
In the limit $B\longrightarrow \infty$ we get the squared-radius $\delta^2=2N/B$ which corresponds to the Gaussian matrix model $B {\rm Tr} M^2$, viz $B=2N/\delta^2$. This can also be obtained by taking the limit  $B\longrightarrow \infty$ of the one-cut solution. The Wigner's semi-circle law is thus also obtained for the free quadratic matrix model $B{\rm Tr} M^2$ with a squared-radius given simply by
.\begin{eqnarray}
\delta^2=\frac{2N}{B}.\label{pred0}
\end{eqnarray}
In fact, the eigenvalues distribution of a free scalar field theory on the fuzzy sphere ${\bf S}^2_N$ with an arbitrary kinetic term, viz $S={\rm Tr}(M{\cal K}M+BM^2)/2$, where ${\cal K}(0)=0$ and ${\cal K}$ is diagonal in the basis of polarization tensors $T_l^m$,  is always given by a Wigner semicircle law \cite{Nair:2011ux}. In this case the radius-squared is given by
\begin{eqnarray}
\delta^2=\frac{4f(B)}{N}~,~f(B)=\sum_{l=0}^{N-1}\frac{2l+1}{{\cal K}(l)+B}.
\end{eqnarray}



\begin{figure}[htbp]
\begin{center}
\includegraphics[width=13cm,angle=-0]{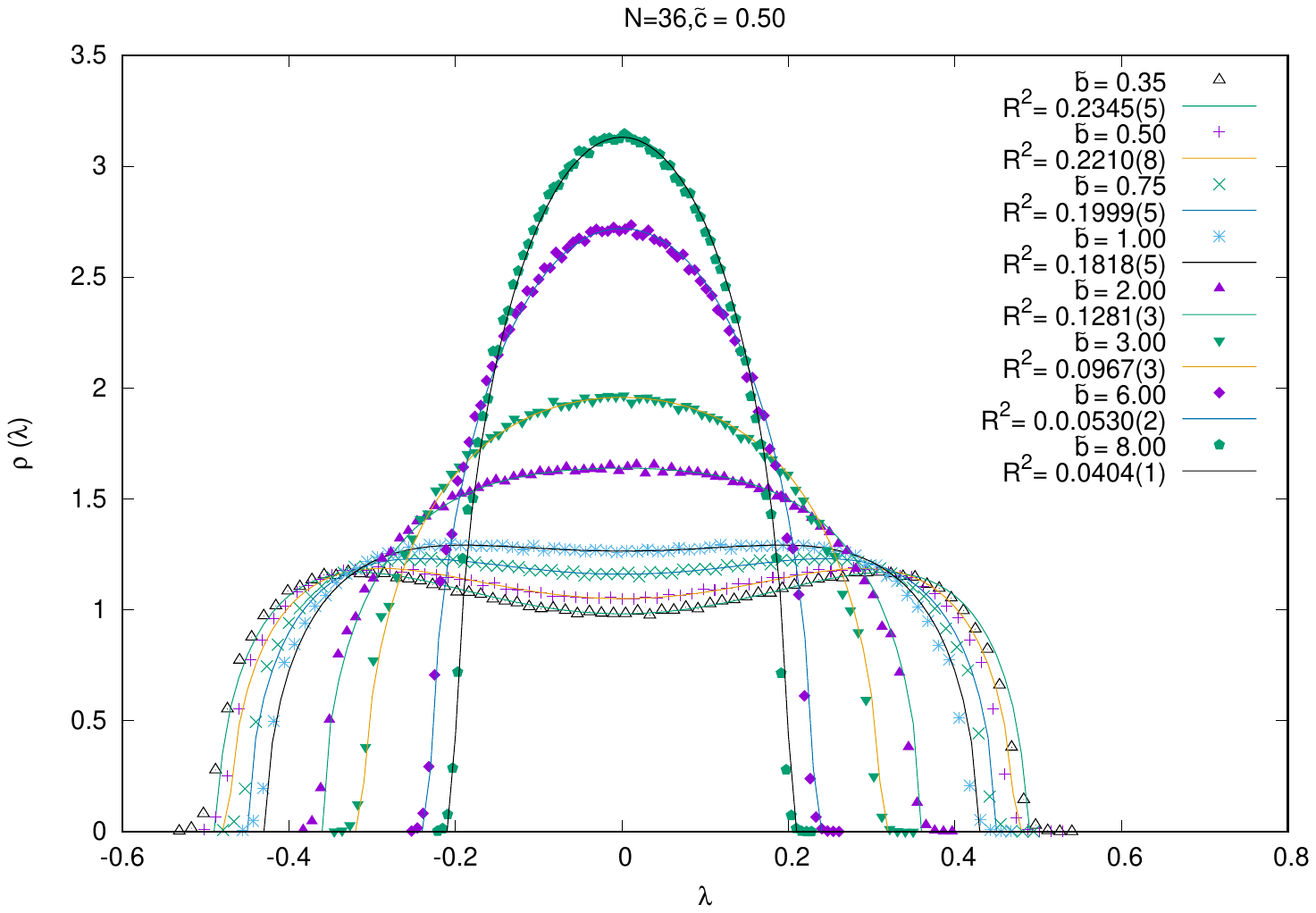} 
  \includegraphics[width=13cm,angle=-0]{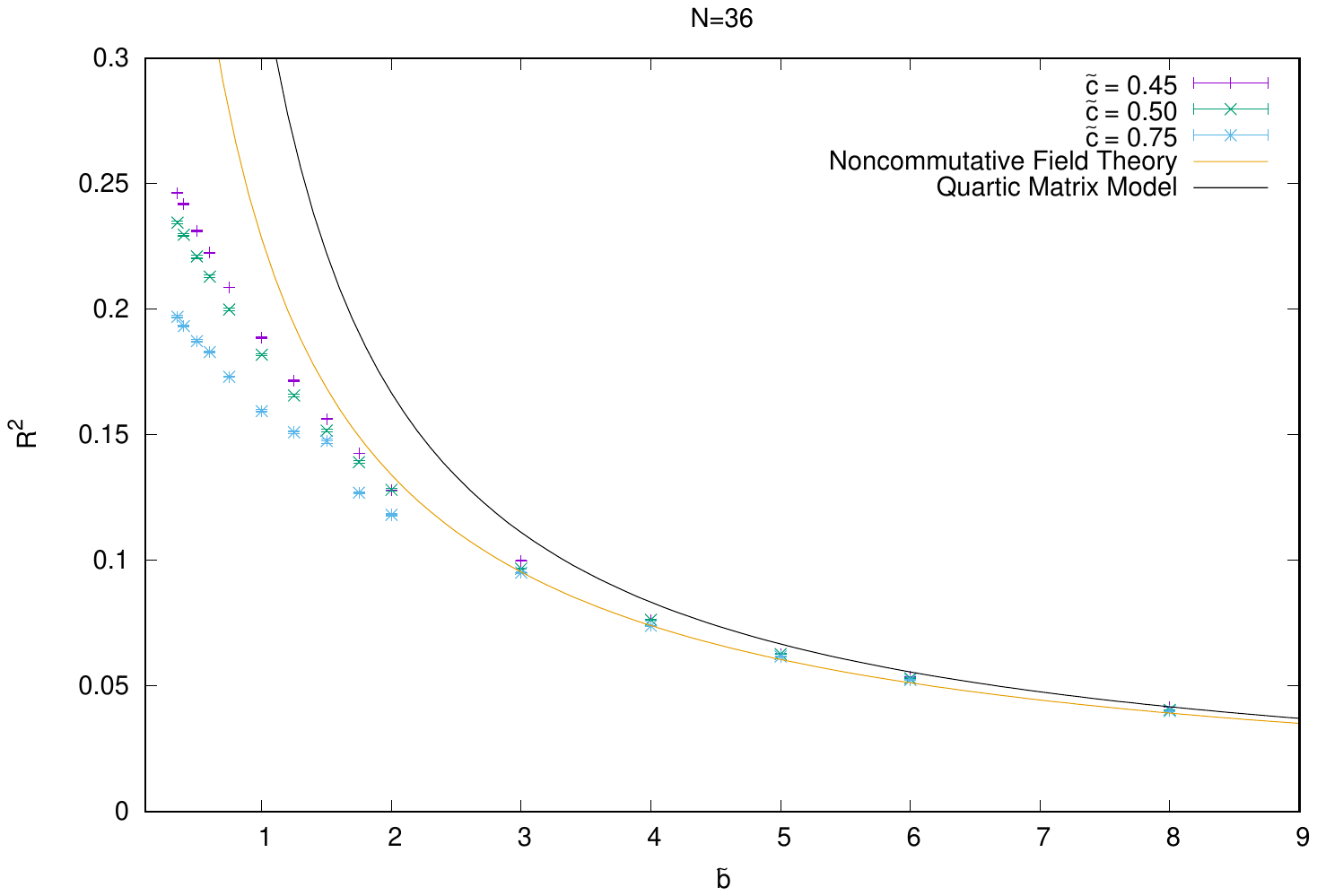}
\end{center}
\caption{Wigner's semicircle law.}\label{wigner}
\end{figure}

The Wigner's semi-circle law computed near the origin using the multitrace matrix model (\ref{multitrace}) should then be compared with the noncommutative field theory prediction (\ref{pred1})+(\ref{pred2}) for negative values of the multitrace coupling $C^{\prime}$. But for positive values of $C^{\prime}$ the behavior should be compared with the prediction of pure matrix model (\ref{pred0}). This will allow us to confirm directly that the underlying space or emergent geometry is indeed a fuzzy sphere ${\bf S}^2_N$. See figure (\ref{wigner}). But for more detail see \cite{Ydri:2015zsa}.

\subsection{Emergent Yang-Mills matrix model $\Rightarrow$ Noncommutative gauge theory}
The emergent geometry of the model (\ref{multitrace}) can be exhibited  in a drastic way by exhibiting an emergent non-commutative gauge theory on a fuzzy sphere ${\bf S}^2_N$ in the uniform-ordered phase \cite{Ydri:2016daf}. This gauge theory is an $SO(3)$-symmetric Yang-Mills matrix model of the IKKT-type, i.e. it is a matrix model which involves three covariant Hermitian matrix coordinates $X_a$ with $SO(3)$ rotational symmetry and $U(N)$ gauge invariance.

In order to see this result we need to assume that we are in the uniform-ordered phase and thus the matrix $M$ can be expanded around the identity matrix ${\bf 1}$. For simplicity, we further assume that the dimension is even, i.e. the matrix $M$ is $2N\times 2N$. Then, without any loss of generality, we can expand the matrix $M$ as
\begin{eqnarray}
M=M_0{\bf 1}_{2N}+M_1~,~Tr M_1=0.
\end{eqnarray}
Hence
\begin{eqnarray}
M_1=\sigma_aX_a~,~
M_0=a+m,
\end{eqnarray}
where $\sigma_a$ are the standard Pauli matrices, $m$ is the fluctuation in the zero mode, and $X_a$ are three hermitian $N\times N$ matrices. By substitutin, we obtain immediately the model
\begin{eqnarray}
Z&=&\int {\cal D} X_a \exp(-V[X]))\int dm\exp(-f[m]).
\end{eqnarray}
The potential $V$ is given by an $SO(3)$-symmetric three-matrix model which is precisely equal to the $D=3$ Yang-Mills matrix model defined by 
\begin{eqnarray}
V=-CTr[X_a,X_b]^2+2CTr(X_a^2)^2+2(B+6Na^2C^{\prime})TrX_a^2+4iN aC^{\prime}\epsilon_{abc} TrX_aX_bX_c.\label{three-matrix}
\end{eqnarray}
The integration over $m$ can be done and the result consists of some function of $TrX_a^2$ and $i\epsilon_{abc}TrX_aX_bX_c$. This next-to-leading contribution (in $1/N$)  is essentially the one-loop result and it is by construction subleading compared to $V$.

The Chern-Simons term is proportional to the value $a$ of the order parameter. Thus, it is non-zero only in the Ising phase, and as a consequence, by tuning 
the parameters appropriately to the region in the phase diagram where the Ising phase exists, we will induce a non-zero value for the Chern-Simons. This Chern-Simons term is effectively the Myers term responsible for the condensation of the geometry \cite{Myers:1999ps,Azuma:2004zq}.

The above three-matrix model is then precisely a random matrix theory describing noncommutative gauge theory on the fuzzy sphere, where the first term is the Yang-Mills piece, whereas the second and third terms combine to give mass and linear terms for the normal scalar field on the sphere (recall that the index $a$ runs from $1$ to $3$). This is essentially the random matrix theory describing noncommutative gauge theory on the fuzzy sphere  found in   \cite{Steinacker:2003sd}. However, we should emphasis here that we have obtained dynamically this gauge theory on the fuzzy sphere by going to the phase where a non-zero uniform order persists and thus securing a non-zero Chern-Simons term crucial for the condensation of the fuzzy sphere geometry. In \cite{Steinacker:2003sd}, this was achieved by constraining the matrix $M$ directly by hand in a particular way.

In summary, by expanding the above three-matrix model (\ref{three-matrix}) around its background solution (which is a fuzzy sphere solution)  we obtain a noncommutative $U(1)$ gauge theory on the fuzzy sphere \cite{Castro-Villarreal:2004ulr,OConnor:2006iny}.  Thus, this Yang-Mills matrix model sustains itself emergent geometry \cite{Delgadillo-Blando:2007mqd,Delgadillo-Blando:2008cuz}. Furthermore, we know that noncommutative $U(1)$ gauge theory is effectively a gravity theory. Thus, this Yang-Mills matrix model sustains also emergent gravity \cite{Steinacker:2010rh,Steinacker:2016vgf,Steinacker:2007dq,Yang:2008fb,Yang:2006dk}.

\subsection{Emergent geometry}

In summary, the non-uniform-ordered phase is important for noncommutative geometry and matrix models whereas the uniform-ordered phase  is essential to commutative geometry. Indeed, the central proposal of this chapter is to turn the original logic of noncommutative geometry and its matrix models on its head by starting from matrix models and attempt to reach noncommutative geometry and not the other way around. This will be precisely/explicitly done by searching in the phase diagram for a uniform-ordered phase.

Hence, we will take as first principle the random multitrace matrix model (\ref{multitrace}), or a generalization  thereof, then attempt to reach the noncommutative field theory (\ref{matrix})+(\ref{phi-four}) by searching for a uniform-ordered phase in the phase diagram of the former.  This relies on three facts:
\begin{itemize}
\item First, multitrace matrix models such as  (\ref{multitrace}) do not involve in their definition a spectral triple specifying the geometry like those implicitly defining noncommutative field theories such as (\ref{matrix})+(\ref{phi-four}).
\item Second, the uniform-ordered phase is a commutative order requiring the existence of an underlying space which is a priori commutative.
\item Third, the underlying geometry turns out to be quantized, i.e. noncommutative geometry because of the existence of a triple point in which a commutative order (uniform/Ising), a noncommutative order (non-uniform/stripe) and a matrix order coexist. 
\end{itemize}
Thus, from the existence of a uniform-ordered phase  and a corresponding Ising phase transition, for some values of  the coefficient  of the multitrace coupling, we can infer the existence of an underlying space and an emergent geometry transition. The dimension of this emergent space can be determined from the critical exponents of the uniform-to-disordered phase transition by virtue of scaling and universality properties of second order phase transitions \cite{Wilson:1973jj}. This exercise was done in great detail in \cite{Ydri:2015zsa}.

Furthermore, the existence of a uniform-ordered phase and an Ising transition from a disordered phase to this uniform-ordered phase in a pure matrix model such as the multitrace matrix model (\ref{multitrace}) is indicative that this multitrace matrix model falls (for negative values of the multitrace coupling $C^{\prime}$) in the universality class of the noncommutative phi-four theory (\ref{matrix})+(\ref{phi-four}).  In other words, the multitrace matrix model (\ref{multitrace}) captures the same geometry as the geometry specified by the spectral triple which went implicitly into the definition of the noncommutative phi-four theory (\ref{matrix})+(\ref{phi-four}). For positive values of the  multitrace coupling $C^{\prime}$ the multitrace matrix model (\ref{multitrace}) does not sustain a uniform-ordered phase and hence it falls in the universality class of the real quartic matrix model (\ref{purematrix}). We have then an emergent geometry as we vary $C^{\prime}$ from positive to negative values. The dimension of the underlying space, or more precisely this emergent geometry, are determined from the measurement/calculation of the critical exponents characterizing the Ising transition.

Hence, it is the statistical physics of the commutative (uniform/Ising) phase of the multitrace matrix model that captures the geometry.  And, as it turns out,   this geometry is quantum in the sense that it is emergent (reminiscent of second quantization of geometry) but it is also quantized in the sense that it is noncommutative (first quantization of geometry).

Indeed, this emergent quantum geometry is necessarily noncommutative since the multitrace matrix model (\ref{multitrace}) includes necessarily a stripe or non-uniform-ordered phase which is a noncommutative order by excellence (a transition from a disordered phase to a non-uniform-ordered phase is necessarily generated by the  single trace terms of the multitrace matrix model (\ref{multitrace})). This important fact can be further confirmed by studying the transition from non-uniform-ordered to uniform-ordered and verifying that this transition is also $2$nd order (similarly to the Ising transition) and it is in fact a continuation of the Ising transition to larger values of the quartic coupling.  

The precise metric on the emergent geometry can be fixed by studying the Wigner's semi-circle law near the origin, i.e. by studying the eigenvalue distribution of the matrix $M$ for vanishingly small values of the quartic coupling. This eigenvalue distribution captures clearly the properties of the free propagator.

This fourth ingredient in our proposal allows us, for example if the dimension is determined to be two from the critical exponents, to discriminate  between the noncommutative torus ${\bf T}^2_{\theta}$ and the fuzzy sphere ${\bf S}^2_N$ which can be both used to regularize non-perturbatively the Moyal-Weyl plane. In fact, these two spaces lead to different behavior of the radius $ \delta$ of the Wigner's semi-circle law as a function of the matrix size $N$  and the mass parameter $B$. This issue was discussed in the previous section.



The fifth and final ingredient in our emergent geometry proposal consists in verifying explicitly the geometrical content of the multitrace matrix model (\ref{multitrace}) by expanding the matrix $M$ around the uniform-ordered configuration $M_0=m {\bf 1}$. We obtain an $SO(3)-$symmetric three-matrix model with a Chern-Simons term proportional to the value $m$ of the order parameter in the uniform-ordered phase, i.e. to the magnetization. This three-matrix model describes a noncommutative gauge theory on the fuzzy sphere \cite{Steinacker:2003sd}. The Chern-Simons term is precisely Myers term responsible for the condensation of the geometry \cite{Myers:1999ps,Azuma:2004zq}.

In summary, our proposal for "emergent geometry from random multitrace matrix models" consists of five ingredients  \cite{Ydri:2020efr,Ydri:2017riq,Ydri:2016daf,Ydri:2015zsa}:
\begin{enumerate}
\item The random multitrace matrix model (\ref{multitrace}) is taken as  "first principle". There is no Laplacian entering the definition of this action and thus there is no geometry a priori. First, we need to determine whether or not this pure matrix model contains in its phase diagram a uniform-ordered phase and an Ising transition from this uniform-ordered phase to disordered phase.
\item The existence of a uniform-ordered phase in a multitrace matrix model such as (\ref{multitrace}) signals the existence of an underlying space and an emergent geometry as we vary the multitrace coupling $C^{\prime}$ from positive to negative values. By using scaling and universality of second order phase transitions we can determine the dimension of the underlying space or emergent geometry from the critical exponents of the Ising line.
\item From the existence of a stripe or non-uniform-ordered phase with a transition line between uniform-ordered and non-uniform-ordered phases which is second order and a continuation of the Ising line we can infer that the underlying space or emergent geometry is in fact noncommutative.
\item The study of the Wigner's semi-circle law  of the multitrace matrix model (\ref{multitrace}) at the origin will allow us to discriminate decisively between the behavior of the noncommutative field theory and the real quartic matrix model.
\item In some interesting cases the expansion around the uniform-ordered phase exhibits the geometrical content of the multitrace matrix model as a gauge theory on a noncommutative background.
\end{enumerate}

\subsection{Wilsonian renormalization group equation}
The partition function of the multitrace matrix model ${\rm Tr}M{\rm Tr}M^3$ can be rewritten in the usual form

               \begin{eqnarray}
 Z&=&\int {\cal D}M~\exp(-N Tr_{N} V(M))\nonumber\\
 V(M)&=&\frac{g_2}{2}M^2+\frac{g_4}{4}M^4+\frac{g}{3N}(Tr_NM)M^3.
               \end{eqnarray}
               We have 
               \begin{eqnarray}
                 g_2=\pm 1~,~g_4=\frac{\tilde{C}}{\tilde{B}^2}~,~g=\frac{3}{4}\frac{\tilde{D}}{{\tilde{B}^2}}.
               \end{eqnarray}
               We choose $g_2=-1$ since $B$ is taken to be negative which is the region of interest for noncommutative scalar phi-four theories. Note also that the multitrace term is of the same order as the quartic term and stability requires that $\tilde{C}>-\tilde{D}$ which was noted previously on several occasions.

Now, by employing repeatedly large $N$ factorization of multi-point functions  of $U(N)-$invariant objects into product of one-point functions \cite{Higuchi:1994rv,Higuchi:1993nq,Higuchi:1993tg,Higuchi:1994dv,Kawamoto:2013laa} we can effectively convert multitrace terms into singletrace terms. In addition, by expanding the multitrace and quartic terms and using the Schwinger-Dyson identities in the cubic potential we can rewrite the above multitrace matrix model as a cubic potential of the form \cite{Ydri:2020efr}
                \begin{eqnarray}                 
                 &&V(M)=\frac{g_2^{\prime}}{2}M^2+\frac{g_3}{3}M^3\nonumber\\
                 &&g_2^{\prime}=\pm\sqrt{(g_2+\frac{g_4}{g_3^2})^2-2g_4}\nonumber\\
                 &&g_3=ga_1=\frac{g}{N}\langle {\rm Tr}M\rangle.
  \end{eqnarray}
  This is essentially a mean-field-approximation which seems to be exact in the large $N$ limit. Remark that the cubic coupling is proportional to the magnetization, i.e. $a_1\equiv m$.
  
\begin{figure}[htbp]
\begin{center}  
  \includegraphics[width=9cm,angle=-0]{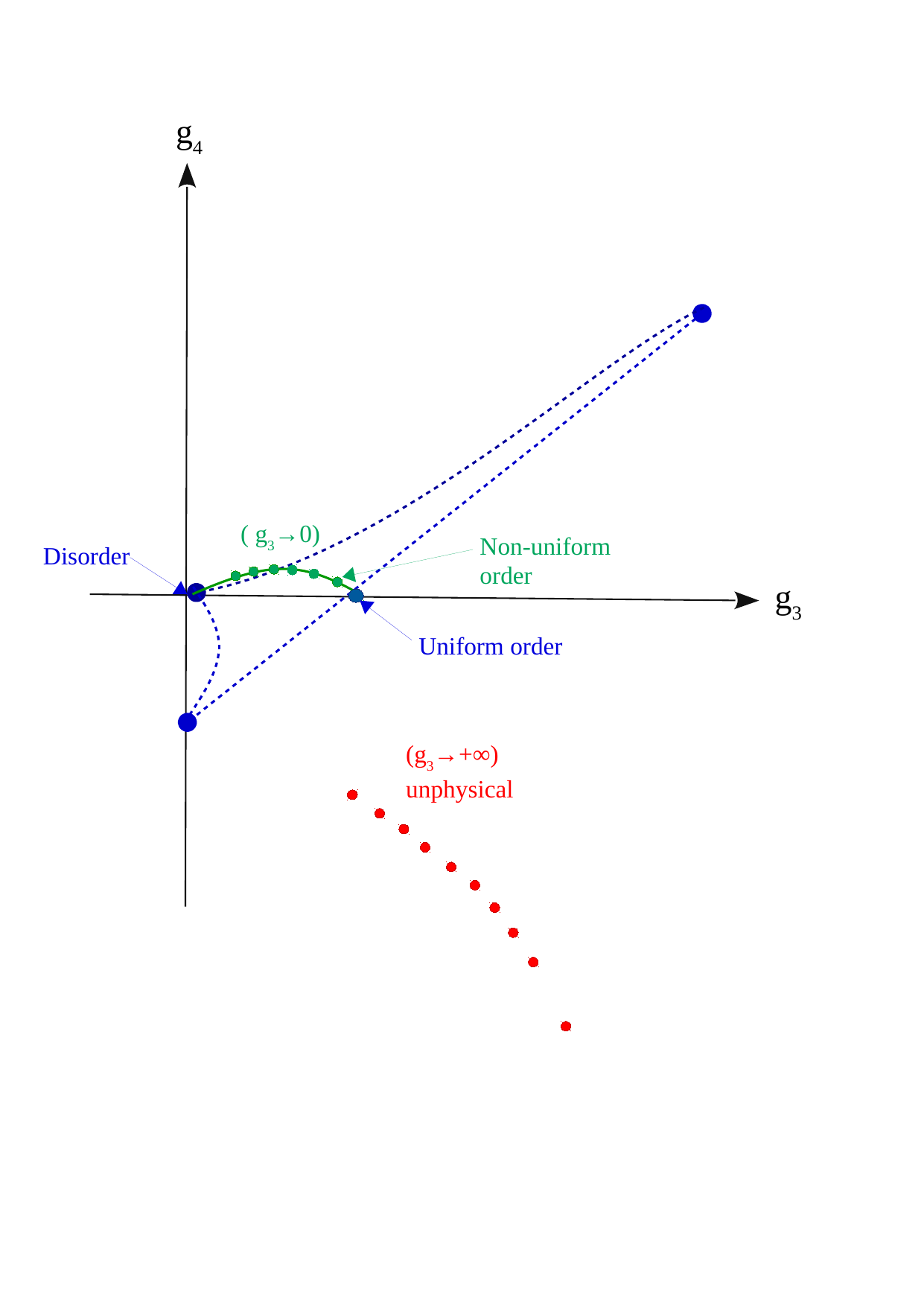}
\end{center}
\caption{The RG fixed points of the multitrace matrix model (\ref{multitrace}) (green and red lines) compared with the RG fixed points of the singletrace cubic-quartic matrix model (blue line)  \cite{Ydri:2020efr}.}\label{sketch2}
\end{figure}

  This cubic potential admits the standard Liouville quantum gravity fixed point  given by \cite{Higuchi:1994rv,Higuchi:1993nq,Higuchi:1993tg,Higuchi:1994dv}
  \begin{eqnarray}                 
                 \frac{g_3}{(g_{2*}^{\prime})^{\frac{3}{2}}}=\epsilon\equiv\frac{1}{432^{1/4}}.
  \end{eqnarray}
  The solution which goes through the two standard fixed points $(0,0)$ and  $(0,\epsilon)$ of the pure cubic potential is given explicitly by  \cite{Ydri:2020efr}
  \begin{eqnarray}                 
                 g_{4*}=g_3^2\bigg[g_3^2-\sqrt{g_3^4-2g_2g_3^2+\big(\frac{g_3}{\epsilon}\big)^{4/3}}-g_2\bigg].\label{result}
  \end{eqnarray}
  The behavior for $g_3^2\longrightarrow \infty$ and $g_3^2\longrightarrow 0$ is given explicitly by
   \begin{eqnarray}                 
                 g_{4*}=-\frac{1}{2}(\frac{g_3}{\epsilon})^{4/3}~,~g_3^2\longrightarrow \infty.
   \end{eqnarray}
   And
   \begin{eqnarray}                 
                 g_{4*}=g_3^2(-g_2-(\frac{g_3}{\epsilon})^{2/3})~,~g_3^2\longrightarrow 0.
   \end{eqnarray}
   We have then a critical line of fixed points (\ref{result}) which interpolate smoothly between the fixed point $(0,0)$ (interpreted as the fixed point of the $3$rd order matrix  coexistence line)  and  the fixed point $(0,\epsilon)$  (interpreted as the fixed point of the $2$nd order commutative/Ising  coexistence line)  going through a maximum in between. This critical line, restricted to the positive quadrant since the behavior $g_{4*}\longrightarrow -\infty$ as $g_3\longrightarrow + \infty$ is unphysical for noncommutative field theory, is interpreted as the critical line of fixed points associated with the noncommutative/stripe coexistence line. The RG fixed points are sketched on figure (\ref{sketch2}).

\section{Emergent geometry from Yang-Mills matrix models}

\subsection{IKKT model and  other Yang-Mills matrix models}


The one-dimensional reduction of ${\cal N}=1$ supersymmetric Yang-Mills theory in $D=10$ dimensions gives the so-called BFSS matrix model known also as M-(atrix) theory \cite{Banks:1996vh2}. This is a Yang-Mills matrix quantum mechanics which describes the low energy dynamics of a system of $N$ type IIA D0-branes \cite{Witten:1995im2}. The BFSS matrix model is also conjectured to describe the discrete light-cone quantization (DLCQ) of  M-theory. This is therefore our most fundamental matrix model as it gives a quantum mechanical theory of gravity.

The zero-dimensional reduction of the BFSS matrix model is precisely the IKKT matrix model \cite{Ishibashi:1996xs2}. This model is supposed to give a non-perturbative regularization of type IIB superstring theory in the Schild gauge. The IKKT model is the $D=10$ Yang-Mills matrix model.

The IKKT model itself is equivalent to Connes' approach to geometry.

Recall that a commutative/noncommutative space in Connes' approach to geometry is given in terms of a spectral triple $({\cal A},\Delta,{\cal H})$  \cite{Connes:1996gi2}. Here, ${\cal A}$ is the algebra of functions or bounded operators on the space, $\Delta$  is the Laplace operator which encodes the metric aspects, and ${\cal H}$ is the Hilbert space on which the algebra of bounded operators and the differential operator $\Delta$ are represented. The Laplacian $\Delta$ should be replaced with the Dirac operator ${\cal D}$ in the case of a spin structure.

In the IKKT matrix model we have, in a precise sense, an emergent matrix geometry which makes it essentially equivalent to Connes' noncommutative geometry. Here, the algebra ${\cal A}$ is given, in the large $N$ limit, by Hermitian matrices with smooth eigenvalue distributions and bounded square traces \cite{Sochichiu:2000ud2}. The Laplacian/Dirac operator is given in terms of the background solutions while the Hilbert space ${\cal H}$ is given by the adjoint representation of the gauge group $U(N)$ of the IKKT matrix model .

The BMN Yang-Mills matrix quantum mechanics is the plane-wave deformation of the BFSS matrix model \cite{Berenstein:2002jq2}. In fact, it is the unique one-parameter deformation of the BFSS Yang-Mills matrix quantum mechanics which preserves all ${\cal N}=16$ supersymmetries. This deformation is  given in terms of a mass parameter $\mu$.  It is conjectured that this matrix model describes the light-cone quantization of superparticles and supermembranes in the maximally supersymmetric pp-wave background of $11$-dimensional M-theory and $11$-dimensional supergravity.

The zero-dimensional reduction of the BMN matrix model gives again the IKKT matrix model but with additional mass deformation terms.

In the following, we will derive the IKKT-type IIB matrix model, with and without mass deformation terms, from the dimensional reduction of the BFSS/BMN matrix model.

\begin{itemize}
\item First, the action of the BFSS matrix model is written in terms of nine $N\times N$ bosonic Hermitian matrices $X_a$ ($a=1,...,9$), a $U(N)$ gauge field $A_t$ and sixteen $N\times N$ fermionic Majorana matrices $\psi_{\alpha}$ ($\alpha=1,...,16$). The bosonic and fermionic fields $X_a$ and $\psi_{\alpha}$ transform in the adjoint representation of the gauge group, viz $D_tX_a=\partial_tX_a-i[A_t,X_a]$ and $D_t\psi_{\alpha}=\partial_t\psi_{\alpha}-i[A_t,\psi_{\alpha}]$. The Euclidean BFSS action, on a circle of circumference $\beta$, is given explicitly by 

      \begin{eqnarray}
S_{\rm BFSS}=N\int_0^{\beta}dt{\rm Tr}\bigg[\frac{1}{2}(D_tX_a)^2-\frac{1}{4}[X_a,X_b]^2+\frac{1}{2}{\psi}^{T}C_9D_t\psi-\frac{1}{2}{\psi}^{T}C_9\gamma^a[X_a,\psi]\bigg].\label{rea}
      \end{eqnarray}
 \item The Euclidean BMN action, on a circle of circumference $\beta$, is a one-parameter deformation of the BFSS model which reads 
      \begin{eqnarray}
        S_{\rm BMN}&=&S_{\rm BFSS}+S_{\rm MASS}.\label{rea0}
        \end{eqnarray}
        \begin{eqnarray}
        S_{\rm MASS}&=&N\int_0^{\beta}dt{\rm Tr}\bigg[-i\frac{\mu}{6}\epsilon_{ijk} X_i[X_j,X_k]+\frac{\mu^2}{18}X_i^2+\frac{\mu^2}{72}X_a^2-i\frac{\mu}{8}\psi^{T}C_9\gamma^{789}\psi\bigg].\label{rea1}
      \end{eqnarray}

\item The local minimum of this model is a maximally supersymmetric configuration preserving all dynamical supersymmetries given by a static fuzzy sphere spanning the 7, 8, 9 directions, viz
  \begin{eqnarray}
[X_i,X_j]=i\alpha\epsilon_{ijk}X_k~,~X_{a\ne i}=0~,~D_tX_a=0~,~\alpha=-\frac{\mu}{3}.
  \end{eqnarray}
 \end{itemize}
The IKKT-type IIB matrix model is obtained from the actions (\ref{rea}), (\ref{rea0}) and (\ref{rea1}) by considering their zero-dimensional reduction.

The BFSS model, in Lorentzian signature, can also be re-written in terms of a $32$-component  Majorana-Weyl  spinor $\Psi$. The corresponding action is given explicitly by 

\begin{eqnarray}
S_{\rm BFSS}=N\int dt~ {\rm Tr}\bigg[\frac{1}{2}(D_{t}X_a)^2+\frac{1}{4}[{X}_a,{X}_b]^2+\frac{i}{2}\bar{\Psi}\Gamma^0D_t\Psi+\frac{1}{2}\bar{\Psi} \Gamma^a[{X}_a,\Psi]\bigg].
      \end{eqnarray}
 Here, $\Gamma^a$ are the Dirac matrices in $D$ dimensions in the Weyl representation and $\bar{\Psi}=\Psi^TC_{10}$. The zero-dimensional reduction of this action is immediately given by (with $A_t\equiv X_0$)      
 \begin{eqnarray}
        S_{\rm IKKT}=\frac{1}{g^2}{\rm Tr}\bigg[\frac{1}{4}[X_A,X_B][X^A,X^B]+\frac{1}{2}\bar{\Psi}\gamma^A[X_A,\Psi]\bigg].\label{re1}
      \end{eqnarray}
  The IIB superstring action can be obtained from this matrix model in the double scaling limit $N\longrightarrow\infty$ and $g^2\longrightarrow 0$ keeping $Ng^2$ is kept fixed. In Euclidean signature this action reads 
\begin{eqnarray}
        S_{\rm IKKT}=\frac{1}{g^2}{\rm Tr}\bigg[-\frac{1}{4}[X_A,X_B]^2-\frac{1}{2}\bar{\Psi}\gamma^A[X_A,\Psi]\bigg].\label{re2}
      \end{eqnarray}
    Equations (\ref{re1}) and (\ref{re2}) define the IKKT matrix model, without mass deformation, in Lorentzian and Euclidean signatures respectively.

Equation (\ref{re2}) defines in fact a supersymmetric matrix model in dimensions $D=3,4,6$ where, similarly to the case $D=10$, can be obtained from the zero-dimensional reduction of ${\cal N}=1$ supersymmetric Yang-Mills theory in $D$ dimensions. These are the basic $SO(D)$-symmetric Yang-Mills matrix models of the IKKT-type. The convergence properties of their corresponding partition functions are studied in \cite{Krauth:1998xh,Austing:2001bd,Austing:2001pk}. The partition function exists only in $D=10,6,4$. The $D =3$ supersymmetric partition function is divergent while the bosonic truncation is convergent.

The most generic mass deformation of the Yang-Mills matrix model (\ref{re2}) is of the general form

      \begin{eqnarray}
        S_{\rm MASS}=\frac{1}{g^2}{\rm Tr}\bigg[-i\frac{\mu_1}{6}\epsilon_{ijk} X_i[X_j,X_k]+\frac{\mu_2^2}{18}X_i^2+\frac{\mu_3^2}{72}X_a^2-i\frac{\mu_4}{8}\bar{\Psi}\Gamma\Psi\bigg].
      \end{eqnarray}
      For example, the mass deformation for $D=10$ is obtained from the zero-dimensional reduction of the action (\ref{rea1}) with $\Gamma=\Gamma^{789}$. Clearly, the Dirac matrix $\Gamma$ depends on the dimension $D$. We have then obtained a four-parameter action where the parameters $\mu_i$ are expected to be considerably constrained by the requirement of supersymmetry, e.g. in $D=10$ we have $\mu_i=\mu$.

      The fermionic determinant in $D=4$ is real positive while the determinants in $D=6$ and $D=10$ are complex which means in particular that in $D=10$ and $D=6$ we can have spontaneous symmetry breaking of rotational invariance \cite{Anagnostopoulos:2001yb3,Nishimura:2001sq3,Nishimura:2004ts3,Anagnostopoulos:2011cn3} while in $D=4$ there is no spontaneous symmetry breaking \cite{Ambjorn:2000bf3,Ambjorn:2000dx3,Burda:2000mn3,Ambjorn:2001xs3}.

      The cases $D=10$ and $D=6$ will then enjoy $SO(D)$ rotational symmetry which can be shown, by means of the so-called Gaussian expansion method  \cite{Nishimura:2011xy3}, to become in the large $N$ limit spontaneously broken down to $SO(3)$ due precisely to the phase of the Pfaffian \cite{Nishimura:2000ds3,Nishimura:2000wf3}. This spontaneous symmetry breaking corresponds effectively to a dynamical compactification to $3$ dimensions.

The $D=3$ Yang-Mills matrix model seems therefore to be the most fundamental Yang-Mills matrix model while the case $D=4$, as we have already discussed albeit briefly, is actually a $D=3$ model coupled to a scalar fluctuation.

\subsection{The effective potential of the $D=3$ Yang-Mills matrix model}
We are therefore led to consider the bosonic truncation of the $D=3$ Yang-Mills matrix model. Thus, the fundamental dynamical variables are three $N\times N$ Hermitian matrices $X_i$ which can be rotated into each other by $SO(3)$ transformations. The most general action (up to quartic power in the matrices $X_i$) is then given by
\begin{eqnarray}
S[X]=N{\rm Tr}\bigg[-\frac{1}{4}[X_i,X_j]^2+\frac{2i\alpha}{3}{\epsilon}_{ijk}X_iX_jX_k+BX_i^2+C(X_i^2)^2\bigg].\label{actionbeta0}
\end{eqnarray}
The last term is not induced by maximally supersymmetric mass deformation of the BFSS/IKKT matrix models. However, it is a central ingredient of noncommutative gauge theory on the fuzzy sphere as we will discuss shortly. The inclusion of this quartic term may even improve the convergence properties of the partition function of the supersymmetric $D=3$ Yang-Mills matrix model.

The above action can also be rewritten as
\begin{eqnarray}
  S[D]&=&\frac{1}{g^2N}{\rm Tr}\bigg[-\frac{1}{4}[D_i,D_j]^2+\frac{2i}{3}{\epsilon}_{ijk}D_iD_jD_k+B \alpha^2 D_i^2+C(D_i^2)^2\bigg]~,~X_i=\alpha D_i\nonumber\\
  &=&\frac{1}{g^2N}{\rm Tr}\bigg[-\frac{1}{4}[D_i,D_j]^2+\frac{2i}{3}{\epsilon}_{ijk}D_iD_jD_k+b  (D_i^2-c_2)+C(D_i^2-c_2)^2\bigg]\nonumber\\
  &&\frac{1}{g^2}=\alpha^4N^2\equiv \tilde{\alpha}^4~,~c_2=\frac{N^2-1}{4}~,~b=B\alpha^2+2c_2C~.\label{actionbeta}
\end{eqnarray}
This action enjoys a $U(N)$ gauge invariance. Indeed, gauge transformations are implemented here by the unitary
transformations $U\in U(N)$ as follows: $D_i{\rightarrow}D_i^{\prime}=UD_{i}U^{-1}$, i.e. the $N\times N$ Hermitian matrices $D_i$ play the role of the covariant derivatives of this gauge symmetry. The field strength or curvature tensor $F_{ij}$ associated with these covariant matrices $D_i$ is then explicitly defined by
\begin{eqnarray}
  F_{ij}=[D_i,D_j]-i\epsilon_{ijk}D_k.
\end{eqnarray}
The equations of motion derived from the action (\ref{actionbeta}) read in terms of the gauge-covariant current $J_i$ as follows
\begin{eqnarray}
[D_j,F_{ij}]=2C[D_i,D_j^2-c_2]_{+}+2b D_i+J_i~.\label{eom}
\end{eqnarray}
The local minimum of the model is given by the generators $L_i$ of the spin $s\equiv (N-1)/2$ irreducible representation of $SU(2)$ from which the fuzzy sphere configuration can be constructed. Indeed, for $J_i=0$, we have the solution 
\begin{eqnarray}
b=0~:~D_i=L_i~\text{where}~[L_i,L_j]=i\epsilon_{ijk}L_k~,~L_i^2=c_2.
\end{eqnarray}
\begin{eqnarray}
b\ne 0~:~D_i=\varphi L_i~\text{where}~(1+m^2)\varphi^2-\varphi-\mu=0~,~C=\frac{m^2}{2c_2}~,~-\mu=b-m^2=B\alpha^2.
\end{eqnarray}
The matrix coordinates $\hat{x}_i$ on the fuzzy sphere can then be constructed from the generators $L_i$ in the usual way, viz
\begin{eqnarray}
\hat{x}_i=\frac{RL_i}{\sqrt{c_2}}~\text{where}~[\hat{x}_i,\hat{x}_j]=i\theta \epsilon_{ijk}\hat{x}_k~,~\hat{x}_i^2=R^2~,~\theta=\frac{R}{\sqrt{c_2}}.\label{S2fuzzy}
\end{eqnarray}
This is the origin of the phenomena of emergent geometry from Yang-Mills matrix models. Thus, the covariant derivatives/matrices $D_i$ act really as covariant matrix coordinates which is their correct interpretation.

The partition function of the theory is given by
\begin{eqnarray}
Z_{N}[J]=\int {\prod}_{i=1}^3\left[dD_i\right]e^{-S[D]-\frac{1}{Ng^2}{\rm Tr}J_iD_i}.
\end{eqnarray}
Now we adopt the background field method to the problem of quantization of this theory \cite{Castro-Villarreal:2004ulr}. This method consists in making a perturbation of the field around the
classical solution and then quantizing the fluctuation. Towards this end we first separate the field as  $ D_i=B_i+Q_a$ and write the action in the form
\begin{eqnarray}
S[D_i]&=&S[B_i]+\frac{1}{g^2N}{\rm Tr}\big(\hat{J}_i-J_i\big)Q_i
-\frac{1}{2g^2N}{\rm Tr}[B_i,Q_j]^2+ \frac{1}{2g^2N}(1+2C){\rm Tr}[B_i ,Q_i]^2\nonumber\\
&+&\frac{1}{g^2N}{\rm Tr} Q_i[F_{ij}^B,Q_j]+\frac{1}{2g^2N}{\rm Tr}Q_i\bigg[2b {\delta}_{ij}+4C (B_k^2-c_2){\delta}_{ij}+8 C B_iB_j\bigg]Q_j+...\nonumber\\
&&\hat{J}_a=-[B_j,F_{ij}^B]+2C[B_i,B_j^2-c_2]_{+}+2b B_i+J_i. \label{action1}
\end{eqnarray}
Here, $F_{ij}^B=[B_i,B_j]-i{\epsilon}_{ijk}B_k$. This action is invariant under the gauge transformations $B_i{\longrightarrow}B_i$, $Q_i{\longrightarrow}UQ_iU^{\dagger}+U[B_i,U^{\dagger}]$. This means in particular that in order to fix the gauge in a consistent way the gauge fixing term should be covariant with respect to the background field. We impose here the covariant Lorentz gauge $[B_i,Q_i]=0$. The gauge fixing term and the Faddeev-Popov term are therefore given by

\begin{eqnarray}
 S_{\text g.f}+S_{\text gh}
& = &-\frac{1}{2g^2N}{\rm Tr}\frac{[B_i,Q_i]^2}{\xi}+\frac{1}{g^2N}{\rm Tr}b^{\dagger}[B_i,[B_i,b]].
\label{ghost}
\end{eqnarray}
We will choose now for simplicity the gauge ${\xi}^{-1}=1+2C$ which will cancel the $4$th term in (\ref{action1}). This gauge becomes the Feynman gauge ${\xi}=1$ in the limit $N{\longrightarrow}{\infty}$ and the Landau gauge ${\xi}=0$ in the limit $C{\longrightarrow}{\infty}$. Furthermore, we will assume that the background field $B_i$ satisfies the classical equations of motion (\ref{eom}) and hence the $2$nd term of (\ref{action1}) also vanishes. The partition function
becomes then
\begin{eqnarray}
Z_N[J]=e^{-S[B_i]-\frac{1}{g^2N}{\rm Tr} B_iJ_i}\det\big({\cal B}_i^2\big)\int
{\prod}_{i=1}^3\left[dQ_i\right]~e^{-\frac{1}{2g^2N}{\rm Tr}Q_i{\Omega}_{ij}Q_j+...}.
\end{eqnarray}
Here, $\det\big({\cal B}_i^2\big)$ comes from the integration over the ghost field whereas the gauge field Laplacian ${\Omega}_{ab}$ is defined by
\begin{eqnarray}
{\Omega}_{ij}&=&2b {\delta}_{ij}+ {\cal B
}_k^2{\delta}_{ij}+2{\cal
F}_{ij}^B+4C(B_k^2-c_2){\delta}_{ij}+8C B_iB_j.
\end{eqnarray}
In this equation the notation ${\cal B}_i$ and ${\cal F}_{ij}^B$ means that
the covariant derivatives $B_i$ and their curvature $F_{ij}^B$ act by
commutators, viz ${\cal B}_i(M)=[B_i,M]$ and ${\cal F}_{ij}^B(M)=[F_{ij}^B,M]$ for any $N\times N$ Hermitian matrix $M$. Similarly, ${\cal B}_i^2(M)=[B_i,[B_i,M]]$. By performing the Gaussian path integral we obtain the one-loop effective action
\begin{eqnarray}
{\Gamma}[B_i]=S[B_i]+\frac{1}{2}{\rm Tr}_3{\rm TR}\log{\Omega}-{\rm TR}\log{\cal B}_i^2.\label{effective}
\end{eqnarray}
Here, ${\rm TR}$ is a trace over $4$ indices corresponding to the left and right actions of operators on matrices and ${\rm Tr}_{3}$ is the trace associated with $3$-dimensional rotations. 

We can immediately use the above expression to compute the effective potential in the fuzzy sphere configuration (\ref{S2fuzzy}), i.e. in the covariant matrix coordinates $B_i=\varphi L_i$. We find (by assuming that $|b|,C\ll 1$ and $\varphi\sim 1$) \cite{Castro-Villarreal:2004ulr}
\begin{eqnarray}
  \frac{V_{\rm eff}(\varphi)}{2c_2}&\equiv& \frac{\Gamma[B_i=\varphi L_i]}{2c_2}\nonumber\\
  &=&\tilde{\alpha}^4\bigg[\frac{1}{4}\varphi^4-\frac{1}{3}\varphi^3+\frac{1}{4}m^2\varphi^4-\frac{1}{2}\mu\varphi^2\bigg]+\frac{1}{4c_2}{\rm Tr}_3{\rm TR}\log\bigg[2b\delta_{ij}+\varphi^2{\cal L}_k^2\delta_{ij}+(\varphi^2-\varphi)i\epsilon_{ijk}{\cal L}_k\nonumber\\
    &+&4Cc_2(\varphi^2-1)+8C\varphi^2L_iL_j\bigg]-\frac{1}{2c_2}{\rm TR}\log\varphi^2{\cal L}_i^2\nonumber\\
  &=&\tilde{\alpha}^4\bigg[\frac{1}{4}\varphi^4-\frac{1}{3}\varphi^3+\frac{1}{4}m^2\varphi^4-\frac{1}{2}\mu\varphi^2\bigg]+\frac{1}{4c_2}{\rm Tr}_3{\rm TR}\log\big[\varphi^2{\cal L}_k^2\delta_{ij}\big]-\frac{1}{2c_2}{\rm TR}\log\varphi^2{\cal L}_i^2\nonumber\\
   &=&\tilde{\alpha}^4\bigg[\frac{1}{4}\varphi^4-\frac{1}{3}\varphi^3+\frac{1}{4}m^2\varphi^4-\frac{1}{2}\mu\varphi^2\bigg]+\log\varphi^2.
\end{eqnarray}
\subsection{Emergent geometry: The matrix-to-sphere phase transition}
The basic result summarizing the physics of emergent geometry from Yang-Mills matrix models is the statement that the classical global minimum of the underlying matrix model which realizes the corresponding noncommutative space becomes unstable under quantum fluctuations, as we increase the gauge coupling constant, and then at some critical value evaporates  into a pure matrix configurations without any geometric content.

The existence of a classical background of the Yang-Mills matrix model defines "first quantization" of the geometry whereas the stability of this background under quantum fluctuations defines "second quantization" of the geometry.  In the first instance we have noncommutative geometry whereas in the second instance we have emergent geometry.

\begin{figure}[h]
\begin{center}
\includegraphics[width=12cm]{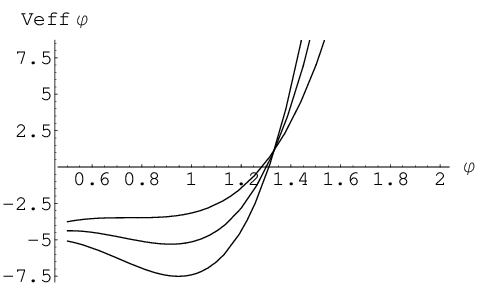}
\caption{Effective potential for different values of $g$ for the Alekseev-Recknagel-Schomerus.}\label{figars}
\end{center}
\end{figure}

For example, in the context of the $D=3$ Yang-Mills matrix model (\ref{actionbeta0}) or (\ref{actionbeta}) the fuzzy sphere configuration  (\ref{S2fuzzy}), although completely stable classically, becomes unstable under the effect of quantum fluctuations as we decrease the inverse gauge coupling constant $1/g^2=\tilde{\alpha}^4$. We will illustrate this effect first in the case of the so-called Alekseev-Recknagel-Schomerus  matrix model \cite{Alekseev:2000fd,Alekseev:1998mc} which is associated with the parameters $b=C=0$. Explicitly, the action is given by 

\begin{eqnarray}
  S[D]=\frac{1}{g^2N}{\rm Tr}\bigg[-\frac{1}{4}[D_i,D_j]^2+\frac{2i}{3}{\epsilon}_{ijk}D_iD_jD_k\bigg].\label{ars}
\end{eqnarray}
The classical potential and the global minimum of the model are then given by 
\begin{eqnarray}
  \frac{V(\varphi)}{2c_2}&=&\tilde{\alpha}^4\bigg[\frac{1}{4}\varphi^4-\frac{1}{3}\varphi^3\bigg]\nonumber\\
  \frac{V^{\prime}(\varphi)}{2c_2}&=&\tilde{\alpha}^4\bigg[\varphi^3-\varphi^2\bigg]=0\Rightarrow \varphi=0~(\text{matrix})~\text{or}~\varphi=1~\text{fuzzy sphere}\nonumber\\
  \frac{V(\varphi=1)}{2c_2}=-\frac{1}{12}\tilde{\alpha}^4< \frac{V(\varphi=0)}{2c_2}=0~&:&~\text{fuzzy sphere is the global minimum}.
\end{eqnarray}
The quantum potential is, on the other hand, given by 
\begin{eqnarray}
\frac{V_{\rm eff}(\varphi)}{2c_2}=\tilde{\alpha}^4\bigg[\frac{1}{4}\varphi^4-\frac{1}{3}\varphi^3\bigg]+\log\varphi^2.
\end{eqnarray}
By taking the first and second derivatives of this potential we obtain
\begin{eqnarray}
&&\frac{V_{\rm eff}^{\prime}}{2c_2}=\frac{1}{g^2}({\varphi}^3-{\varphi}^2)+\frac{2}{\varphi}\nonumber\\
&&\frac{V_{\rm eff}^{\prime\prime}}{2c_2}=\frac{1}{g^2}(3{\varphi}^2-2{\varphi})-\frac{2}{{\varphi}^2}.
\end{eqnarray}
The condition $V^{\prime}(\varphi)=0$, which also reads ${\varphi}^4-{\varphi}^3+2g^2=0$, will give us extrema of the model. These extrema are minima and thus stable if the condition $V^{\prime\prime}(\varphi)>0$ (or equivalently $3{\varphi}^4-2{\varphi}^3-2g^2>0$) is satisfied whereas they are maxima if $3{\varphi}^4-2{\varphi}^3-2g^2<0$. The equation which tell us therefore when we go from a bounded potential to an unbounded potential is given by (see figure (\ref{figars}))
\begin{eqnarray}
3{\varphi}^4-2{\varphi}^3-2g^2=0.
\end{eqnarray}
Solving the above two equations yield immediately the critical values
\begin{eqnarray}
{\varphi}{\equiv}{\varphi}_{*}=\frac{3}{4}~,~g^2{\equiv}g^2_{*}=\bigg(\frac{3}{8}\bigg)^3~,~\tilde{\alpha}\equiv \tilde{\alpha}_*=\bigg(\frac{8}{3}\bigg)^{3/4}=2.08677944.\label{cv}
\end{eqnarray}
This one-loop calculation agrees nicely with  the exact result obtained in \cite{Azuma:2004zq} by means of Monte Carlo simulation of the $D=3$ Yang-Mills matrix model  (\ref{actionbeta0}) with $B=C=0$.

It is not difficult to convince ourselves that the trace part of the covariant coordinate matrices $X_i$ does not enter the Alekseev-Recknagel-Schomerus action (\ref{ars}) and thus it can be removed from the partition function. As it turns out, the removal of this trace part is crucial if we want the Monte Carlo method to thermalize. Hence, we should consider in this case the partition function given by
\begin{eqnarray}
Z=\int \prod_{i=1}^3[dX_i]~\exp(-S)\delta({\rm Tr} X_i).
\end{eqnarray}
The order parameter in this problem is given by the extent of space $\langle {\rm radius}\rangle$, or equivalently by the radius $r$,  defined by
\begin{eqnarray}
  \langle {\rm radius}\rangle =\langle {\rm Tr} X_i^2\rangle~,~\frac{1}{r}=\frac{1}{\tilde{\alpha}^2c_2}\langle {\rm radius}\rangle.
\end{eqnarray}
A more powerful set of order parameters is given by the eigenvalues distributions of the matrices $X_3$, $i[X_1,X_2]$ and $X_i^2$.  We also measure the Yang-Mills and Chern-Simons terms given by
\begin{eqnarray}
{\rm YM}=-\frac{N}{4}{\rm Tr}[X_{i},X_{j}]^2~,~{\rm CS}=\frac{2iN\alpha}{3}\epsilon_{ijk} {\rm Tr} X_iX_jX_k.
\end{eqnarray}
The total action (energy) and the specific heat are given by 
\begin{eqnarray}
S={\rm YM}+{\rm CS}~,~C_v=\langle S^2\rangle-\langle S\rangle^2.
\end{eqnarray}
An exact Schwinger\textendash Dyson identity is given by 
\begin{eqnarray}
\text{identity}~:~ 4\langle {\rm YM}\rangle+3\langle {\rm CS}\rangle\equiv 3 (N^2-1).
\end{eqnarray}
The Monte Carlo results reported here are derived using the Metropolis algorithm in  \cite{DelgadilloBlando:2007vx}.

\begin{figure}[htbp]
\begin{center}
\includegraphics[width=8.0cm,angle=-0]{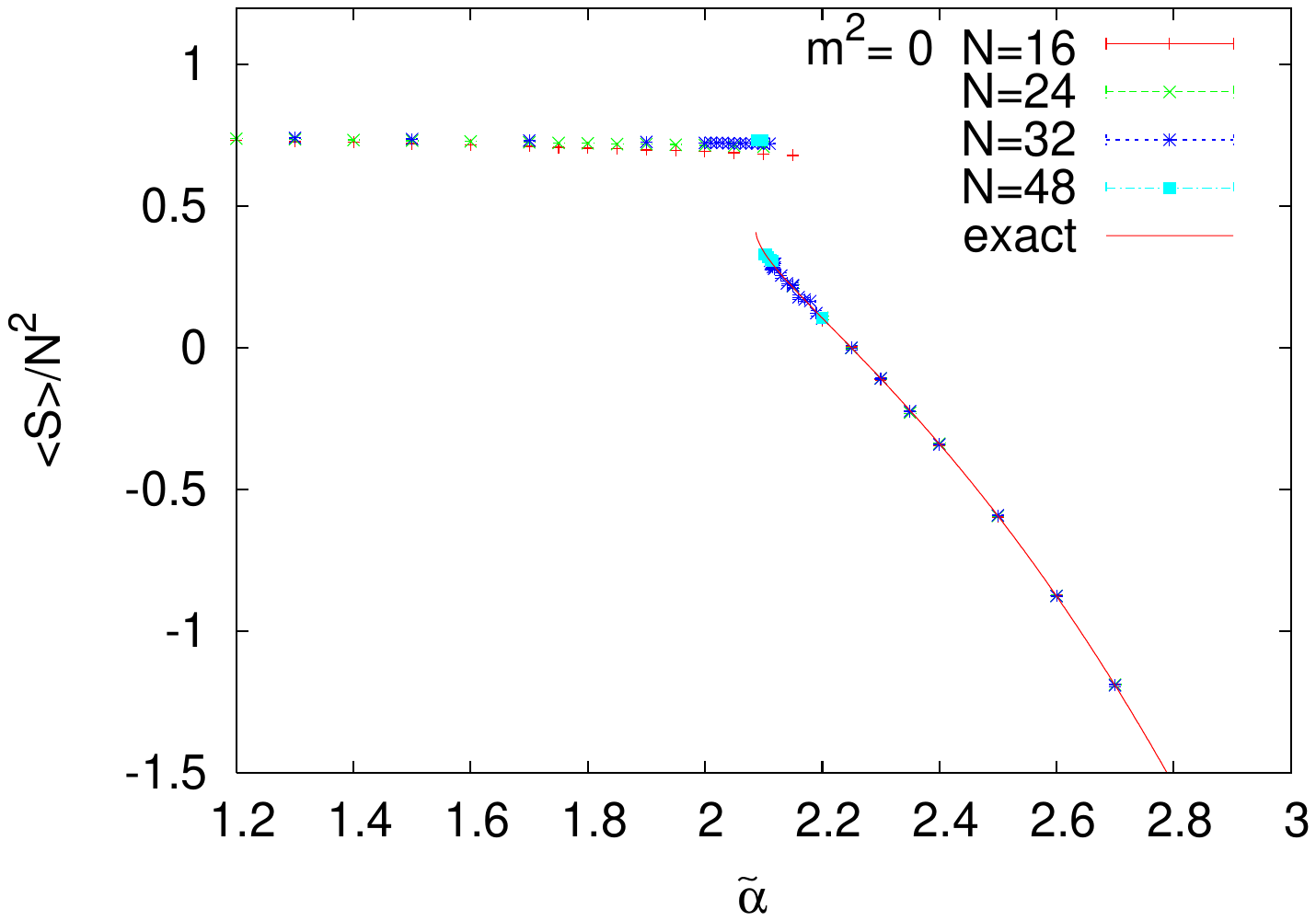}
\includegraphics[width=8.0cm,angle=-0]{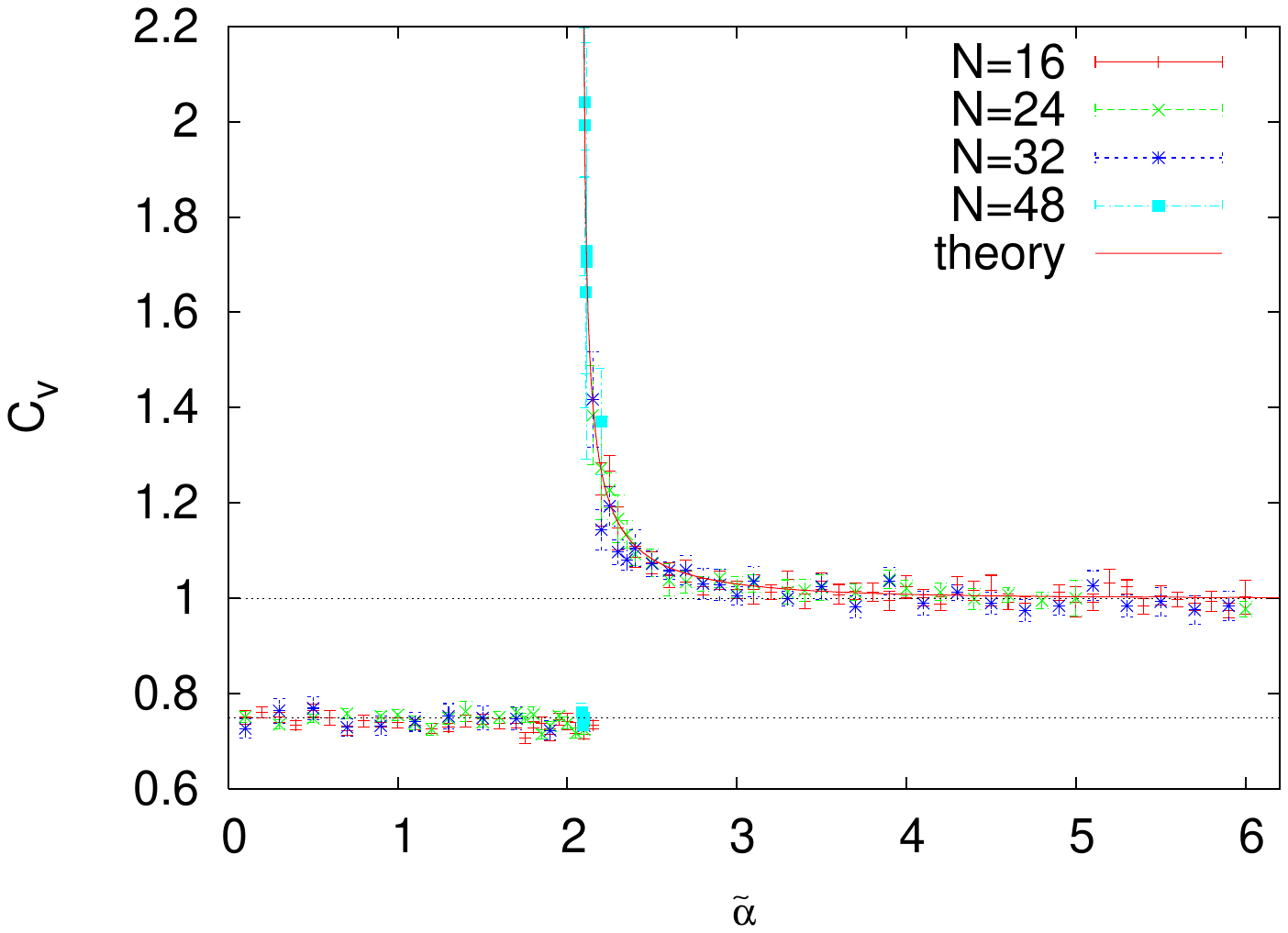}
\includegraphics[width=8.0cm,angle=-0]{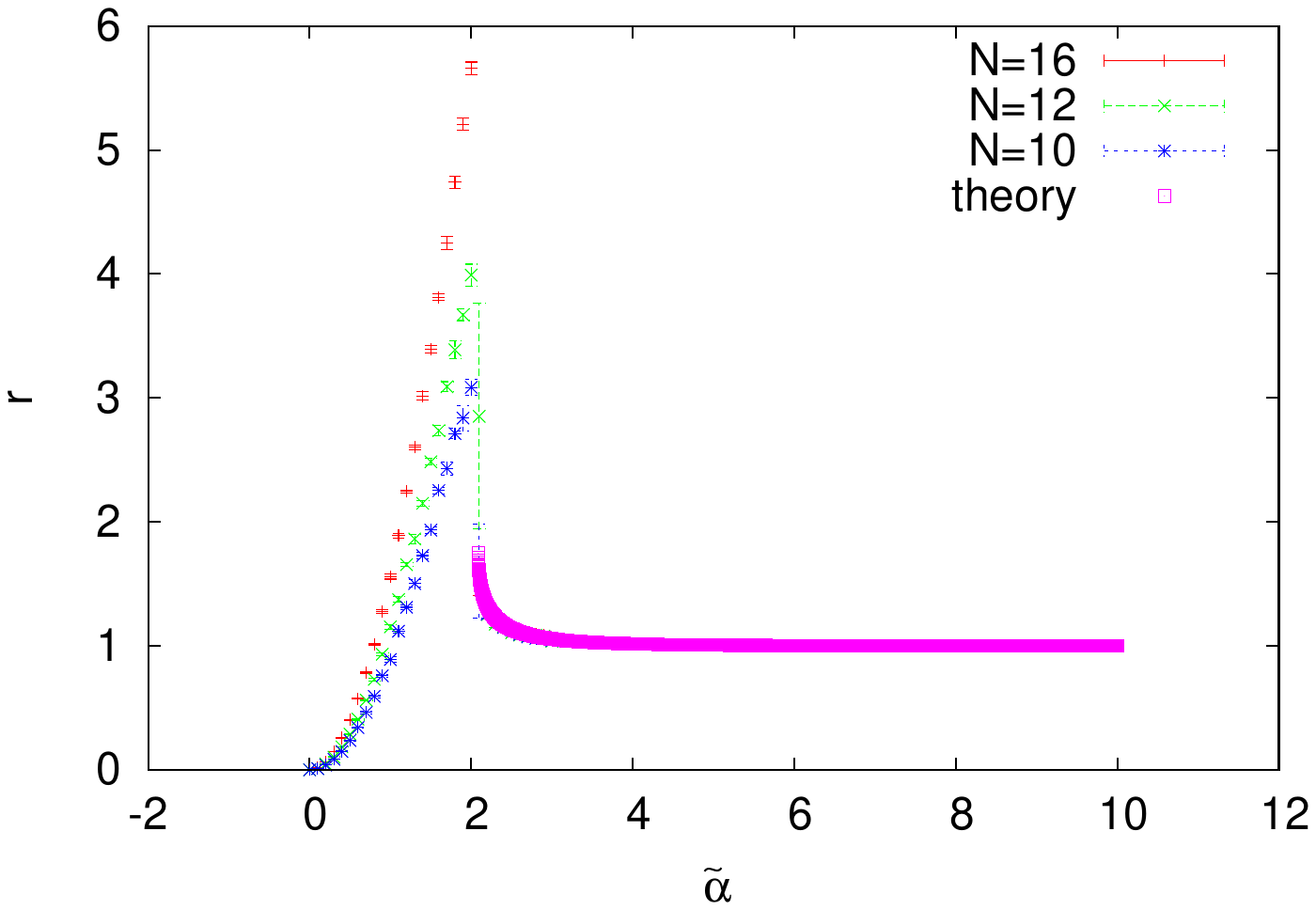}
\includegraphics[width=8.0cm,angle=-0]{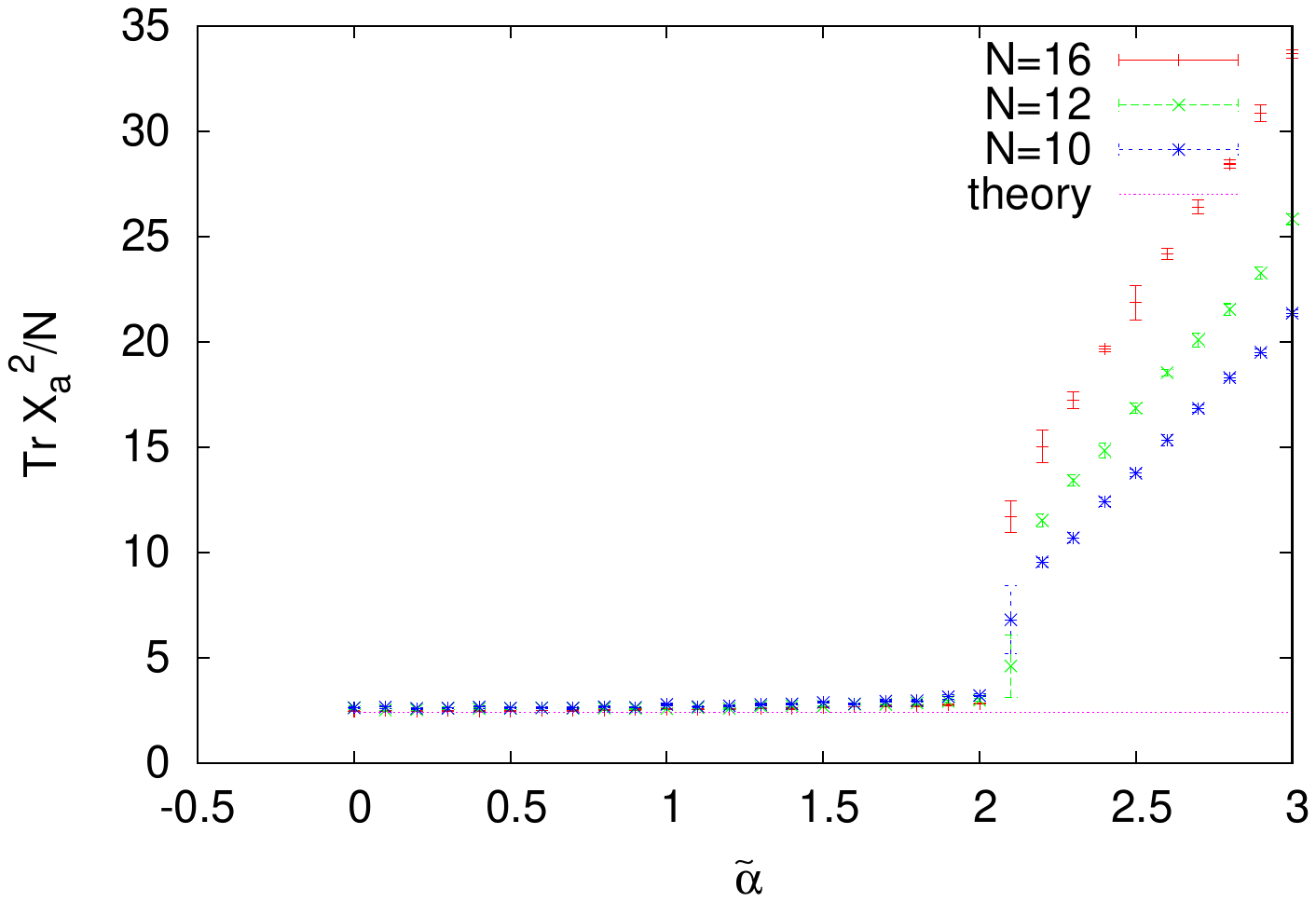}
\caption{The behavior of various observables across the phase transition between the fuzzy sphere and the Yang-Mills phase.}\label{figpublished}
\end{center}
\end{figure}

The Alekseev-Recknagel-Schomerus  model is characterized by two phases: the fuzzy sphere phase and the Yang-Mills phase. Some of the fundamental results are:
\begin{enumerate}
\item{}{\bf The Fuzzy Sphere (Geometric) Phase:} 
\begin{itemize}
\item This appears for large values of $\tilde{\alpha}$. It corresponds to the class of solutions of the equations of motion given by
\begin{eqnarray}
[X_i,X_j]=i\alpha\varphi\epsilon_{ijk}X_k~,~\varphi=1.
\end{eqnarray}
The global minimum is given by the largest irreducible representation of $SU(2)$ which fits in $N\times N$ matrices. This corresponds to the spin $s=(N-1)/2$ irreducible representation, viz
  \begin{eqnarray}
X_i=\varphi\alpha L_i~:~[L_i,L_j]=i\epsilon_{ijk}L_c~,~c_2=L_i^2=s(s+1){\bf 1}_N~,~s=\frac{N-1}{2}.
\end{eqnarray}

The values of the various observables in these configurations are
\begin{eqnarray}
   {\rm YM}=\frac{\varphi^4\tilde{\alpha}^4c_2}{2}~,~{\rm CS}=-\frac{2\varphi^3\tilde{\alpha}^4 c_2}{3}~,~S=\varphi^3\tilde{\alpha}^4c_2(\frac{\varphi}{2}-\frac{2}{3})~,~\langle {\rm radius}\rangle=\varphi^2\tilde{\alpha}^2c_2.
\end{eqnarray}
\item{}The eigenvalues of $D_3=X_3/\alpha$ and $i[D_1,D_2]=i[X_1,X_2]/\alpha^2$ are given by
\begin{eqnarray}
\lambda_i=-s,-s+1,...,s-1,+s~,~s=\frac{N-1}{2}.
\end{eqnarray}
The spectrum of $i[D_1,D_2]$ is a better measurement of the geometry since quantum fluctuations around $L_3$ are more suppressed. Some illustrative data for $\tilde{\alpha}=3$ and $N=4$ is shown in figure (\ref{Emerg1}).

\end{itemize}
\item{}{\bf The Yang-Mills (Matrix) Phase:} 
\begin{itemize}
\item
This appears for small values of $\tilde{\alpha}$. It corresponds to the class of solutions of the equations of motion given by
\begin{eqnarray}
[X_i,X_j]=0.
\end{eqnarray}
This is the phase of almost commuting matrices. It is characterized by the eigenvalues distribution 
\begin{eqnarray}
\rho(\lambda)=\frac{3}{4R^3}(R^2-\lambda^2).\label{for0}
\end{eqnarray}
It is believed that $R=2$. We compute
 \begin{eqnarray}
\langle {\rm radius}\rangle&=&3 \langle {\rm Tr} X_3^2\rangle\nonumber\\
&=&3N\int_{-R}^{R}d\lambda \rho(\lambda)\lambda^2\nonumber\\
&=&\frac{3}{5}R^2 N.\label{for1}
\end{eqnarray}
\item The above eigenvalues distribution can be derived by assuming that the joint eigenvalues distribution of the the three commuting matrices  $X_1$, $X_2$ and $X_3$ is uniform inside a solid ball of radius $R$. This can be actually proven by quantizing the system in the Yang-Mills phase around commuting matrices \cite{Filev:2014jxa}.
\item{}The value of the radius $R$ is determined numerically as follows:
\begin{itemize}
\item{}The first measurement $R_1$ is obtained by comparing the numerical result for $\langle {\rm radius}\rangle$, for the biggest value of $N$, with the formula (\ref{for1}).
\item{}We use $R_1$ to restrict the range of the eigenvalues of $X_3$.
\item{}We fit the numerical result for the density of  eigenvalues of $X_3$, for the biggest value of $N$, to the parabola (\ref{for0}) in order to get a second  measurement $R_2$.
\item{}We may take the average of $R_1$ and $R_2$.
\end{itemize} 
Sample data for $\tilde{\alpha}=0$ with $N=6,8$ and $10$ is shown in figure (\ref{Emerg2}).  
\item{}It is found that the eigenvalues distribution, in the Yang-Mills phase, is independent of $\tilde{\alpha}$. Sample data for $\tilde{\alpha}=0-2$ and $N=10$ is shown in figure (\ref{Emerg3}).  
\end{itemize}

\begin{figure}[htbp]
\begin{center}
\includegraphics[width=12.0cm,angle=-0]{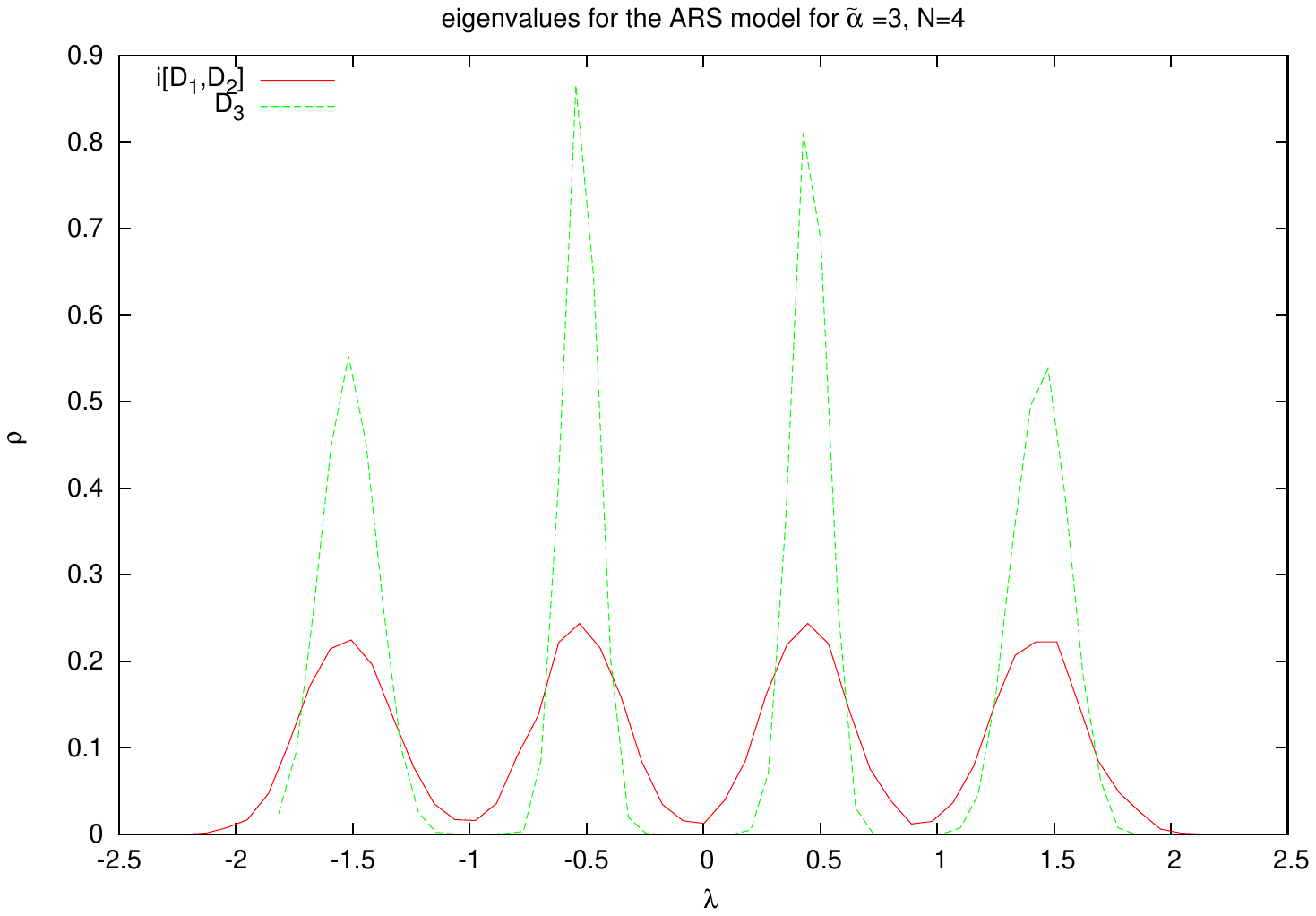}
\end{center}
\caption{The eigenvalue distribution in the fuzzy sphere phase.}\label{Emerg1}
\end{figure}

\begin{figure}[htbp]
\begin{center}
\includegraphics[width=12.0cm,angle=-0]{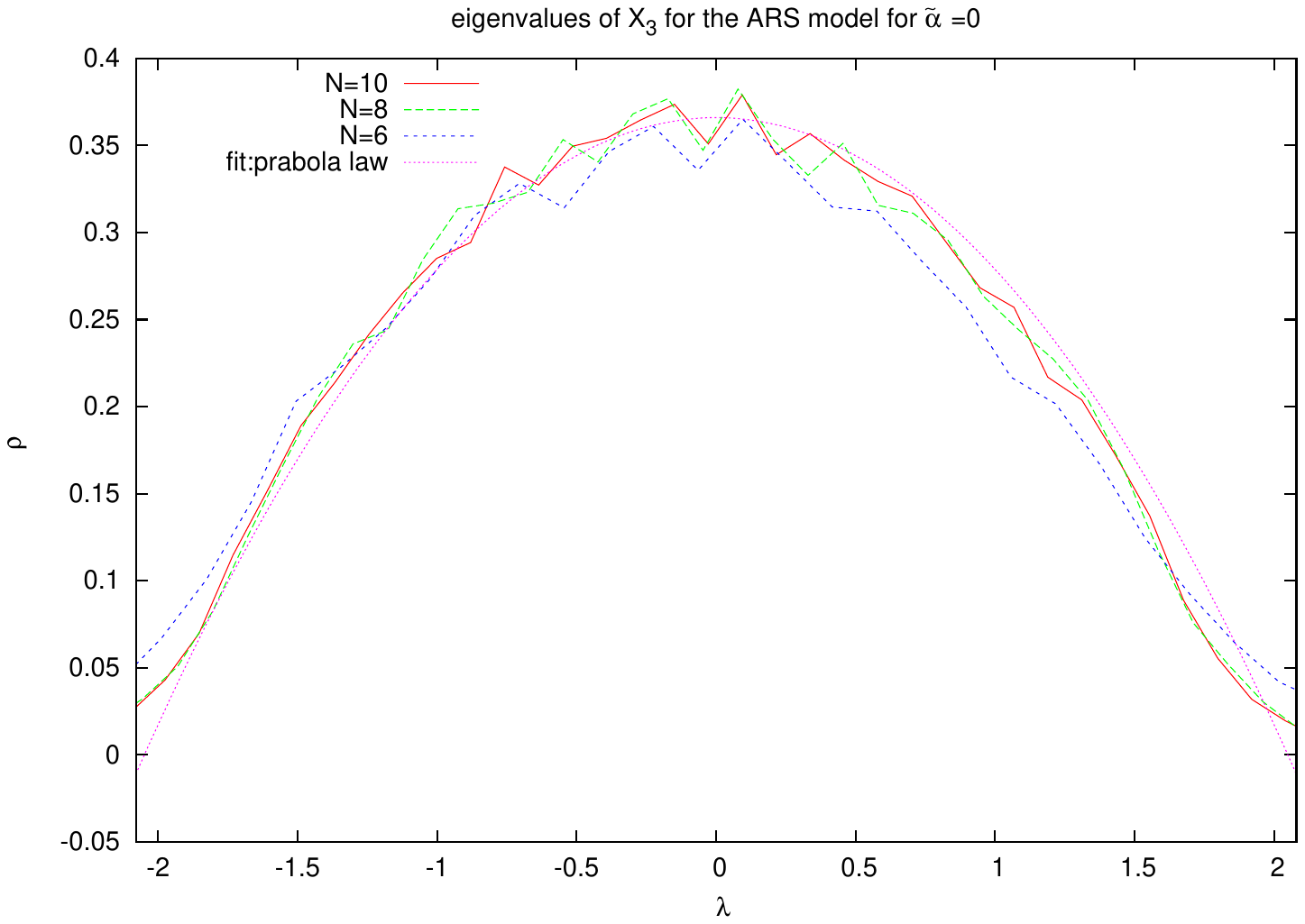}
\end{center}
\caption{The eigenvalue distribution in the Yang-Mills matrix phase as a function of $N$.}\label{Emerg2}
\end{figure}
 
\begin{figure}[htbp]
\begin{center}
\includegraphics[width=12.0cm,angle=-0]{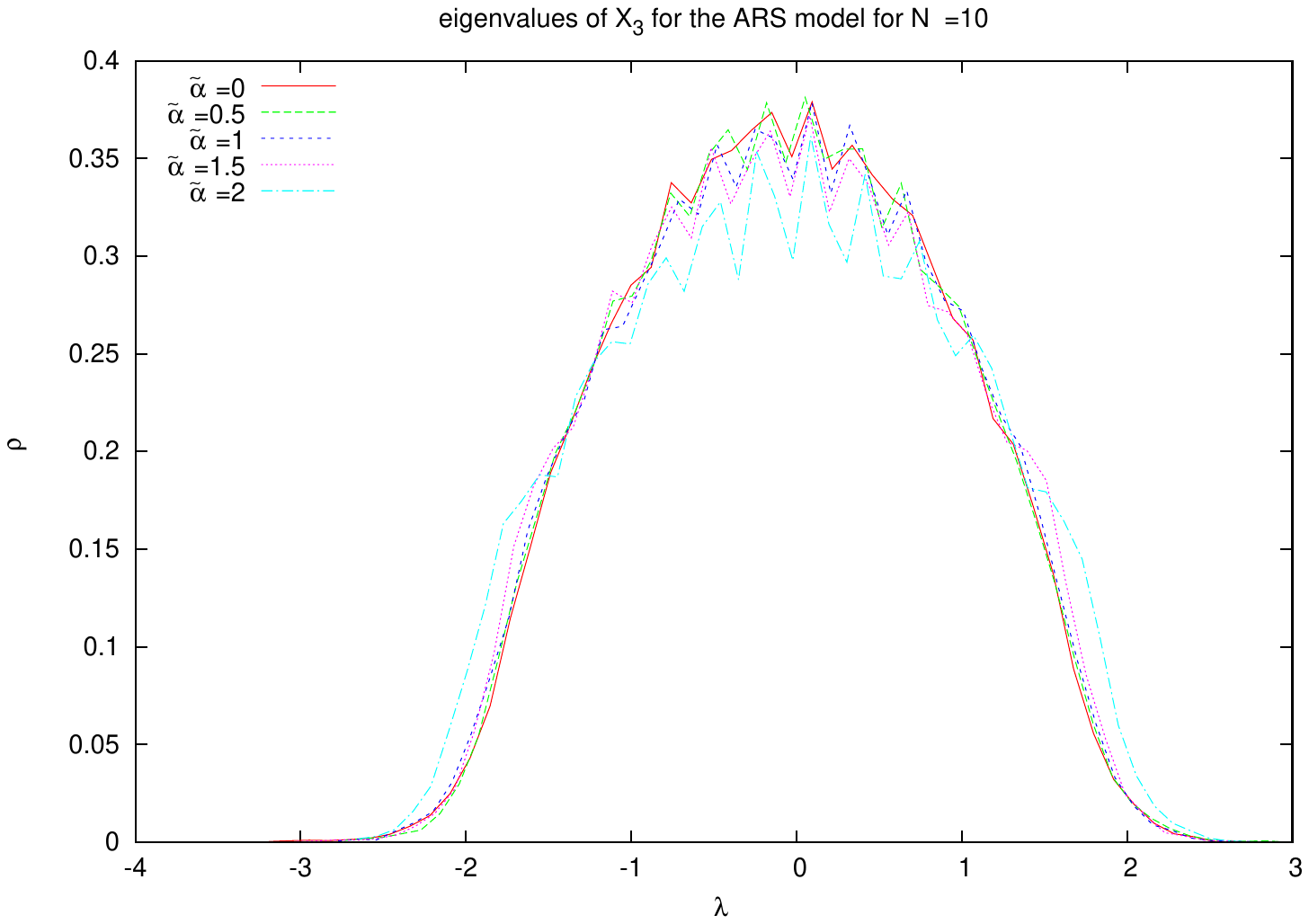}
\end{center}
\caption{The eigenvalue distribution in the Yang-Mills matrix phase as a function of $\tilde{\alpha}$. }\label{Emerg3}
\end{figure}

\item {\bf Critical Fluctuations:}  The transition between the two phases occur at $\tilde{\alpha}=2.1$. The specific heat diverges at this point from the Yang-Mills side while it remains constant from the fuzzy sphere side.  This indicates a second order behaviour with critical fluctuations only from one side of the transition. The Yang-Mills and Chern-Simons actions, and as a consequence the total action, as well as the extent of space $\langle {\rm radius}\rangle$  suffer a discontinuity at this point reminiscent of a first order behavior. In particular, the sphere expands then evaporates at the critical point, i.e. its radius $r$ diverges at the transition point then it starts decreasing fast in the matrix phase until it reaches the value $r=0$. See figure (\ref{figpublished}).

The different phases of the model are characterized by: 
\begin{center}
\begin{tabular}{|c|c|}
\hline
fuzzy sphere ($\tilde{\alpha}>\tilde{\alpha}_*$ )& matrix phase ($\tilde{\alpha}<\tilde{\alpha}_*$)\\
$r=1$ & $
r=0$\\
$C_v=1$  & $C_v=0.75$  \\
\hline
\end{tabular}
\end{center}

\end{enumerate}

\subsection{Noncommutative gauge theory: UV-IR mixing on the fuzzy sphere}
The second basic result summarizing the physics of emergent geometry from Yang-Mills matrix models is the statement that by expanding around the quantum background corresponding to the geometric phase  we obtain a noncommutative gauge theory on the corresponding noncommutative space.

For example, it was shown in \cite{Alekseev:2000fd,Alekseev:1998mc} that, to leading order in the string tension, the dynamics  of open strings moving in a curved space with ${\bf S}^3$ metric, in the presence of a non-vanishing Neveu-Schwarz B-field and with Dp-branes, is equivalent  to a noncommutative gauge theory on the fuzzy sphere with a Chern-Simons term given precisely by the Alekseev-Recknagel-Schomerus matrix model (\ref{ars}). Indeed, this action  (\ref{ars}) can be rewritten as a sum of a Yang-Mills term $S_{\rm YM}$ and a Chern-Simons term $S_{\rm CS}$ as follows

\begin{eqnarray}
 && S[D_i]=S_{\rm YM}[D_i]+S_{\rm CS}[D_i]\nonumber\\
  &&S_{\rm YM}[D_i]=-\frac{1}{4g^2N} {\rm Tr} F_{ij}^2\nonumber\\
  &&S_{\rm CS}[D_i]=-\frac{1}{6g^2N}{\rm Tr}\left[i\epsilon_{ijk}F_{ij}D_k+(D_i^2-c_2)\right].\label{ars2}
\end{eqnarray}
By expanding now the covariant matrix coordinates $D_i$ around the fuzzy sphere configuration given by (\ref{S2fuzzy}) as $D_i=L_i+A_i$ we can rewrite the field strength $F_{ij}=[D_i,D_j]-i\epsilon_{ijk}D_k$ as follows 
\begin{eqnarray}
D_i=L_i+A_i\Rightarrow F_{ij}=F_{ij}^{(0)}+[A_i,A_j]~,~F_{ij}^{(0)}=[L_i,A_j]-[L_j,A_i]-i\epsilon_{ijk}A_k.
\end{eqnarray}
Clearly, we have in the commutative limit the behavior 
\begin{eqnarray}
F_{ij}\longrightarrow F_{ij}^{(0)}~,~F_{ij}^{(0)}\longrightarrow {\cal L}_iA_j-{\cal L}_jA_i-i\epsilon_{ijk}~,~N\longrightarrow\infty.
\end{eqnarray}
In other words, $A_i$ is identified as a noncommutative gauge field on the fuzzy sphere with a field strength given exactly by $F_{ij}$.  In fact, the $U(N)$ gauge symmetry $D_i\longrightarrow UD_iU^{\dagger}$ is equivalent to the  $U(N)$ transformations: $A_i{\longrightarrow}UA_iU^{\dagger}+U[L_i,U^{\dagger}]$. This gauge field, which has three components, describes therefore a two-dimensional noncommutative gauge field tangent to the sphere plus a scalar component normal to the sphere defined by the gauge-covariant prescription \cite{Karabali:2001te}

\begin{eqnarray}
\Phi=R\frac{D_i^2-c_2}{2\sqrt{c_2}}=\frac{1}{2}\left(\hat{x}_iA_i+A_i\hat{x}_i+\frac{RA_i^2}{\sqrt{c_2}}\right)\longrightarrow x_iA_i~,~N\longrightarrow\infty.
\end{eqnarray}
Thus, the second term in the Chern-Simons action $S_{\rm CS}$  is a linear term in the normal scalar field $\Phi$. In fact, the whole Alekseev-Recknagel-Schomerus  action  (\ref{ars2}) can be rewritten in the form
\begin{eqnarray}
S_N[A_i]=-\frac{1}{4g^2N}{\rm Tr}\left[F_{ij}^{(0)}+[A_i,A_j]\right]^2-\frac{i}{2g^2N}{\epsilon}_{ijk}{\rm Tr}\left[\frac{1}{2}F_{ij}^{(0)}A_k+\frac{1}{3}[A_i,A_j]A_k\right].\label{ars3}
\end{eqnarray}
This is indeed a noncommutative gauge theory on the fuzzy sphere ${\bf S}^2_N$ given by the sum of a Yang-Mills term and a Chern-Simons term. In the commutative limit $N{\longrightarrow}{\infty}$, where all commutators vanish, we get the action 
\begin{eqnarray}
S_{\infty}[A_i]=-\frac{1}{4g^2} \int_{S^2} \frac{d{\Omega}}{4{\pi}}(F_{ij}^{(0)})^2-\frac{i}{2g^2}{\epsilon}_{ijk}\int_{S^2}\frac{d{\Omega}}{4\pi}\frac{1}{2}F_{ij}^{(0)}A_k.
\end{eqnarray}
The commutative action $S_{\infty}$ is at most quadratic in the field $A_i$ and as a consequence the corresponding effective action will be essentially given by $S_{\infty}$ itself. However, quantization of the noncommutative action $S_N$ is much more involved and yields a non-trivial effective action. As it turns out, the commutative limit of this noncommutative effective action does not tend to $S_{\infty}$ which is the signature of the UV-IR mixing in this model. In \cite{Castro-Villarreal:2004ulr} the quadratic effective action was computed explicitly and was found to be given in the commutative limit $N{\longrightarrow}{\infty}$ by the expression 

\begin{eqnarray}
{\Gamma}_{2}&=& -\frac{1}{4g^2}\int
\frac{d{\Omega}}{4{\pi}}F_{ij}^{(0)}(1+2g^2{\Delta}_3)F_{ij}^{(0)}-\frac{i}{4g^2}{\epsilon}_{ijk}\int
\frac{d{\Omega}}{4{\pi}}F_{ij}^{(0)}(1+2g^2{\Delta}_3)A_k+4\sqrt{c_2}\int\frac{d{\Omega}}{4{\pi}}\Phi \nonumber\\
&+&\text{non-local quadratic terms}.\label{main1}
\end{eqnarray}
The derivation of this result starts from the effective action (\ref{effective}). First, we should expand the background field as $B_i=L_i+A_i$. The quadratic tree-level action $S_2$, as derives from (\ref{ars3}), explicitly reads
\begin{eqnarray}
  S_2[A_i]=-\frac{1}{4g^2N}{\rm Tr}\big[F_{ij}^{(0)}\big]^2-\frac{i}{2g^2N}{\epsilon}_{ijk}{\rm Tr}\big[\frac{1}{2}F_{ij}^{(0)}A_k\big].\label{lll}
\end{eqnarray}
Second, we will apply directly the result (\ref{effective}) to find quantum
corrections to this quadratic action. This will of course
capture all quantum corrections to the vacuum polarization tensor
as well as tadpole corrections. The quadratic effective action
is given explicitly by
\begin{eqnarray}
{\Gamma}_2  =S_2+\frac{1}{2}{\rm TR}\left(\Delta^{(1)}+\Delta^{(2)}-\frac{1}{2}(\Delta^{(1)})^2\right)-\frac{1}{4}{\rm TR}(\Delta^{(3)})^2_{ii}.
\label{ex}
\end{eqnarray}
Here, ${\Delta}^{(1)}$, ${\Delta}^{(2)}$ and ${\Delta}^{(3)}$ are defined by

\begin{eqnarray}
\Delta^{(1)}  = \frac{1}{{\cal L}^2}\big({\cal L}{\cal A}+{\cal
A}{\cal L}\big)~,~\Delta^{(2)}  =  \frac{1}{{\cal L}^2}{\cal
A}^2~,~{\Delta}^{(3)}_{ij}=\frac{2}{{\cal L}^2}{\cal
F}^{(0)}_{ij}.
\end{eqnarray}
It is obvious, from these expressions, that the propagators of the fluctuation field $Q_i$ and the ghost fields $b^{\dagger}$ and $b$ are given by the inverse of the Laplacian ${\cal L}^2$. In fact, it is a  property of the  Alekseev-Recknagel-Schomerus action given by either (\ref{ars}) or (\ref{ars2}) or (\ref{ars3}) that the propagator in the Feynman gauge $\xi=1$ is given simply by $1/{\cal L}^2$.

\begin{figure}[h]
\begin{center}
\includegraphics[width=10cm]{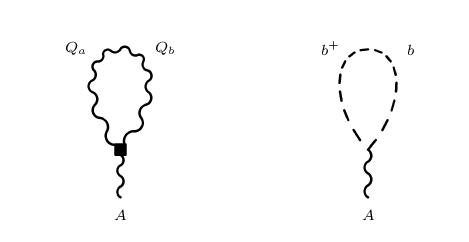}
\caption{Tadpole diagrams.}\label{UVIR1}
\end{center}
\end{figure}

We start with the tadpole contribution. Quantum correction to the tree-level linear term, which is
in fact identically zero, is given by the combination of the two
tadpole diagrams of figure (\ref{UVIR1}). These diagrams are also equal to
the second term in the  expansion (\ref{ex}), viz ${\Gamma}_1 =\frac{1}{2}{\rm TR}\Delta^{(1)}$. We find

\begin{eqnarray}
{\Gamma}_1=4\frac{\sqrt{c_2}}{N}{\rm Tr}\Phi
-\frac{2}{N}{\rm Tr}A_i^2.\label{5.21}
\end{eqnarray}
This identity is exact and as it turns out it is crucial in establishing gauge invariance of the quantum noncommutative gauge theory on the fuzzy sphere.

\begin{figure}[h]
\begin{center}
\includegraphics[width=18cm]{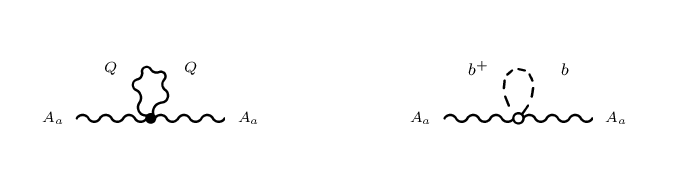}
\caption{Vacuum polarization diagrams ($4$-vertices).}\label{UVIR2}
\end{center}
\end{figure}

\begin{figure}[h]
\begin{center}
\includegraphics[width=8cm]{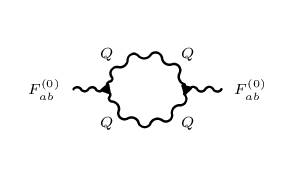}
\caption{Vacuum polarization diagrams ($3$-vertices with the $F$-field).}\label{UVIR3}
\end{center}
\end{figure}

\begin{figure}[h]
\begin{center}
\includegraphics[width=15cm]{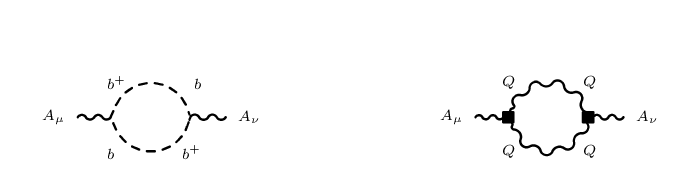}
\caption{Vacuum polarization diagrams ($3$-vertices with the $A$-field).}\label{UVIR4}
\end{center}
\end{figure}

Next, we compute the vacuum polarization tensor. The $4$-vertex contribution to the vacuum polarization tensor is
given by the diagrams of figure (\ref{UVIR2}) which are also equal to the $3$rd term in the expansion (\ref{ex}), viz ${\Gamma}_2^{(4)} =\frac{1}{2}{\rm TR}\Delta^{(2)}$. After some calculation we get the explicit answer
\begin{eqnarray}
{\Gamma}_2^{(4)}=\frac{1}{N}{\rm Tr}A_i{\cal L}^2{\Delta}_4A_i.\label{5.26}
\end{eqnarray}
The operator ${\Delta}_4{\equiv}{\Delta}_4({\cal L}^2)$ is defined by its eigenvalues on polarization tensor $\hat{Y}_{p_1n_1}$ given by (with $L\equiv N-1$)
\begin{eqnarray}
{\Delta}_4(p_1)=\sum_{l_1,l_2}\frac{2l_1+1}{l_1(l_1+1)}\frac{2l_2+1}{l_2(l_2+1)}(1-(-1)^{l_1+l_2+p_1})(L+1)\left\{\begin{array}{ccc}
        p_1 & l_1 & l_2 \\
    \frac{L}{2} & \frac{L}{2} & \frac{L}{2}
    \end{array} \right\}^2\frac{l_2(l_2+1)}{p_1(p_1+1)}.\label{5.27}
\end{eqnarray}
The $3$-vertex contribution comes from three different diagrams. The contribution of the ${\cal F}$ term is given by the diagram of figure (\ref{UVIR3})  and it corresponds to the last term in expansion(\ref{ex}), namely ${\Gamma}_2^{(3F)}=-\frac{1}{4}{\rm TR}({\Delta}^{(3)})^2_{ii}$ whereas
the $4$th term in the  expansion (\ref{ex}), i.e. ${\Gamma}_2^{(3A)}=-\frac{1}{4}{\rm TR}({\Delta}^{(1)})^2$ corresponds
to the combination of the diagrams displayed in figure (\ref{UVIR4}).

The diagram of  figure (\ref{UVIR3})  is found to be given by

\begin{eqnarray}
{\Gamma}_2^{(3F)}=\frac{1}{N}{\rm Tr}F_{ij}^{(0)}{\Delta}_FF_{ij}^{(0)}.\label{5.31}
\end{eqnarray}
The operator ${\Delta}_F{\equiv}{\Delta}_F({\cal L}^2)$ is defined by its spectrum
\begin{eqnarray}
{\Delta}_F(p_1)&=&2\sum_{l_1,l_2}\frac{2l_1+1}{l_1(l_1+1)}\frac{2l_2+1}{l_2(l_2+1)}(1-(-1)^{l_1+l_2+p_1})(L+1)\left\{\begin{array}{ccc}
        l_1 & l_2 & p_1 \\
    \frac{L}{2} & \frac{L}{2} & \frac{L}{2} \end{array}\right\}^2.\label{5.32}
\end{eqnarray}
Similarly, the diagrams of  figure (\ref{UVIR4}) are given by
\begin{eqnarray}
\Gamma_2^{(3A)}&=&-\frac{1}{N}{\rm Tr}A_i{\cal L}_i{\Delta}_3{\cal
  L}_jA_j+\Gamma_2^{(3A)}.\label{5.36}
\end{eqnarray}
The second term is given explicitly by 
\begin{eqnarray}
\Gamma_2^{(3A)}&=&\sum_{p_1n_1}\sum_{p_2n_2}A_{-\mu}(p_1n_1)A_{-\nu}(p_2n_2)(-1)^{n_1+\nu}\bigg[C_{p_1n_11\mu}^{p_1-1m}C_{p_2n_21\nu}^{p_1-1-m}\big({\Lambda}^{(-)}(p_1,p_2)+ {\Sigma}^{(-)}(p_1,p_2)\big)\nonumber\\
&+&C_{p_1n_11\mu}^{p_1+1m}C_{p_2n_21\nu}^{p_1+1-m}\big({\Lambda}^{(+)}(p_1,p_2)+{\Sigma}^{(+)}(p_1,p_2)\big)\bigg].\label{46}
\end{eqnarray}
Again the operator ${\Delta}_3{\equiv}{\Delta}_3({\cal L}^2)$ is defined by its spectrum
\begin{eqnarray}
{\Delta}_3(p_1)&=&\sum_{l_1,l_2}\frac{2l_1+1}{l_1(l_1+1)}\frac{2l_2+1}{l_2(l_2+1)}(1-(-1)^{l_1+l_2+p_1})(L+1)\left\{\begin{array}{ccc}
        p_1 & l_1 & l_2 \\
    \frac{L}{2} & \frac{L}{2} & \frac{L}{2} \end{array}\right\}^2\nonumber\\
&{\times}&\frac{l_2(l_2+1)}{p_1^2(p_1+1)^2}\big(l_2(l_2+1)-l_1(l_1+1)\big).\label{5.37}
\end{eqnarray}
We remark that all quantum corrections to the vacuum polarization tensor given by the equations (\ref{5.26}),
 (\ref{5.31}) and (\ref{5.36}) are written in terms of the
operator
\begin{eqnarray}
{\Delta}(p_1,p_2)&=&\sum_{l_1l_2}\frac{2l_1+1}{l_1(l_1+1)}\frac{2l_2+1}{l_2(l_2+1)}(L+1)\left\{\begin{array}{ccc}
        p_1 & l_1 & l_2 \\
    \frac{L}{2} & \frac{L}{2} & \frac{L}{2} \end{array}\right\} \left\{\begin{array}{ccc}
        p_2 & l_1 & l_2 \\
    \frac{L}{2} & \frac{L}{2} & \frac{L}{2}
    \end{array}\right\}X(l_1,l_2,p_1,p_2),\label{5.47}\nonumber\\
\end{eqnarray}
$X$  is of the form $X(l_1,l_2,p_1,p_2)=\delta_{p_1p_2}\bar{X}(l_1,l_2,p_1)$ and
where the sums are always over $l_1$ and $l_2$ such that
$l_1+l_2+p_1$ is an odd number.

Similarly, the functions ${\Lambda}^{(\pm)}(p_1,p_2)$ and  ${\Sigma}^{(\pm)}(p_1,p_2)$ in the definition of $\Gamma_2^{(3A)}$ (equation (\ref{46})) are of the form
(\ref{5.47}) with some $X$ such that $\Lambda^{\left(\pm\right)}\left(p_1,p_2\right)=\delta_{p_{1},p_{2}}\bar{\Lambda}^{\left(\pm\right)}\left(p_1\right)$ and ${\Sigma}^{(\pm)}(p_1,p_2)={\delta}_{p_1{\pm}2,p_2}\bar{\Sigma}^{(\pm)}(p_1)$ respectively. Furthermore, we can see by inspection that the
Clebsch-Gordan coefficients appearing in the action $
\Gamma_{2}^{(3A_2)}$  are exactly those which appear in  the
scalar mass term 
\begin{eqnarray}
& &\frac{1}{4N}{\rm Tr}[\hat{x}_i,A_i]_{+}^2=\sum_{p_1n_1}\sum_{p_2n_2}A_{-\mu}(p_1n_1)A_{-\nu}(p_2n_2)(-1)^{n_1+\nu}\bigg[C_{p_1n_11\mu}^{p_1-1m}C_{p_2n_21\nu}^{p_1-1-m}\nonumber\\
&\times&\big({\lambda}^{(-)}(p_1,p_2)+ {\sigma}^{(-)}(p_1,p_2)\big)
+C_{p_1n_11\mu}^{p_1+1m}C_{p_2n_21\nu}^{p_1+1-m}\big({\lambda}^{(+)}(p_1,p_2)+{\sigma}^{(+)}(p_1,p_2)\big)\bigg].
\label{47}
\end{eqnarray}
Here, ${\lambda}^{(\pm)}(p_1,p_2)$ and ${\sigma}^{(\pm)}(p_1,p_2)$
are some other functions which are such that
${\lambda}^{(\pm)}(p_1,p_2)={\delta}_{p_1,p_2}\bar{\lambda}^{(\pm)}(p_1)$,
${\sigma}^{(\pm)}(p_1,p_2)={\delta}_{p_1{\pm}2,p_2}\bar{\sigma}^{(\pm)}(p_1)$.

By comparing (\ref{46}) and (\ref{47}) we can immediately deduce
that the action ${\Gamma}_2^{(3A_2)}$, in position space, must
involve anticommutators of $\hat{x}_i$ and $A_i$ instead of commutators
and hence it is a scalar-like contribution. As it turns out, this
action contains (in the commutative limit) gauge-invariant terms which describe
non-local interactions between the normal scalar and the tangent gauge fields on the sphere \cite{Castro-Villarreal:2004ulr}.

  By putting together equations (\ref{5.21}), (\ref{5.26}),
(\ref{5.31}) and (\ref{5.36}) we obtain the full
quadratic effective action of noncommutative gauge theory on the fuzzy sphere ${\bf S}^2_N$. Explicitly, we have
\begin{eqnarray}
{\Gamma}_{2}=S_2+\frac{4\sqrt{c_2}}{N}{\rm Tr}\Phi+\frac{1}{N}{\rm Tr}A_i\big({\cal
L}^2{\Delta}_4-2\big)A_i-\frac{1}{N}{\rm Tr}A_i{\cal L}_i{\Delta}_3{\cal
L}_jA_j+\frac{1}{N}{\rm Tr}F_{ij}^{(0)}{\Delta}_FF_{ij}^{(0)}
+{\Gamma}_2^{(3A_2)}.\label{eff}\nonumber\\
\end{eqnarray}
In  is rather clear that the first $2$ terms and the last $2$ terms are gauge invariant (in the commutative limit $N\longrightarrow\infty$).
Naturally, we also expect that the $3$rd and $4$th terms in (\ref{eff}) to become gauge invariant in the limit. To
check this property explicitly we rewrite these two terms as follows
\begin{eqnarray}
{\rm Tr}A_i\big({\cal L}^2{\Delta}_4-2\big)A_i-{\rm Tr}A_i{\cal L}_i{\Delta}_3{\cal L}_jA_j&=&-\frac{1}{2}{\rm Tr}F_{ij}^{(0)}{\Delta}_3F_{ij}^{(0)}\nonumber\\
&-&\frac{i}{2}{\epsilon}_{ijk}{\rm Tr}F_{ij}^{(0)}\big({\Delta}_3+{\cal L}^2({\Delta}_3-{\Delta}_4)+2\big)A_k\nonumber\\
&+&i{\epsilon}_{ijk}{\rm Tr}{\cal L}_iA_j\big({\cal L}^2({\Delta}_3-{\Delta}_4)+2\big)A_k.
\end{eqnarray}
The first two terms in this expression are now exactly gauge invariant in the commutative limit whereas the third term can not be gauge invariant unless it vanishes identically. We expect therefore by the requirement of gauge invariance alone that we have the asymptotic behaviour ${\cal L}^2({\Delta}_3-{\Delta}_4)+2{\longrightarrow}0$, $N{\longrightarrow}{\infty}$.
As it turns out, this behavior is true for all finite values of $N$. In other words, we have in the identity \cite{Castro-Villarreal:2004ulr}
\begin{eqnarray}
{\cal L}^2({\Delta}_3-{\Delta}_4)+2=0.\label{5.42}
\end{eqnarray}
Thus, the quadratic effective action of noncommutative gauge theory on the fuzzy sphere ${\bf S}^2_N$ reads
\begin{eqnarray}
{\Gamma}_{2}&=&S_2+\frac{4\sqrt{c_2}}{N}{\rm Tr}\Phi+\frac{1}{N}{\rm Tr}F_{ij}^{(0)}\big({\Delta}_F-\frac{1}{2}{\Delta}_3\big)F_{ij}^{(0)}-\frac{i}{2N}{\epsilon}_{ijk}{\rm Tr}F_{ij}^{(0)}{\Delta}_3A_k
+{\Gamma}_2^{(3A_2)}.\label{eff1}
\end{eqnarray}
In summary, the $3$rd and $4$th terms of the effective action (\ref{eff1}) give rise to a non-trivial quantum
contribution to the noncommutative gauge theory on the fuzzy sphere reflecting the existence of a gauge invariant UV-IR mixing which survives the
commutative limit. Indeed, this effective action (\ref{eff1}) goes, in the commutative limit $N\longrightarrow\infty$, to the action (\ref{main1}) where we have also to use the fact that the eigenvalues $\Delta_3$ and $\Delta_F$ are given in the limit by  
\begin{eqnarray}
{\Delta}_F(p_1)\longrightarrow0~,~
{\Delta}_3(p_1)\longrightarrow \frac{2\sum_{l=2}^{p_1}\frac{1}{l}}{p_1(p_1+1)}\ne 0.
\end{eqnarray}

\subsection{The UV-IR mixing in the large mass limit}
The normal scalar fluctuation $\Phi$ can be suppressed, and thus reduce the noncommutative gauge theory to a purely two-dimensional model on the fuzzy sphere ${\bf S}^2_N$, by simply adding a large mass term to the  Alekseev-Recknagel-Schomerus action (\ref{ars3}). In other words, we should consider instead the action
\begin{eqnarray}
S_N[A_i]=-\frac{1}{4g^2N}{\rm Tr}\left[F_{ij}^{(0)}+[A_i,A_j]\right]^2-\frac{i}{2g^2N}{\epsilon}_{ijk}{\rm Tr}\left[\frac{1}{2}F_{ij}^{(0)}A_k+\frac{1}{3}[A_i,A_j]A_k\right]+\frac{2m^2}{g^2N}{\rm Tr}{\Phi}^2.\label{ARS+V}\nonumber\\
\end{eqnarray}
In the presence of this mass term the quadratic tree-level action (\ref{lll}) and the  quadratic effective action (\ref{ex}) become now given  by 
\begin{eqnarray}
S_2=-\frac{1}{2g^2N}{\rm Tr}[L_i,A_j]^2+\frac{1}{2g^2N}{\rm Tr}[L_i,A_i]^2+\frac{m^2}{2g^2N}{\rm Tr}(\hat{x}_iA_i+A_i\hat{x}_i)^2.\label{lllM}
\end{eqnarray}
{\begin{eqnarray}
{\Gamma}_2  &=&S_2+{\rm TR}\bigg[\frac{1}{2}\bigg(\frac{1}{\Delta}\bigg)_{ij}{\omega}^{(1)}_{ji}-\frac{1}{{\cal
L}^2}({\cal L}{\cal A}+{\cal A}{\cal L})\bigg]\nonumber\\
&+&{\rm TR}\bigg[\frac{1}{2}\bigg(\frac{1}{\Delta}\bigg)_{ij}{\omega}_{ji}^{(2)}-\frac{1}{4}\bigg(\frac{1}{\Delta}\bigg)_{ik}{\omega}_{kl}^{(1)}\bigg(\frac{1}{\Delta}\bigg)_{lj}{\omega}^{(1)}_{ja}
-\frac{1}{{\cal L}^2}{\cal A}^2+\frac{1}{2}\frac{1}{{\cal
L}^2}({\cal L}{\cal A}+{\cal A}{\cal L})\frac{1}{{\cal
L}^2}({\cal L}{\cal A}+{\cal A}{\cal L})\bigg].\label{exM}\nonumber\\
\end{eqnarray}
The operators ${\omega}^{(1)}$ and ${\omega}^{(2)}$ contain the linear and quadratic vertices respectively and they are given explicitly by

\begin{eqnarray}
{\omega}^{(1)}_{ij}&=&({\cal L}{\cal A}+{\cal A}{\cal
L}){\delta}_{ij}+2{\cal
F}^{(0)}_{ij}+\frac{2m^2}{c_2}(LA+AL){\delta}_{ij}+\frac{4m^2}{c_2}(L_iA_j+A_iL_j)\nonumber\\
{\omega}^{(2)}_{ij}&=&{\cal A}^2{\delta}_{ij}+2[{\cal A}_i,{\cal
A}_j]+\frac{2m^2}{c_2}A^2{\delta}_{ij}+\frac{4m^2}{c_2}A_iA_j.
\end{eqnarray}
The Laplacian $\Delta$ is given, on the other hand,  by  ${\Delta}_{ij}={\cal L}^2\delta_{ij}+4m^2P^N_{ij}$ where $P^N_{ij}=\hat{x}_i\hat{x}_j$
is the normal projector on the fuzzy sphere ${\bf S}^2_N$. The tangent projector on the fuzzy sphere  is then obviously given by the orthogonal complement  defined  by $P^T_{ij}=\delta_{ij}-\hat{x}_i\hat{x}_j$. These two projectors correspond to the normal and tangent projective modules  on the fuzzy sphere respectively. These projective modules play in noncommutative geometry the same role played by fiber bundles in differential geometry.

The propagator, in the large mass limit $m^2\longrightarrow\infty$, takes the relatively simple form
\begin{eqnarray}
\frac{1}{\Delta}=P^T\frac{1}{{\cal L}^2}P^T+O(\frac{1}{m^2}).
\end{eqnarray}
It is then straightforward to see that the combined contribution of the tadpole diagrams and the $4$-vertex corrections to the vacuum polarization tensor vanishes in this limit, viz

\begin{eqnarray}
  {\Gamma}_1+{\Gamma}^{(4)}_2={\rm TR}\frac{1}{2}\bigg(\frac{1}{\Delta}\bigg)_{ij}\bigg({\cal L}{\cal A}+{\cal A}{\cal
L}+{\cal A}^2\bigg){\delta}_{ij}-{\rm TR}\frac{1}{2}\bigg(\frac{1}{{\cal L}^2}\bigg)\bigg({\cal L}{\cal A}+{\cal A}{\cal
L}+{\cal A}^2\bigg)\equiv 0.
\end{eqnarray}
This result should  be compared with the
commutative limit of the sum of the two actions (\ref{5.21}) and
(\ref{5.26}) which as we have shown does depend in the limit on the $2$-dimensional gauge field. In other words, suppressing the normal component of the gauge field, by giving it
a large mass, allowed us to suppress in the limit the contribution
of the tangent gauge field to the tadpole and to the $4$-vertex
correction of the vacuum polarization tensor. By the requirement
of gauge invariance this suppression will also occur in the other
contributions to the vacuum polarization tensor and as
consequence the large mass of the scalar field regulates
effectively the UV-IR mixing.
\subsection{Phase  diagram in the large mass limit and stability of the geometric phase}
The massive  Alekseev-Recknagel-Schomerus action (\ref{ARS+V}) can also be rewritten as a $D=3$ Yang-Mills matrix model given by 
\begin{eqnarray}
  S[D]&=&S_{\rm ARS}[D]+V[D]\nonumber\\
  S_{\rm ARS}&=&\frac{1}{g^2N}{\rm Tr}\bigg[-\frac{1}{4}[D_i,D_j]^2+\frac{2i}{3}{\epsilon}_{ijk}D_iD_jD_k\bigg]\nonumber\\
    V[D]&=&\frac{1}{g^2N}{\rm Tr}\bigg[-\mu D_i^2+\frac{m^2}{2c_2}(D_i^2)^2\bigg]~,~\mu=m^2.\label{massivears}
\end{eqnarray}
The case $m^2=0$ is studied in \cite{Azuma:2004zq,Azuma:2004ie} whereas the case $m^2\ne 0$ is studied in \cite{Delgadillo-Blando:2008cuz,OConnor:2006iny,Delgadillo-Blando:2007mqd}.

By following the standard Faddeev-Popov procedure \cite{Faddeev:1967fc} and taking 
the background field configuration to be $D_a=\varphi L_a$ one finds that the free energy, the effective potential and the equation of motion of the order parameter $\varphi$ to be given by
\begin{eqnarray}
  F=-\log Z~,~\frac{F}{N^2}=\frac{3}{4}\log\tilde{\alpha}^4+\frac{\tilde{\alpha}^4}{2}
\bigg[\frac{{\varphi}^4}{4}-\frac{{\varphi}^3}{3}
+m^2\frac{{\varphi}^4}{4}-\mu\frac{\varphi^2}{2}\bigg]+
\log\tilde{\alpha}{\varphi}.\label{formula1}
\end{eqnarray}
\begin{equation}
\label{V_eff}
\frac{V_{\rm eff}}{2c_2}=\tilde{\alpha}^4\left[\frac{\varphi^4}{4}
-\frac{\varphi^3}{3}+m^2\frac{\varphi^4}{4}-\mu\frac{\varphi^2}{2}\right]+\log \varphi^2.
\end{equation}

\begin{eqnarray}
\label{equ-for-phi}
\frac{\tilde{\alpha}^4}{2}
\bigg[{\varphi}^4-{\varphi}^3
+m^2{\varphi}^4-\mu{\varphi^2}\bigg]+1=0 \Rightarrow \bar{\varphi}^4-\bar{\varphi}^3-t\bar{\varphi}^2+\frac{2}{a^4}=0~,~\bar{\varphi}=(1+m^2)\varphi.
\end{eqnarray}
Here, $t$ and $a$ are the actual parameters of the model  defined explicitly by
\begin{eqnarray}
t=\mu(1+m^2)~,~ a^4=\tilde{\alpha}^4/(1+m^2)^3.
  \end{eqnarray}
Thus, the critical point found for the   Alekseev-Recknagel-Schomerus model with $\mu=m^2=0$ will be  replaced by a critical line in the  $a-t$ plane.

First, we compute the expected radius of the sphere (extent of space) as follows 
\begin{equation}
\label{sphere_expected_radius}
\frac{1}{r}=\frac{\langle Tr D_i^2\rangle}{Nc_2}=
-\frac{2}{\tilde{\alpha}^4}\frac{dF}{d \mu}=\varphi^2.
\end{equation}
We also compute the average value of the total action (energy) by 
\begin{equation}
{\cal S}=\langle S\rangle/N^2
=\tilde{\alpha}^4\frac{d}{d\tilde{\alpha}^4}\bigg(\frac{F}{N^2}\bigg)=\frac{3}{4}-\frac{\tilde{\alpha}^4{\varphi}^3}{24}-\frac{\tilde{\alpha}^4\mu{\varphi}^2}{8}. \label{u} 
\end{equation}  
Scaling the covariant matrix coordinates $D_i$ as $D_i\longrightarrow (1-\epsilon )D_i$ in both the
 total action $S[D]$ and the measure $\int \prod_{i=1}^3[dD_i]$  leaves the partition function $Z=\int \prod_{i=1}^3[dD_i]\exp(-S[D])$ invariant which in turn leads to a non-trivial Ward identity given by 
\begin{eqnarray}\label{wardid}
\frac{\tilde{\alpha}^4}{N}<K_m>=3N^2~,~K_m={\rm Tr}\bigg(-[D_i,D_j]^2 +2 i \epsilon_{ijk} D_i D_j D_k-2 m^2D_i^2+\frac{2m^2}{c_2} (D_i^2)^2\bigg).\label{wa1}
\end{eqnarray}
Using this identity we can express the energy as
\begin{eqnarray}
{\cal S}=\frac{3}{4}+\frac{\tilde{\alpha}^4}{6N^3}\langle i{\epsilon}_{ijk}{\rm Tr}
D_iD_jD_k\rangle-\frac{\tilde{\alpha}^4}{2N^3}m^2\langle {\rm Tr} D_i^2\rangle.\label{wa2}
\end{eqnarray}
Thus, we can immediately compute the  specific heat to be given by
\begin{eqnarray}
C_v=\frac{\langle S^2\rangle-\langle S\rangle^2}{N^2}=\frac{\langle S\rangle }{N^2}-\tilde{\alpha}^4\frac{d}{d\tilde{\alpha}^4}\bigg(\frac{\langle S\rangle}{N^2}\bigg)=\frac{3}{4}+\frac{\tilde{\alpha}^5\varphi}{32}(\varphi +2m^2)\frac{d\varphi}{d\tilde{\alpha}}.\label{cv}
\end{eqnarray}
We also define the Yang-Mills and Chern-Simons term by
\begin{eqnarray}
4{\rm YM}=-\frac{\langle {\rm Tr}[D_i,D_j]^2\rangle}{2Nc_2}={\varphi}^4+\frac{8}{\tilde{\alpha}^4}~,~\text{and }~3{\rm CS}=\frac{\langle i{\epsilon}_{ijk}{\rm Tr} D_iD_jD_k\rangle}{Nc_2}=-{\varphi}^3.
\end{eqnarray} 
Let us now examine the predictions for the quantum transition as
determined by the effective potential given in (\ref{V_eff}).

The extrema of
the classical potential occur at
\begin{equation}
\varphi=\frac{1}{1+m^2}\left\{0,~ \varphi_{\pm}=\frac{1\pm \sqrt{1+4t}}{2}\right\}.
\end{equation}
For $\mu$ positive the global minimum is ${\varphi}_+$. The solution $\varphi=0$ is a local maximum and ${\varphi}_-$ is a local minimum. In particular for $\mu=m^2$  we obtain the global minimum ${\varphi}_+=1$. For $\mu$ negative the global minimum is still ${\varphi}_+$ but $0$ becomes a local minimum and ${\varphi}_-$ a local maximum. If $\mu$ is sent more negative then 
the global minimum  ${\varphi}_+=1$ becomes degenerate with  ${\varphi}=0$ 
at $t=-\frac{2}{9}$ and the maximum height of the barrier
is given by $V_-={\tilde\beta^4}/324 $ which occurs at ${\varphi}_-=\frac{1}{3}$. The model has a first order transition at $t=-2/9$ where the classical ground states switches from ${\varphi}_+$ for $t>-2/9$ to $0$ for $t<2/9$.

Let us now consider the effect of quantum fluctuations. The condition $V^{'}_{\rm eff}=0$ 
gives us extrema of the model. For large enough $\tilde{\alpha}$,  and large enough $m$ and $\mu$, it 
admits two positive solutions. The largest solution can be identified
with the ground state of the system. It will
determine the radius of the sphere. 
The second solution is the local maximum  of $V_{\rm eff}$ and will determine the height of the 
barrier (figure
\ref{figVeff}). As the coupling is decreased these two solutions merge and
the barrier disappears. This is the critical point of the model. For smaller couplings
than the critical value $\tilde\alpha_*$ the fuzzy sphere
solution $D_i= {\varphi}L_i$ no longer exists. Therefore, the classical transition described above is
significantly affected by quantum fluctuations.

\begin{figure}
\begin{center}
\includegraphics[width=10.0cm]{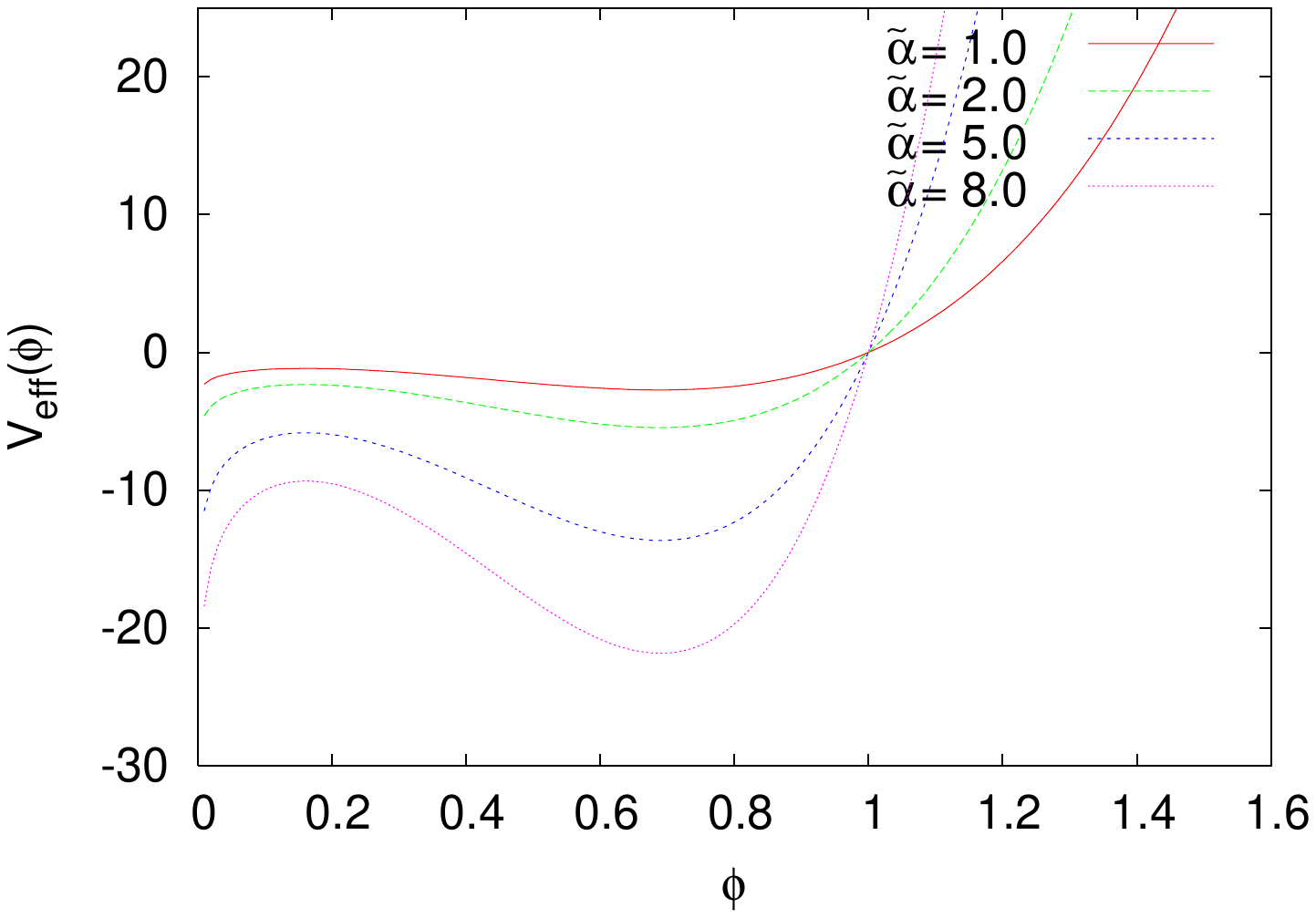}
\caption{{The effective potential for $m^2=20$.}}\label{figVeff}
\end{center}
\end{figure}

The condition when the barrier disappears is $V_{\rm eff}^{\prime\prime}=0$. At this point the local minimum  merges with the local maximum. Solving the two equations  $V_{\rm eff}^{\prime}=V_{\rm eff}^{\prime\prime}=0$ yield the critical value
\begin{eqnarray}
g_*^2=\frac{1}{\tilde{\alpha}_*^4}=\frac{{\varphi}_{*}^2({\varphi}_*+2\mu)}{8}~,~{\varphi}_{*}=\frac{3}{8(1+m^2)}\bigg[1+\sqrt{1+\frac{32\mu (1+m^2)}{9}}\bigg].\label{critline}
\end{eqnarray}
If we take $\mu$ negative we see that $g_*$ goes to zero at $\mu(1+m^2)=-1/4$ and the critical 
coupling $\tilde{\alpha}_* $ is sent to infinity and therefore
for $\mu(1+m^2)<-\frac{1}{4}$ the model has no fuzzy sphere phase. However, in the region $-\frac{1}{4}<\mu(1+m^2) <-\frac{2}{9}$ the action $S_{\rm ARS}+V$ is completely positive. It is therefore
not sufficient to consider only the configuration $D_i=\varphi L_i$,
but rather all $SU(2)$ representations must be considered. 
Furthermore, for large $\tilde{\alpha}$ the ground state will be dominated
by those representations with the smallest Casimir. This means that
there is no fuzzy sphere solution for $\mu(1+m^2)<-\frac{2}{9}$.

The limits of interest are the limit $\mu=m^2{\longrightarrow}0$ and  the limit $\mu=m^2{\longrightarrow}\infty$. 
In these cases, we have the critical values
\begin{eqnarray}
{\varphi}_{*}=\frac{3}{4}~,~\tilde{\alpha}_*^4=(\frac{8}{3})^3~,~\mu=m^2\longrightarrow 0.\label{cv0}
\end{eqnarray}
\begin{eqnarray}
{\varphi}_{*}=\frac{1}{\sqrt{2}}~,~\tilde{\alpha}^4_{*}=\frac{8}{m^2}~,~\mu=m^2\longrightarrow \infty.\label{pre2}
\end{eqnarray}
This means that the phase transition, in the limit of large mass, is located at a smaller value of
the coupling constant $\tilde{\alpha}$ as $m$ is increased.  In other
words, the region where the fuzzy sphere is stable is extended to lower
values of the gauge coupling.

The critical behavior for the model $\mu=m^2=0$ is given explicitly by
\begin{eqnarray}
\varphi =\frac{1}{4}\bigg[3+\sqrt{6}\sqrt{\frac{\tilde{\alpha}-\tilde{\alpha}^{*}}{\tilde{\alpha}^{*}}}-\frac{4}{3}\frac{\tilde{\alpha}-\tilde{\alpha}^{*}}{\tilde{\alpha}^{*}}+O({\epsilon}^{\frac{3}{2}})\bigg]~,~\epsilon=\frac{\tilde{\alpha}-\tilde{\alpha}^{*}}{\tilde{\alpha}^{*}}.\label{sq1}
\end{eqnarray}

\begin{eqnarray}
{\cal S}=\frac{5}{12}-\frac{1}{3^{\frac{1}{8}}2^{\frac{5}{8}}}\sqrt{\tilde{\alpha}-\tilde{\alpha}_*}-\frac{7}{3^{\frac{5}{4}}2^{\frac{5}{4}}}(\tilde{\alpha}-\tilde{\alpha}_*)+O((\tilde{\alpha}-\tilde{\alpha}_*)^{\frac{3}{2}}).
\end{eqnarray}
\begin{eqnarray}
C_v=\frac{29}{36}+\frac{1}{2^{\frac{11}{8}}3^{\frac{7}{8}}}\frac{1}{\sqrt{\tilde{\alpha}-\tilde{\alpha}_*}} +O((\tilde{\alpha}-\tilde{\alpha}_*)^{\frac{1}{2}}).\label{sq2}
\end{eqnarray}
This gives immediately a divergent specific heat with critical exponent equals $1/2$.

The generic model with $\mu=m^2\ne 0$ is given by the critical behavior

\begin{eqnarray}
\bar\phi=\bar{\phi}_*+
\frac{4}{a_*^{\frac{5}{2}}}\frac{1}{\sqrt{3\bar{\varphi}_*+4t}}\sqrt{a-a_*}+...
\end{eqnarray}
\begin{eqnarray}
{\cal S}={\cal S}_*
-a_c^4\frac{\bar{\phi}_*(\bar{\phi}_*+2t)}{\sqrt{3\bar{\phi}_*+4t}}
\sqrt{\frac{a-a_c}{a_c}}+...~,~{\cal S}_*=\frac{3}{4}-\frac{(\bar{\phi}_*+3t)}{3(\bar{\phi}_*+2t)}.
\end{eqnarray}
\begin{eqnarray}
C_v=C_v^B+\frac{1}{8\sqrt{1+\frac{\tilde{\alpha}_*^4{\phi}_*^3}{16}}}\frac{\sqrt{\tilde{\alpha}_*}}{\sqrt{\tilde{\alpha}-\tilde{\alpha}_*}}+...~,~C_v^B=\frac{3}{4}+\frac{(3+4t)\bar{\phi}_*+2t}{8(3\bar{\phi}+4t)^2}.\label{pft}
\end{eqnarray}
The basic prediction here is that the critical exponent of the 
specific heat for this model 
is given precisely by
\begin{eqnarray}
\alpha=\frac{1}{2}.
\end{eqnarray}
If we  extrapolate these results to large $m^2$ 
where we know that ${\phi}_*\longrightarrow
1/\sqrt{2}$ and $\tilde{\alpha}_*^4\longrightarrow 8/m^2$  we get the specific heat in the fuzzy sphere phase to be given by
\begin{eqnarray}
C_v=\frac{3}{4}+\frac{1}{32\sqrt{2}m^2}+\frac{1}{2^{\frac{21}{8}}m^\frac{1}{4}}\frac{1}{\sqrt{\tilde{\alpha}-\tilde{\alpha}_*}}+...
\end{eqnarray}
The massive  Alekseev-Recknagel-Schomerus action (\ref{massivears}) is studied by means of the Monte Carlo method (Metropolis algorithm) in \cite{Delgadillo-Blando:2008cuz}.

For example, the critical value $\tilde{\alpha}_s$ is measured as
follows. We observe that different actions $\langle S\rangle$ which correspond to
different values of $N$ (for some fixed value of $m^2$) intersect at
some value of the coupling constant $\tilde{\alpha}$ which we define to be the critical point 
$\tilde{\alpha}_s$ (figure \ref{Savm200}). The quantities ${\cal S}$, ${\rm YM}$, ${\rm CS}$ and the radius $1/r$ are 
all continuous across the transition point in this regime. We observe that the Monte Carlo data tend to approach the theoretical prediction as $N$ is increased (figure (\ref{obsm200}).

The specific heat in the fuzzy sphere phase is constant equal to $1$,
it starts to decrease at $\tilde{\alpha}_{\rm max}$, goes through a
minimum at $\tilde{\alpha}_{\rm min}$, and then goes up again to the
value $0.75$ when $\tilde{\alpha}\longrightarrow 0$ (figure
\ref{cvmlarge}).  Extrapolating 
$\tilde{\alpha}_{\rm max}$ and $\tilde{\alpha}_{\rm min}$ to
$N=\infty$ (figure \ref{figex}) we obtain our estimate for
the critical coupling $\tilde{\alpha}_c$ which agree with $\tilde{\alpha}_s$ within
statistical errors.

Thus, it seems that the specific heat in the regime of large values of
$m^2$ becomes constant in the matrix phase equal to
$C_v=0.75$. There remains the question of whether or not the specific
heat has critical fluctuations and diverges with a critical exponent
$\alpha=1/2$ as predicted by equation (\ref{pft}).

In fact, it seems that the {matrix}-to-sphere phase transition is $3$rd
order in the limit of large mass which is a behavior reminiscent of pure matrix models.

Indeed, the essential ingredient in producing this geometric transition is clearly the Chern-Simons term  which is due to the Myers effect \cite{Myers:1999ps}.  For example, the $SO(3)$ matrix model corresponding to the potential $V$ (given in the last line of equation (\ref{massivears})) does not have any transition but when the Chern-Simons term is added we reproduce the one-cut-to-the two-cut transition of the real quartic matrix model. By adding then the Yang-Mills terms, i.e. by considering the full model we should get a  generalization of  the one-cut-to-the two-cut transition. As it turns out, the matrix-to-sphere transition is in a clear sense a one-cut to N-cut transition as shown explicitly for the model with $\mu=m^2=0$. 

\begin{figure}
\begin{center}
  \includegraphics[width=8.0cm]{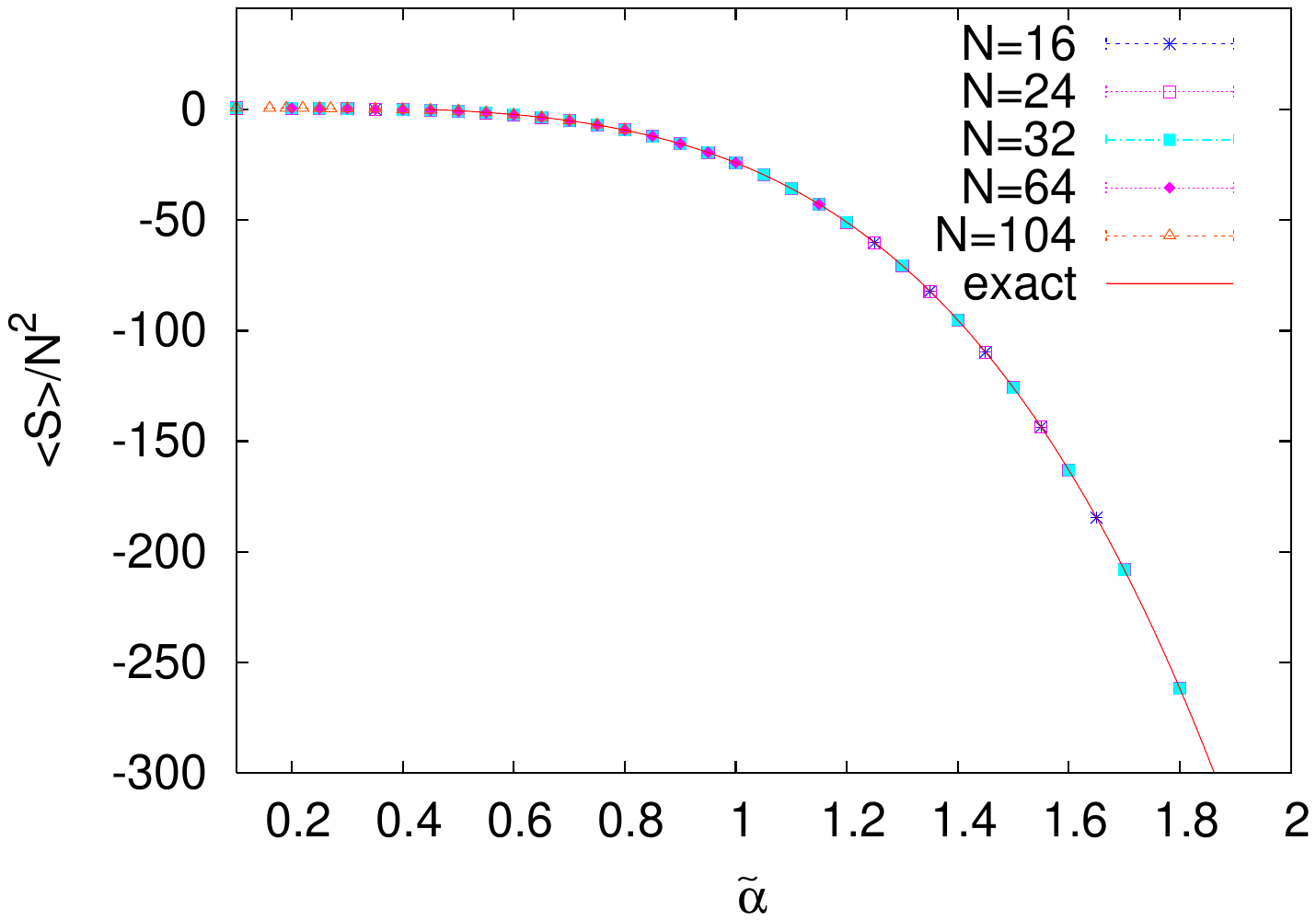}
  \includegraphics[width=8.0cm]{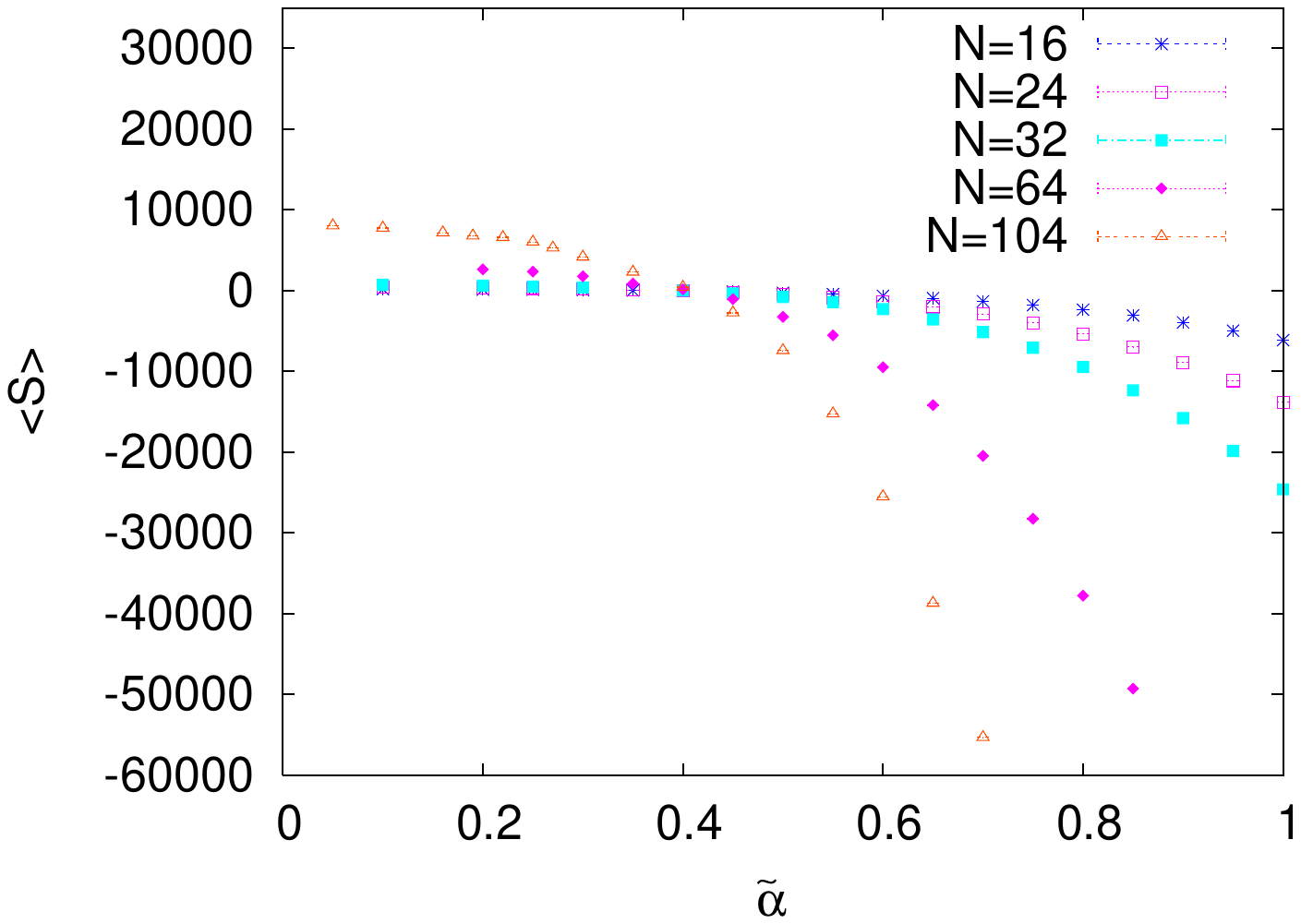}
\caption{The critical value $\tilde{\alpha}_s$ is where the
  curves $\langle S\rangle $ for different values of $N$ intersect.}\label{Savm200}
\end{center}
\end{figure}

\begin{figure}
\begin{center}
\includegraphics[width=8.0cm]{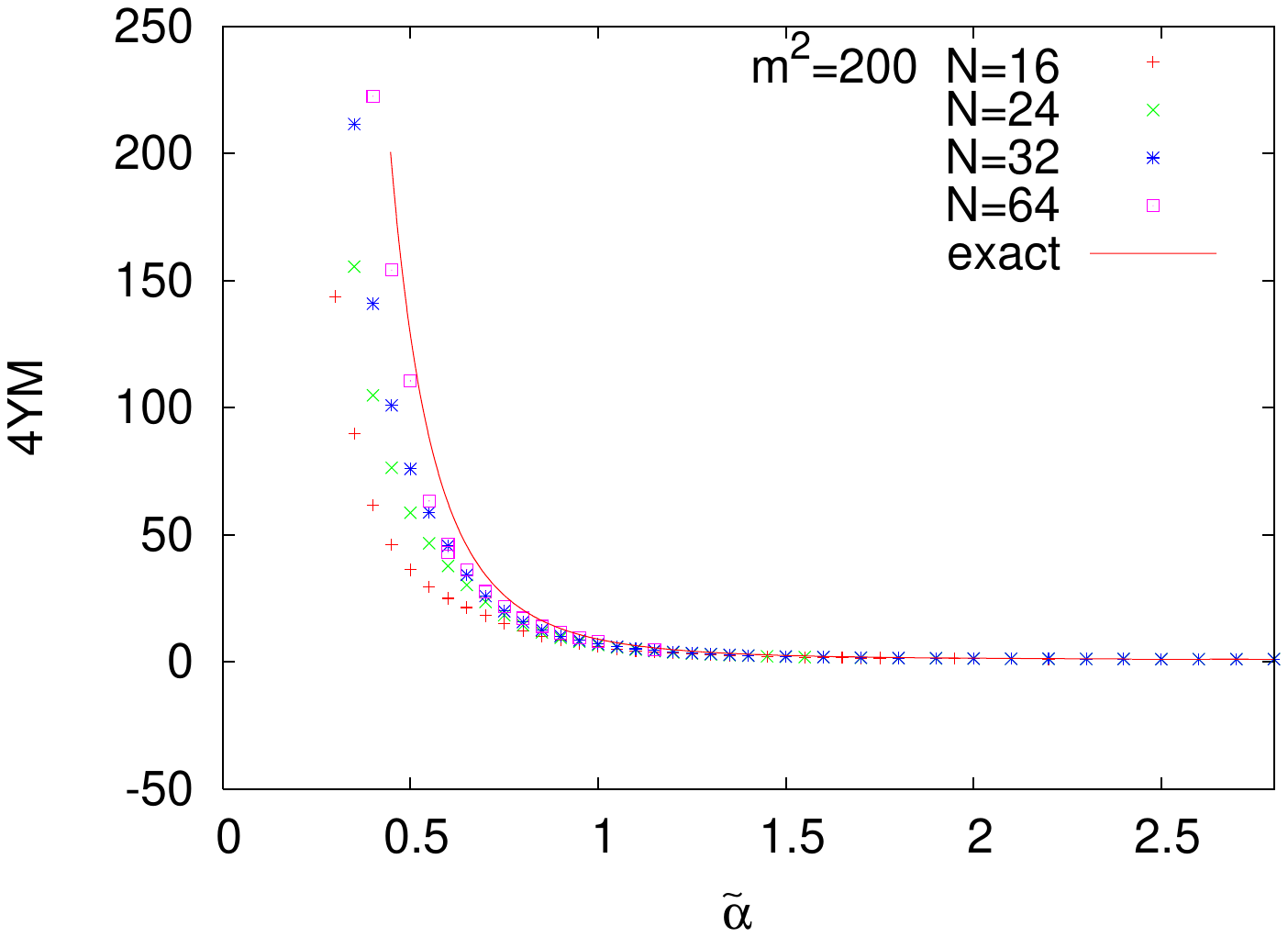}
  \includegraphics[width=8.0cm]{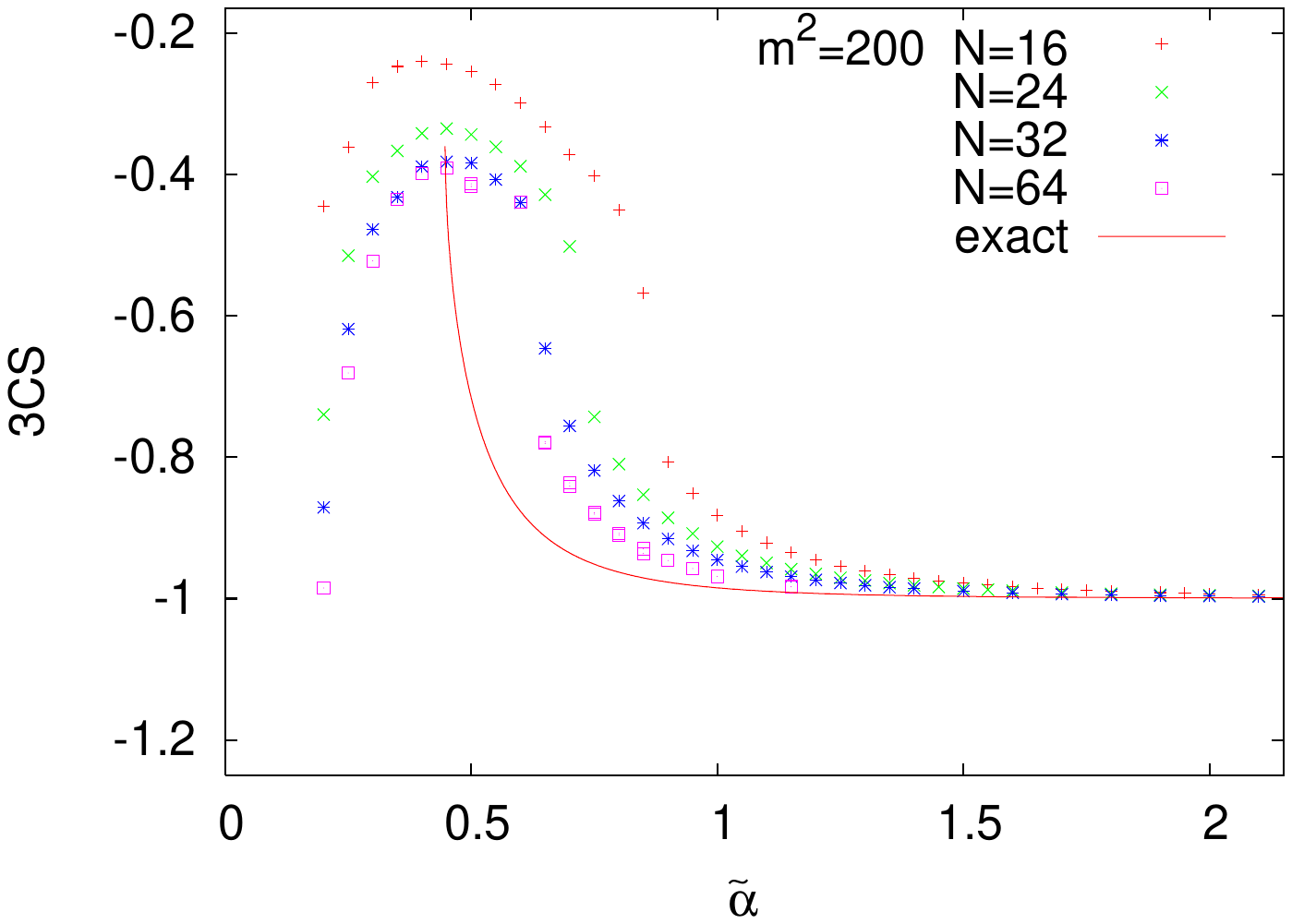}
\includegraphics[width=8.0cm]{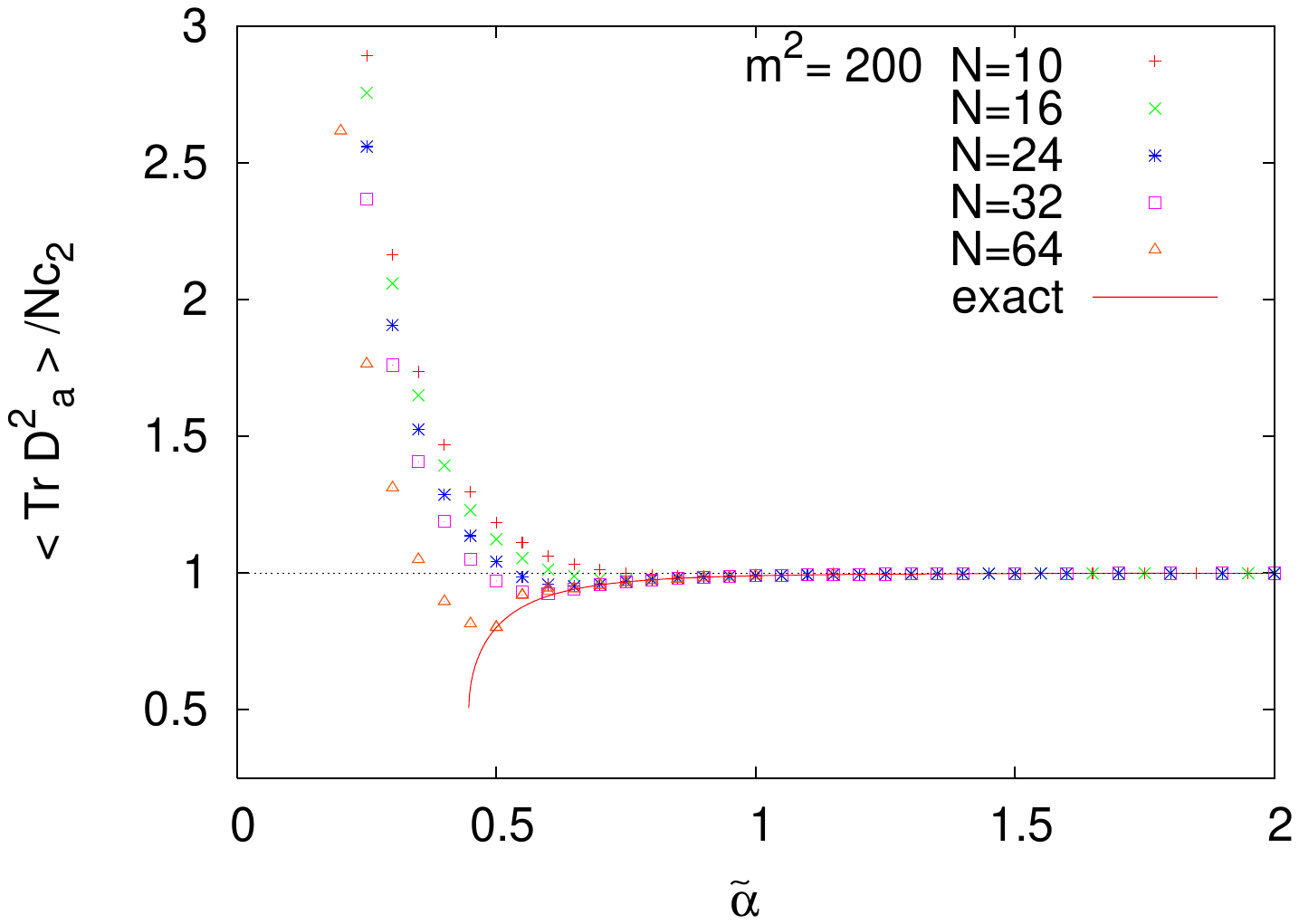}
\caption{The observed data approaches the theoretical prediction as $N$ is increased.}\label{obsm200}
\end{center}
\end{figure}

\begin{figure}
\begin{center}
\includegraphics[width=8cm]{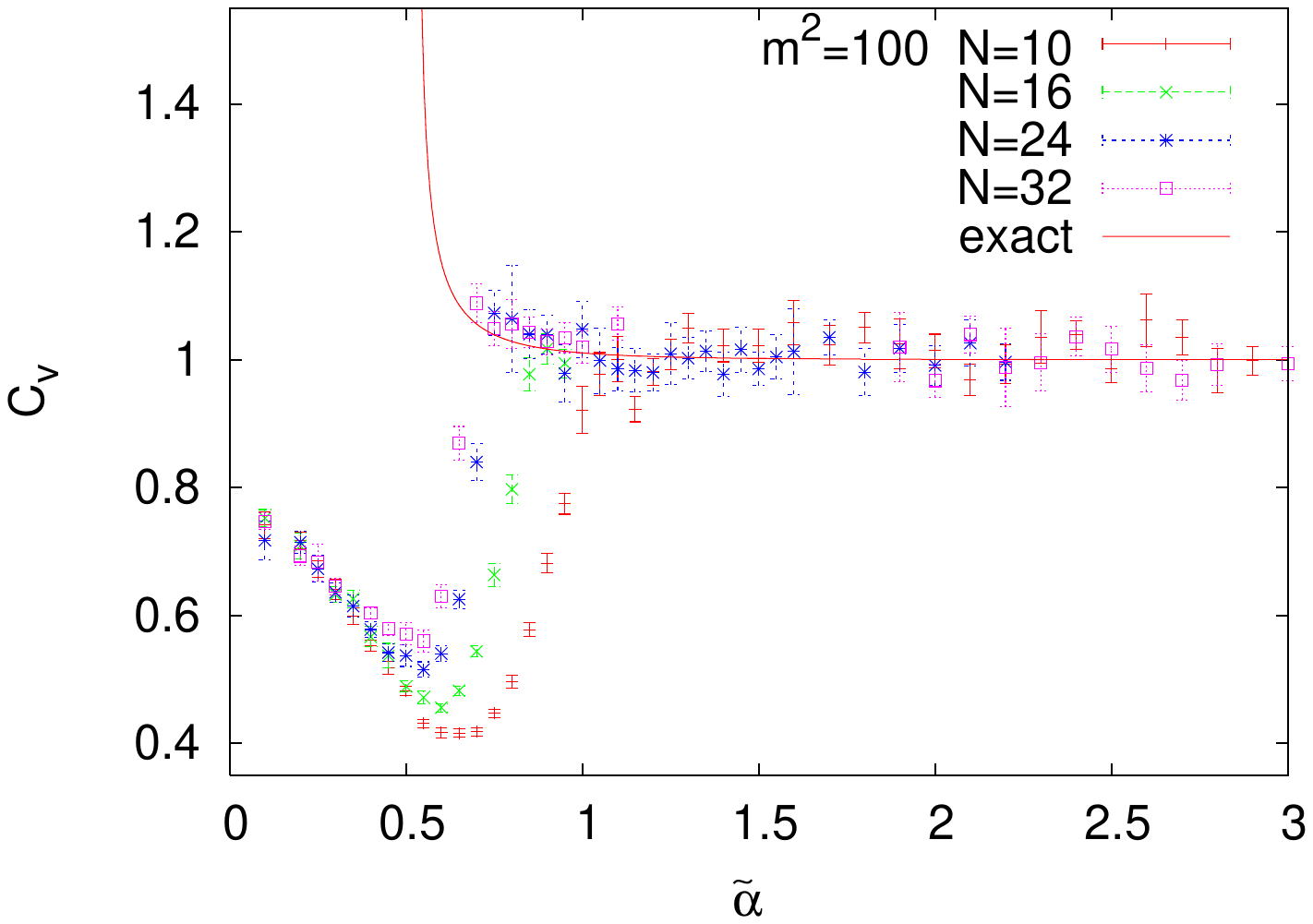}   
\includegraphics[width=8cm]{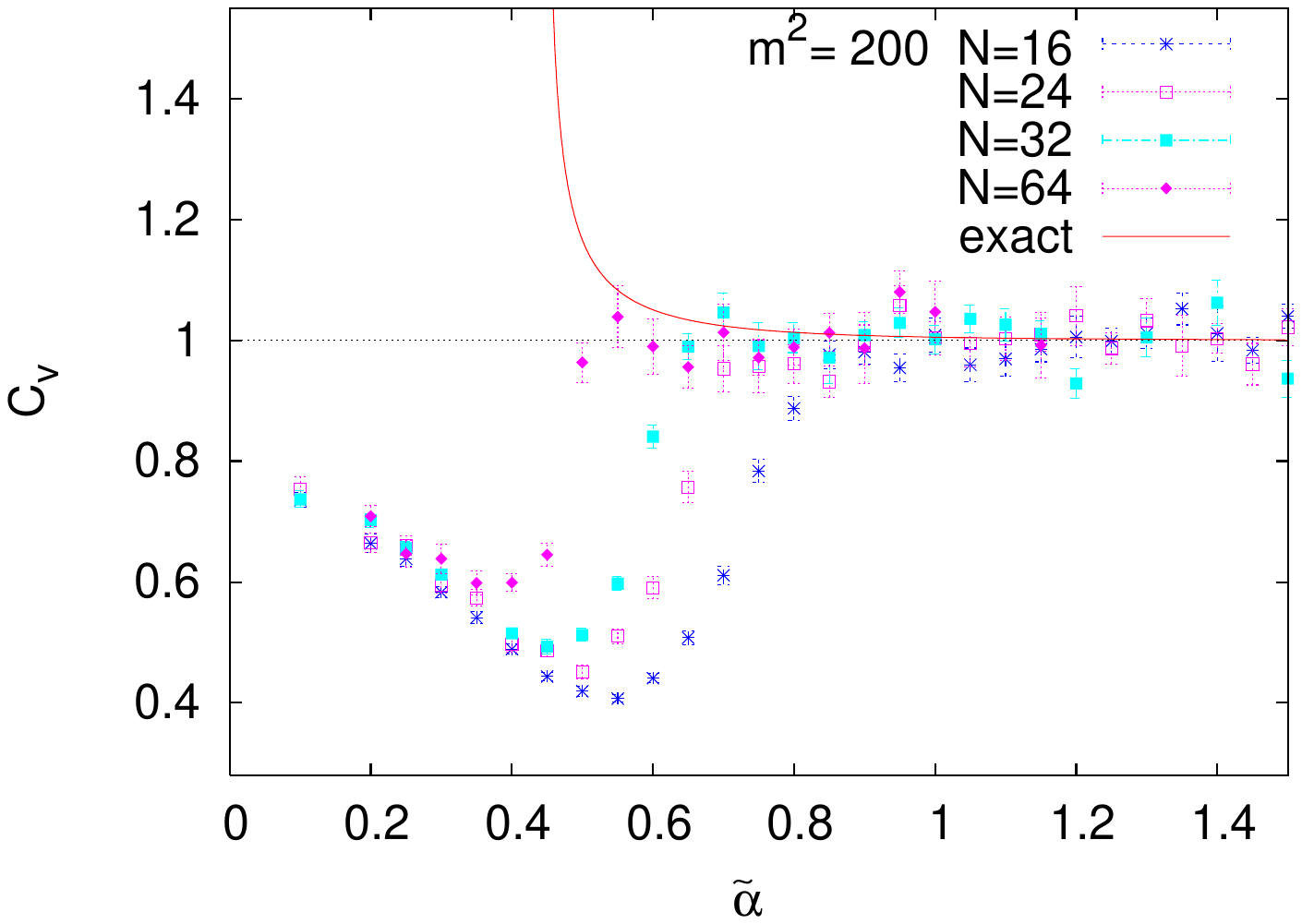} 
\caption{The specific heat of the matrix model (\ref{massivears}) at large mass.}\label{cvmlarge}
\end{center}
\end{figure}

\begin{figure}
\begin{center}
\includegraphics[width=8cm]{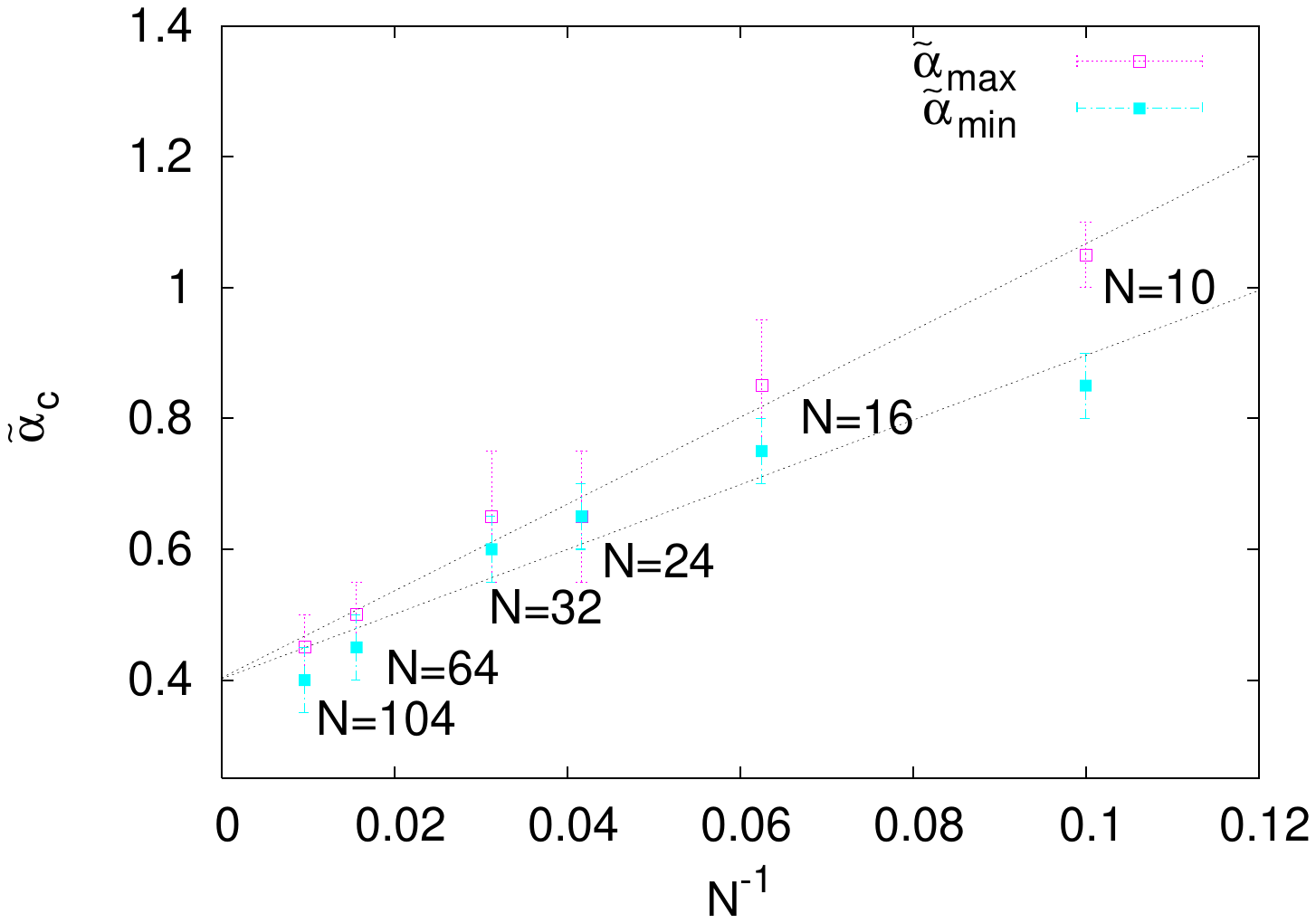}
\caption{By extrapolating the measured values of $\tilde{\alpha}_{\rm
    max}$ and $\tilde{\alpha}_{\rm min}$ to $N=\infty $ we obtain the
    critical value $\tilde{\alpha}_{c}$.}\label{figex}
\end{center}
\end{figure}

 The phase digaram is given in figure (\ref{phadia}). The Yang-Mills matrix model (\ref{massivears}) is characterized by an exotic line of
discontinuous transitions with a jump in the entropy, characteristic of a $1$st order transition, yet
with divergent critical fluctuations and a divergent specific heat with critical exponent $\alpha=1/2$.
The low temperature phase (small values of the gauge coupling constant) is a geometrical one
with gauge fields fluctuating on a round sphere. As the temperature is increased the sphere
evaporates in a transition to a pure matrix phase with no background geometrical structure.
This model presents an appealing picture of a geometrical phase emerging as the system cools
and suggests a scenario for the emergence of geometry in the early universe.

\begin{figure}
\begin{center}
\includegraphics[width=15cm]{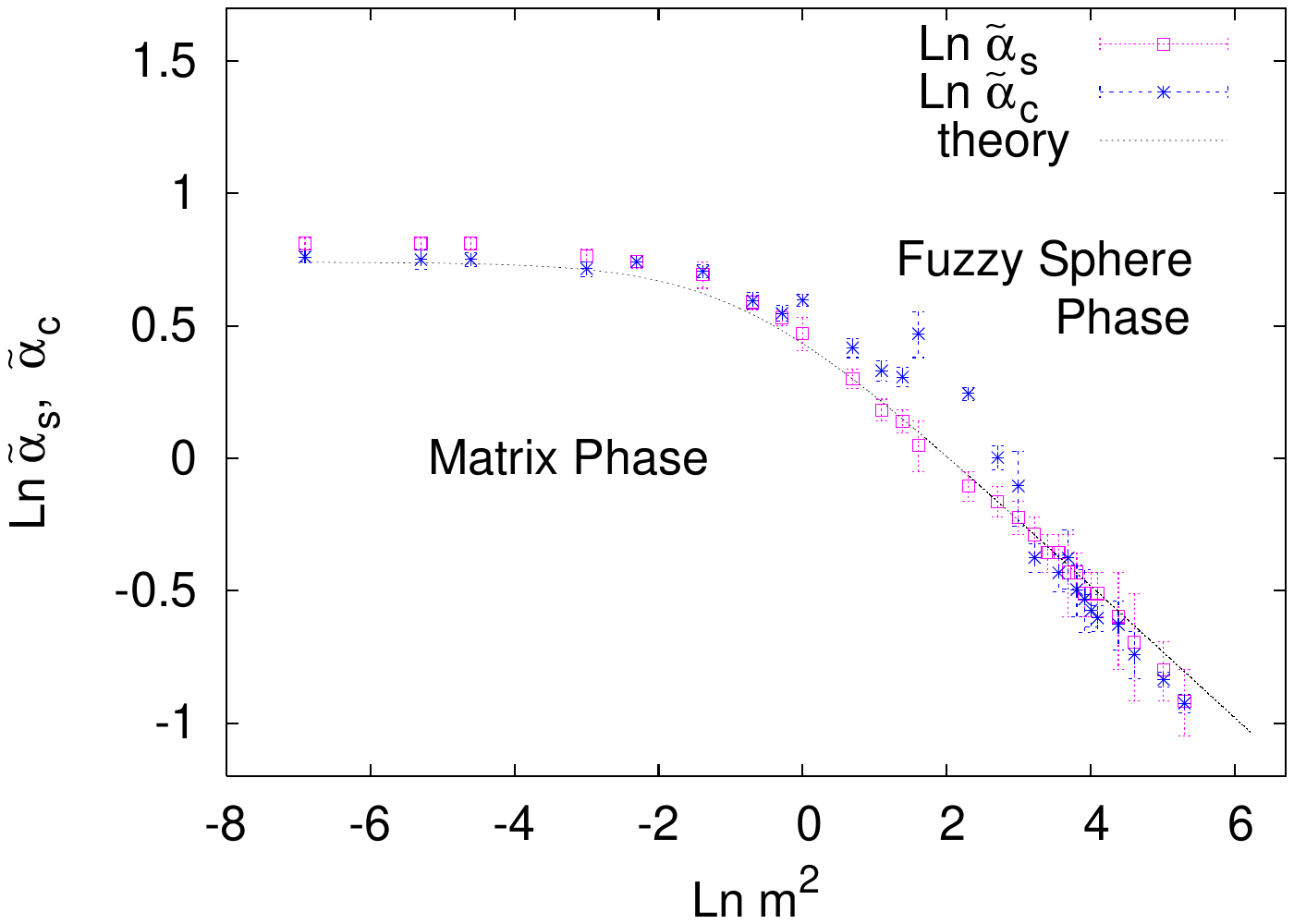}
\caption{Phase diagram showing the measured critical line (theoretical prediction given by equation (\ref{critline})) separating the geometrical
  and matrix phases of the model (\ref{massivears}).   }\label{phadia}
\end{center}
\end{figure}
\subsection{Emergent geometry and phase diagram with a cosmological term}

The Alekseev-Recknagel-Schomerus model with a  cosmological term is a $D=3$ Yang-Mills matrix model obtained by adding only a mass term ${\rm Tr}X_i^2$ to the basic  action (\ref{ars}). The action is given explicitly by (with $b\equiv \tau=\beta/\alpha^2$)
\begin{eqnarray}
  S[D]&=&\frac{1}{g^2N}{\rm Tr}\bigg[-\frac{1}{4}[D_i,D_j]^2+\frac{2i}{3}{\epsilon}_{ijk}D_iD_jD_k+\tau D_i^2\bigg].\label{action1var}
\end{eqnarray}
The effective potential in this case is give by
\begin{eqnarray}
  \frac{V_{\rm eff}(\varphi)}{2c_2}
     &=&\tilde{\alpha}^4\bigg[\frac{1}{4}\varphi^4-\frac{1}{3}\varphi^3+\frac{1}{2}\tau \varphi^2\bigg]+\log\varphi^2.\label{effe1var}
\end{eqnarray}
Here, the scalar field $\varphi$ play the role of the order parameter characterizing the phase diagram while the role of the temperature $T$ is played by the gauge coupling constants squared, i.e. $T\equiv 1/\tilde{\alpha}^4=g^2$.

The model on the fuzzy sphere is extensively studied by analytical and Monte Carlo methods for both $\tau=0$ and $\tau\neq 0$ in \cite{Azuma:2004zq,Delgadillo-Blando:2007mqd,Delgadillo-Blando:2008cuz, Azuma:2005bj,Delgadillo-Blando:2012skh}. The phase structure in this case can be summarized as follows:

\begin{itemize}
\item We start by setting the logarithmic quantum correction to zero. The classical equation of motion admits three solutions:
\begin{eqnarray}
\varphi_0=0~,~\varphi_{\pm}=\frac{1\pm\sqrt{1-4\tau}}{2}.
\end{eqnarray}
The solution $\varphi_0=0$ (the Yang-Mills or matrix phase) is the global minimum (ground state) of the system in the regime $\tau>1/4$. The solution $\varphi_-$  (the geometric or fuzzy sphere phase) is the global minimum in the regime $0<\tau<1/4$. The model has no ground state for $\tau<0$, i.e. $\beta<0$. The two global minima $D_i=0$ and $D_i=\varphi_-L_i$ are separated by a potential barrier whose maximum height is reached at the local maximum $\varphi_+$.

\item We should also mention here that the configuration $D_i=\varphi J_i$ is also a local minimum of the system. The $J_i$ are the generators of $SU(2)$ in a reducible representation  characterized by the spin quantum numbers $j_i< s=(N-1)/2$ satisfying $\sum_i(2j_i+1)=N$. More precisely, we find that the configuration  $D_i=\varphi_-L_i$ has a negative energy and thus lower than the zero energy of the configuration $D_i=0$ only in the regime $0<\tau<2/9$. This negative energy is minimized when $J_i=L_i$. In this regime the fuzzy sphere is indeed stable and the expansion of the matrix model around the fuzzy background $D_i=\varphi_-L_i$ gives a noncommutative gauge theory which also includes coupling to a normal scalar field, i.e. a noncommutative Higgs system. 

\item At $\tau=2/9$ the two configurations  $D_i=\varphi_-L_i$ and  $D_i=0$ become degenerate. Thus, in the regime $2/9<\tau<1/4$ the fuzzy sphere becomes unstable. The coexistence curve between the geometric fuzzy sphere phase and the Yang-Mills matrix phase asymptotes therefore to the line $\tau=2/9$ (and not to the line $\tau=1/4$) where the energy functional becomes a complete square.
\item If we include the logarithmic quantum correction the potential becomes unbounded from below near $\varphi=\varphi_0=0$, i.e. the effective potential   (\ref{effe1var}) is really valid only in the fuzzy sphere phase $\varphi=\varphi_-\neq 0$. But the Yang-mills phase can still be accessed by Monte Carlo simulation of the matrix model (\ref{action1var}).
\item In the quantum case the minimum $\varphi_-$ (corresponding to the geometric fuzzy sphere phase) becomes a function of both $\tau$ and $\tilde{\alpha}^4$. The critical coexistence curve exists therefore in the $(\tau,\tilde{\alpha})$ plane where the local minimum $\varphi_-$ disappears. The conditions determining this curve are obviously given by $V_{\rm eff}^{\prime}=0$ and $V_{\rm eff}^{\prime\prime}=0$. Explicitly, we obtain the curve $\tilde{\alpha}_*=\tilde{\alpha}_*(\tau)$ defined by the equations
\begin{eqnarray}
\frac{1}{\tilde{\alpha_*^4}}=\frac{\varphi_*^2(\varphi_*-2\tau)}{8}~,~\varphi_*=\frac{3}{8}(1+\sqrt{1-\frac{32\tau}{9}}).\label{ec}
\end{eqnarray}
Thus, as we increase $\tau$ from $0$ to $1/4$ the critical value $\tilde{\alpha}_*$ increases from around $2$ to infinity. Thus,  the critical temperature $T_*\equiv 1/\tilde{\alpha}_*^4=g_{S*}^2$ decreases towards zero as we increase $\tau$ to $1/4$. In other words, the geometric fuzzy sphere phase exists in the region of low temperatures $T$ (or large $\tilde{\alpha}$) and $\tau<1/4$.
\item Hence, as the temperature is increased the fuzzy sphere phase evaporates to a pure matrix phase with no background geometrical structure. In this model the geometry condenses or emerges only as the system cools.
\item These predictions, which are based on the one-loop effective potential (\ref{effe1var}), are confirmed by Monte Carlo simulation only for $\tau<2/9$. It is observed (in Monte Carlo simulation) that the coexistence curve between the geometric fuzzy sphere phase (low temperatures) and the Yang-Mills matrix phase (high temperatures) for $2/9<\tau<1/4$ asymptotes very rapidly to the line $\tau=2/9$ for $\tilde{\alpha}>\tilde{\alpha}_*=4.02$ \cite{Delgadillo-Blando:2012skh}. In other words, the region $2/9<\tau<1/4$ corresponds to the Yang-Mills matrix phase for all values of $\tilde{\alpha}$.

\item In fact for $\tau>2/9$ the geometric fuzzy sphere background is a metastable state with an observable decay to the Yang-Mills matrix background.  This decay is not observable for $\tau=2/9$ although the fuzzy sphere is not the true ground state even here.
\item In the Yang-Mills matrix phase the ground state is given by  $\varphi=\varphi_0=0$ and fluctuations are insensitive to the value of  $\tilde{\alpha}$ and are dominated by commuting matrices. In fact, in this phase the matrix model (\ref{action1var}) is dominated by the Yang-Mills term \cite{OConnor:2012vwc}.
\item More precisely, the Yang-Mills matrix phase is characterized by a joint eigenvalue distribution, for the three matrices $D_1$, $D_2$ and $D_3$, which is uniform inside a solid ball of some radius $R=2.0$ in ${\bf R}^3$. The eigenvalue distribution of a single matrix is then given by the so-called parabolic law, viz \cite{Berenstein:2008eg,Filev:2013pza,OConnor:2012vwc,Filev:2014jxa}
\begin{eqnarray}
\rho(x)=\frac{3}{4R^3}(R^2-x^2).
\end{eqnarray}
\item The transition from the geometric fuzzy sphere phase to the Yang-Mills matrix phase is of an exotic character in the sense that by crossing the coexistence curve at fixed $\tau$ from the fuzzy sphere side we encounter divergent specific heat with critical exponent equal $1/2$.  However, by crossing the coexistence curve at fixed $\tilde{\alpha}>\tilde{\alpha}_*=4.02$ we find no critical fluctuations and the transition is associated with a continuous internal energy and discontinuous specific heat.
\end{itemize}
The phase diagram is shown on figure (\ref{emerg}).

\begin{figure}[htbp]
\begin{center}
\includegraphics[width=15.0cm,angle=0]{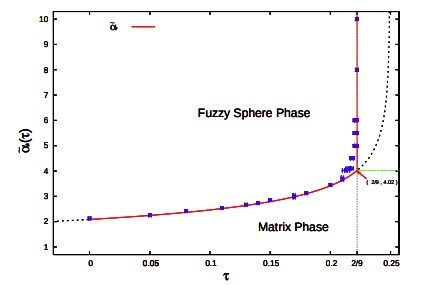}
\end{center}
\caption{The phase diagram of the Yang-Mills matrix model (\ref{action1var}).}\label{emerg}
\end{figure}
\subsection{Generalization}
Emergent geometry from Yang-Mills matrix models can be generalized to the following systems:
\begin{itemize}
  \item Noncommutative scalar field coupled to a gauge theory on the fuzzy sphere can be described by $D=4$ Yang-Mills matrix models \cite{Ydri:2014rea,Ydri:2012bq}. See also \cite{Ambjorn:2000bf,Anagnostopoulos:2005cy}.
  \item Fuzzy projective spaces ${\bf CP}^n_F$ are the most natural generalization of the fuzzy sphere ${\bf S}^2_N$ \cite{Balachandran:2001dd,Dolan:2006tx}. Fuzzy  ${\bf CP}^2_F$ is considered for example in \cite{Grosse:2004wm,Dou:2007in,Azuma:2004qe}.
  \item Fuzzy ${\bf S}^2_{N_1}\times {\bf S}^2_{N_2}$ is described by $D=6$ Yang-Mills matrix models \cite{Ydri:2016osu,Ydri:2016kua}. See also \cite{DelgadilloBlando:2006dp,Behr:2005wp,CastroVillarreal:2005uu}.
  \item Fuzzy ${\bf S}^4_F$ and ${\bf S}^3_F$ are obtained from fuzzy ${\bf CP}^3_N$ by squashing the unwanted modes \cite{Medina:2002pc,Dolan:2003kq}. As a consequence, these three spaces can all be embedded in the same Yang-Mills matrix model.
  \item The noncommutative ${\bf AdS}^2_{\theta}$ space and  noncommutative ${\bf AdS}^2_{\theta}$ black hole are described by Lorentzian $D=3$ and $D=4$ Yang-Mills matrix models respectively \cite{Bouraiou:2021tpv,Ydri:2021pcz,Ydri:2021xpw}. The Cartesian products of spheres and AdS spaces are then obtained by taking tensor products of their respective matrix models.

  \item Emergent gravity from Euclidean/Lorentzian Yang-Mills matrix models \cite{Steinacker:2010rh,Steinacker:2016vgf,Steinacker:2007dq}.
  \item Emergent time (cosmology) from Lorentzian $D=10$ Yang-Mills matrix model \cite{Kim:2011cr,Ito:2015mxa,Kim:2011ts,Kim:2012mw}.
    \item Emergent geometry, gravity and cosmology from supersymmetric Yang-Mills matrix quantum mechanics \cite{Kim:2006wg,Park:2005pz}.
\end{itemize}

\end{document}